\documentclass{aa}
\usepackage{graphicx}
\usepackage{xcolor}
\usepackage{txfonts}
\def\be{\begin{equation}}
\def\ee{\end{equation}}
\def\bea{\begin{eqnarray}}
\def\eea{\end{eqnarray}}

\begin{document} 
      \title{Correlations of r-Process Elements in Very Metal-Poor Stars as Clues to their Nucleosynthesis Sites}
      
    \titlerunning{Correlation of the r-Process Elements}

   \author{K. Farouqi
          \inst{1}, F-K. Thielemann\inst{2,3}, S. Rosswog\inst{4} 
          \and
           K.-L. Kratz\inst{5,6}}

     \authorrunning{K. Farouqi et al.}
     \institute{ZAH, Landessternwarte, University of Heidelberg,
              K\"onigstuhl 12, D-69117 Heidelberg, Germany\\
              \email{kfarouqi@lsw.uni-heidelberg.de}
    \and
     University of Basel, Department of Physics, 
	 Klingelbergstrasse 82, CH-4056 Basel, Switzerland \\
	 \email{f-k.thielemann@unibas.ch}
	 \and
      GSI Helmholtz Center for Heavy Ion Research, Planckstraße 1,
      64291 Darmstadt, Germany  
	 \and 
	  The Oskar Klein Centre, Department of Astronomy, Stockholm University, 
	  Stockholm, Sweden	\\
      \email{stephan.rosswog@astro.su.se}
	\and
	  University of Mainz, Department of Chemistry, Pharmacy \& Geosciences, 
	  D-55126 Mainz, Germany	\\
     \email{klk@uni-mainz.de}
     \and
	  Max-Planck Institut f\"ur Chemie (Otto-Hahn Institut), D-55128 Mainz, Germany
             }

   \date{Received April 9, 2021}

 
  \abstract
   {}
   {Various nucleosynthesis studies have pointed out that the r-process elements in very metal-poor (VMP) halo stars might have different origins. By means of familiar concepts {from statistics (correlations, cluster analysis, rank tests of elemental abundances)}, we look for causally correlated elemental abundance patterns and attempt to link them to astrophysical events. Some of these events produce the r-process elements jointly with iron, { while} others do not have any significant iron contribution. We try to (a) characterize these different types of events by their abundance patterns and (b) identify them among the existing set of suggested r-process sites.}
   {The Pearson and Spearman correlation coefficients are used in order to investigate the relationship { between} iron and the r-process elements (X) in VMP halo stars. We gradually track the evolution of those coefficients in terms of the element-enrichment with respect to Fe [X/Fe] and the metallicity [Fe/H]. This approach, { aided by cluster analysis to find different structures of abundance patterns and rank tests to identify whether several events contributed to the observed pattern} is new and provides deeper insights into the abundances of VMP stars.}
   {In the early stage of our Galaxy, at least three r-process nucleosynthesis sites have been active. The first two produce and eject iron and the majority of the lighter r-process elements. We assign them to two different types of core-collapse events, not identical to regular core-collapse supernovae (CCSNe), which produce only light trans-Fe elements. The third category is characterized by a strong r-process and responsible for the major fraction of the heavy main r-process elements without a significant co-production of Fe. It does not appear to be connected to CCSNe, in fact the Fe found in the related r-process enriched stars must come from previously occurring CCSNe. The existence of actinide boost stars indicates a further division among strong r-process sites. We assign these two strong r-process sites to neutron star mergers without fast black hole formation and to events where the ejecta are dominated by black hole accretion disk outflows. Indications from the lowest-metallicity stars hint at a connection to massive single stars (collapsars) forming black holes in the early Galaxy.}
   {}


   \maketitle
   \authorrunning{K. Farouqi et al.}
%

\section{Introduction}
Where were all the heavy elements, beyond iron and up to the actinides, created?
This is still one of the not clearly answered questions of modern physics. 
Since the early geochemical abundance determinations of \citet{Suess.Urey:1956},
the seminal works of \citet{Burbidge.Burbidge.ea:1957} and \citet{Cameron:1957},
and the neutron shell-structure investigations of the nuclear chemistry 
community \citep{Coryell:1953,Coryell:1961}, many efforts have been undertaken
to understand the mechanisms and the sites that forge the chemical elements.
Much progress has been made since then for the light and intermediate mass elements
\citep[e.g.][]{matteucci86,Timmes.Woosley.Weaver:1995,Kobayashi.Umeda.Nomoto.ea:2006,nomoto13,Kobayashi.Karakas.Lugaro:2020}. 
While the  pioneering papers of \citet{Burbidge.Burbidge.ea:1957} and \citet{Cameron:1957} laid out the underlying nuclear physics of the rapid neutron capture r-process, responsible for the heaviest elements in the Universe, the site was still unclear. Many years of improving nuclear input, astrophysical modeling, observational efforts, and interpreting the isotopic composition of meteoritic grains followed \citep[see e.g.][]{seeger.fowler.clayton:1965,Hillebrandt:1978,Kratz:1988,Cowan.Thielemann.Truran:1991,Kratz.Bitouzet.ea:1993,hoffman.woosley.qian:1997,Cowan.Pfeiffer.ea:1999,Freiburghaus.Rembges.ea:1999,Pellin.Davis.ea:1999,Pfeiffer.Kratz.ea:2001,Moeller.Pfeiffer.Kratz:2003,Pellin.Savina.ea:2006,arnould07,Qian.Wasserburg:2007,Kratz.Farouqi.ea:2007,Farouqi.Kratz.ea:2010,Roederer.Cowan.ea:2010,Thielemann.Arcones.ea:2011,kratz14,Hill.Christlieb.ea:2017,Ott:2017,Cowan.Sneden.ea:2021}.
Only in recent years
 a number of concrete proposals for producing the heaviest nuclei in Nature have come forward. These include neutron star mergers \citep[e.g.][]{Freiburghaus.Rosswog.Thielemann:1999,rosswog99,Just.Bauswein.ea:2015,Bauswein.Just.ea:2017,Thielemann.Eichler.ea:2017,rosswog18}, magneto-rotational jet supernovae
 \citep[e.g.][]{Winteler.Kaeppeli.ea:2012,Moesta.ea:2015,Nishimura.Sawai.ea:2017,moesta18,Reichert.Obergaulinger.ea:2021}, and collapsars \citep[e.g.][]{Siegel.Barnes.Metzger:2019,Siegel:2019}. The first site is related to stellar evolution (and explosions) in binary systems\footnote{Compact binaries can also be assembled dynamically \citep{benacquista13}, but recent studies \citep{ye20} conclude that the contribution to the overall merger rate is very small.},  while the latter two options are both related to the final collapse of  massive stars.
 The original idea was that regular core collapse supernovae could be responsible 
 for a strong r-process, i.e. reproducing solar r-process abundances \citep[e.g.][]{Woosley.Wilson.ea:1994,Takahashi.Witti.Janka:1994}, also up to the heaviest nuclei, within a high-entropy wind \citep[e.g.][]{Farouqi.Kratz.ea:2010}. Recent supernova simulations, however, do not support such high entropies and it seems that, if at all, 
 supernovae could only lead to a weak r-process, not producing the heavy r-process nuclei in solar proportions \citep{Roberts.Reddy.Shen:2012,Martinez-Pinedo.Fischer.ea:2012,Martinez-Pinedo.Fischer.Huther:2014,Wu.Fischer.ea:2014,mirizzi15,Fischer.Guo.ea:2020}. Other options for a weak r-process include 
 so-called electron capture (EC) supernovae \citep{Wanajo.Janka.Mueller:2011}
 that originate from progenitor stars in the mass range from 8-10 M$_\odot$ \citep[but see recent investigations arguing for their possible non-existence, e.g.][]{Jones.Roepke.ea:2016,Kirsebom.Jones.ea:2019}. Very recently, quark-deconfinement (QD) supernovae have  been suggested \citep{Fischer.Wu.ea:2020} as another weak r-process scenario.
 
\noindent While most of the suggested sites are based on modeling alone (possibly permitting indirect identifications in low-metallicity stars), only neutron star mergers are robustly and by direct observations of the event itself
connected to r-process production. The follow-up of the gravitational wave event GW170817 \citep{Abbott.Abbott.ea:2017} revealed strong electromagnetic emission in the aftermath of the merger \citep{kasliwal17,evans17,villar17,kilpatrick17} and showed in particular 
 the expected signatures of an r-process powered kilonova. The decay of its 
 bolometric lightcurve agreed well with the expectations for
radioactive heating rates from a broad range of r-process elements
\citep[e.g.][]{Metzger.Martinez.ea:2010,rosswog18,Zhu.Wollaeger.ea:2018,Metzger:2019}, therefore there cannot be a reasonable 
doubt that neutron star mergers are indeed a major r-process source.
 The blue emission, that was observed after one day, points to the production of a
light (lanthanide-free) r-process \citep{evans17}, while the late ($\sim$ 1 week)
red emission is the natural expectation for heavy (lanthanides and beyond)
r-process ejecta. This heavy r-process is the unavoidable result of decompressing
neutron star matter from its initial, very low  ($Y_e < 0.1$) $\beta$-equilibrium electron fraction \citep{lattimer77,Freiburghaus.Rosswog.Thielemann:1999,Korobkin.Rosswog.ea:2012} 
and it is also supported by observational evidence from late-time near-infrared observations \citep{Wu.Barnes.ea:2019,Kasliwal.Kasen.ea:2019}. The early blue emission, in turn,
shows that a substantial fraction of the ejecta has been re-processed via weak
interactions to larger $Y_e$-values which resulted in a light, lanthanide-free r-process \citep[for the variation in nucleosynthesis conditions see e.g.][]{Wanajo.Sekiguchi.ea:2014,Just.Bauswein.ea:2015,martin15,Wu.Fernandez.Martinez.ea:2016,Bauswein.Just.ea:2017,Miller.Ryan:2019}.
This is also supported by the identification of the light r-process element
strontium \citep{Watson.Hansen.ea:2019}. In summary, there is strong evidence that this
neutron star merger event has produced at least a broad, and maybe the whole,
r-process range. However, based on the observed lanthanide fraction $X_{La}$, \citet{Ji.Drout.Hansen:2019} find that at least for the neutron star merger GW170817 this does not represent a typical solar r-process pattern.
 
 \noindent Observations of low metallicity stars indicate the existence of a weak r-process site \citep[see e.g.][]{honda06,Honda.Aoki.ea:2007,Hansen.Primas.ea:2012}, while most r-process enhanced stars show a solar r-process pattern \citep[e.g.][]{sneden08,Hansen.Holmbeck.ea:2018}. This goes together with a variation of e.g. the Sr/Eu ratio, ranging from about 1120 down to 0.5 \citep{Hansen.Holmbeck.ea:2018}, and indicating the different decline of the abundance curve as a function of $A$. { \citep[This led to suggestions for the contribution from a different r-process site, explaining abundance features of light r-process elements like Sr, Y, Zr, Mo, Ru, Ag, Pd, see e.g.][] {Cowan.Pfeiffer.ea:1999,Kratz.Farouqi.ea:2007,Montes.Beers.ea:2007,Qian.Wasserburg:2007,Farouqi.Kratz.Pfeifer:2009,Hansen.Primas:2011,Hansen.Andersen.ea:2014,Hansen.Montes.ea:2014,Mishenina.Pignatari.ea:2019}}. Some of the r-process enriched stars show an "actinide boost", i.e. their Th or U to Eu ratio is supersolar \citep[e.g.][]{Roederer.Cowan.ea:2010,Holmbeck.Beers.ea:2018,Holmbeck19a}. If interpreted as having been born with a solar-type r-process pattern, their age determination would lead to absurd results \citep{Cowan.Pfeiffer.ea:1999,Schatz.ea:2002,hill02,Kratz2004,Roederer.Kratz.ea:2009,Hayek.Wiesendahl.ea:2009,Mashonkina:2014,Hill.Christlieb.ea:2017, Holmbeck19a}. Another result from the observation of low-metallicity stars is that especially Eu, with reasonably easy to detect spectroscopic features, shows a large scatter in comparison to Fe, much larger than the alpha elements (from O to Ti) which go back to core-collapse supernova nucleosynthesis
 \citep[e.g.][]{Hansen.Holmbeck.ea:2018}. This points to the strong r-process being a very rare event, occurring with a frequency smaller than that of supernovae by a factor of 100 to 1000. This is also underlined by the detection/non-detection of $^{244}$Pu in deep-sea sediments \citep[e.g.][]{Wallner.Faestermann.ea:2015,hotokezaka15}. 
 
 \noindent Summarizing the discussion above: we have a number of suggested r-process sites, but only one of them is proven by a direct observation of the explosive event. Observations of low metallicity stars show essentially three types of patterns, a weak or limited r-process, a strong solar-type r-process, and an actinide-boosted r-process (in some publications also referred to as weak, main, and strong r-process). Whether the latter two types are produced in different sites or result from variations within the same site  (e.g. neutron star mergers) is still debated. The question is now how such observations can point back to the r-process sites, and whether it is possible to identify  features which can provide additional insight. A promising approach is to look for correlations among different elements, which might directly identify the nucleosynthesis of a specific site \citep[see e.g.][]{Barklem.Christlieb.ea:2005,Francois.Depagne.ea:2007,Mashonkina.Vinogradova.ea:2007,Kratz.Farouqi.ea:2008}. \citet{Cowan.Sneden.ea:2005} compared the abundances of Fe, Ge, Zr, and
r-process Eu in low metallicity stars. They found a strong correlation of Ge with Fe, indicating
the same nucleosynthesis origin (core-collapse supernovae), a weak correlation of Zr with Fe,
indicating that other sites than regular core-collapse supernovae (without or low Fe-ejection) contribute as
well, and no correlation between Eu and Fe, pointing essentially to a pure r-process origin
with negligible Fe-ejection. More recent data from the SAGA and JINA databases \citep{Sagadatabase,JINAbase:2018} permit a correlation between Eu and Fe for
[Eu/Fe]$<$0.3, i.e. for stars with lower than average r-process enrichment. Interpreted in a straight-forward way
this would point to a negligible Fe/Eu ratio (in comparison to solar ratios) in the major r-process sources, while a noticeable co-production
of Fe with Eu is possible in less strong r-process sources, e.g. with a weak r-process. Such cases could again be identified with the limited-r entry in observations \citep{Hansen.Holmbeck.ea:2018}. 

\noindent In the upcoming sections we will concentrate on studying correlations among different r-process elements, as well as in relation to Fe (a supernova product), with the aim to obtain additional indicators for  the responsible r-process sites. In the appendices \ref{statistics}, \ref{appsinglemultible}, and \ref{kmeans} we briefly 
summarize basic statistical concepts, such as the Pearson and Spearman correlation coefficients, the coefficients of determination of a linear regressions, the effects of superpositions of data sources { via rank tests} that we use to analyze such correlations, and tests for a clustering of correlations. Sections \ref{application1} to \ref{corrTh} will apply these tools to low metallicity star observations from light trans-Fe elements via the lanthanides to elements of the third r-process peak and up to actinides. In section \ref{interpr} we summarize these findings and attempt to link the observational features to the suggested sites. We further
use  statistical tools  to estimate the frequencies of these individual event sites, before presenting our conclusions in section \ref{concl}.

\section{Observed r-Process Abundance Patterns in Very Metal-Poor Halo Stars and the Use of Statistical Methods to Test for Correlations}
\label{application1}
\subsection{The Variety of r-process abundance patterns in low-metallicity stars} 
\label{sec:2.1}
In this paper we want to investigate whether and how the abundances of different elements in very metal-poor (VMP) and extremely metal-poor (EMP) halo stars can point back towards their originating astrophysical site. Here, we compare abundance patterns, similar attempts with the aim to identify key components/nucleosynthesis sites contributing to galactic evolution have been undertaken before with different methods \citep[e.g.][]{Ting.Freeman.ea:2012}. In addition, we want to use correlations between elements to interpret whether they originate from identical or different sources.   
But before addressing these questions with  statistical methods 
we want to first have a look at these patterns and identify elemental features which can help in these investigations. Over the past decades, thousands of VMP (and EMP) halo stars with [Fe/H]$\leq -2$ have been detected in the galactic halo and several dwarf galaxies by a number of large-scale surveys, e.g. a series of papers resulting from the HK survey \citep[][and references therein]{Bonifacio.Monai.Beers:2000,Roederer.Preston.ea:2014}, the Hamburg/ESO survey \citep[][and references therein]{Hill.Christlieb.ea:2017}, up to the r-Process Alliance surveys \citep{Hansen.Holmbeck.ea:2018,Sakari.Placco.ea:2018,Ezzedine.Rasmussen.ea:2020,Holmbeck.Hansen.ea:2020}.
For the further discussion it is important to consider that the so-called solar r-process abundances, obtained by subtracting solar s-process abundances from solar abundances
\citep[see e.g.][]{arlandini99,Kaeppeler.Gallino.ea:2011,Prantzos.Abia.ea:2020}, might combine a number of different contributions. Especially the light trans-Fe elements like Sr, Y, Zr, seem to have also other origins aside from the typical r-process \citep{Travaglio.Gallino.ea:2004,Froehlich.Martinez-Pinedo.ea:2006,Farouqi.Kratz.Pfeifer:2009,Hansen.Primas.ea:2012,Hansen.Andersen.ea:2014,Eichler.Nakamura.ea:2018,Akram.Farouqi.ea:2020} which could possibly be attributed to regular core-collapse supernovae. When looking at the SAGA database (only for Milky Way stars) in this metallicity window, one recognizes that one finds Fe and Sr detections in 1277 of them, but Fe and Eu detections only in 498 stars (combined with a strong scatter in the Eu abundances). While this is not a proof that Eu could not have been detected in all stars, it is { an} indication that an r-process, producing the typical r-process element Eu { in detectable quantities}, is less frequent than the majority of light trans-Fe element producing events {(including Sr)}. However, from GW170817 we know that strong r-process events like neutron star mergers produce { light r-process elements like} Sr as well \citep{Watson.Hansen.ea:2019}.

\noindent For stars with a clear r-process contribution the [Eu/Fe] ratio is an important indicator.
\citet{Hansen.Holmbeck.ea:2018} and \citet{Roederer.Preston.ea:2014} introduced different ranges,  [Eu/Fe]$<$0.3 for so-called incomplete or limited-r stars, which show apparently a weak r-process (we will discuss this below, also whether an upper limit close to [Eu/Fe]=0-0.1 is more appropriate). Furthermore so-called complete r-process stars (also called r-rich) with subdivisions in r-I (0.3$<$[Eu/Fe]$<$1) and r-II ([Eu/Fe]$>$1)
were introduced with these different r-process enrichments.
One very striking aspect of the incomplete stars is the { apparent} observational absence of the third r-process peak elements (Z>72), including also Pb and the actinides like Th and U. Another feature of those stars is the gradual depletion of the elements beyond Zr (Z=40). The ratio of Sr and Eu in those stars is at least ten times higher than in the complete stars, whereas the ratio of Sr, Y and Zr among each other is roughly the same in both categories \citep{Mashonkina.Vinogradova.ea:2007,Kratz.Farouqi.ea:2008}. This behavior will later be discussed in more detail in comparison to the complete r-process stars. 

\begin{figure*}[h!]
\centerline{
\includegraphics[width=10cm]{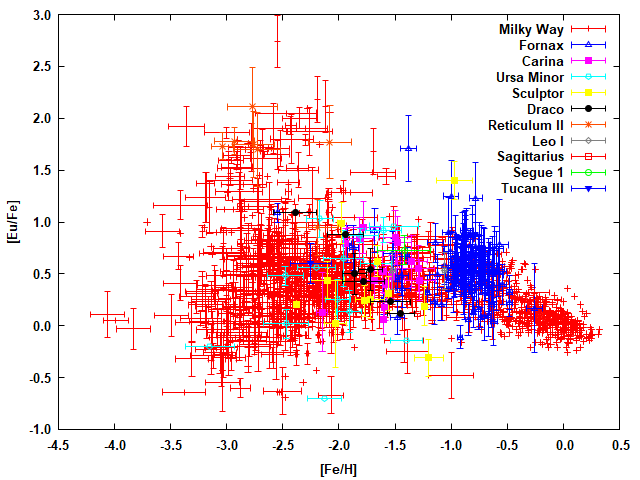}
\includegraphics[width=10cm]{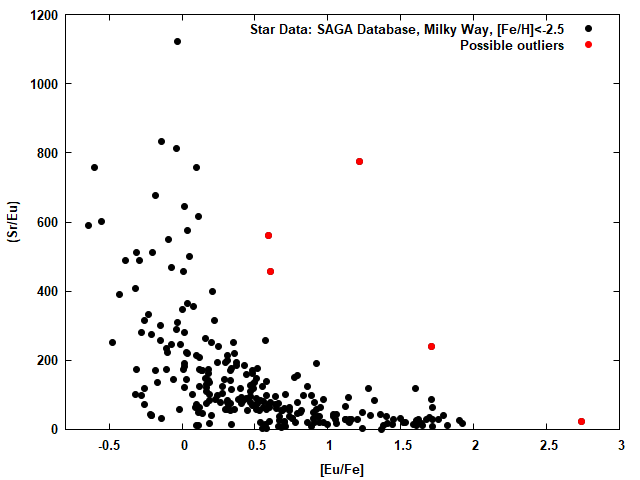}
}
\caption{{ Left panel:} [Eu/Fe] ratios for all 1572 stars with Eu detections from the SAGA Database as a function of metallicity [Fe/H]. An enormous scatter can be recognized at low metallicities before at about [Fe/H]$\approx$-2 an averaging effect sets in. { Right panel:} individual linear Sr/Eu ratios for all stars with [Fe/H]$<$-2.5 as a function of [Eu/Fe]. The change from very high ratios for limited-r stars to low values for complete r-process stars is extreme. One might even recognize two groups within the limited-r stars.}
\label{fig:fullEuFe_SrEuFe}
\end{figure*}

\noindent The whole sample of observed [Eu/Fe] values is plotted in Fig.~\ref{fig:fullEuFe_SrEuFe} (left panel) based on the SAGA database \citep{Sagadatabase}. We see at low metallicities a huge scatter. This has been discussed widely as an indication for a rare r-process site \citep[e.g.][]{wehmeyer15,cescutti15,Haynes.Kobayashi:2019,VandeVoort.ea:2020} with a highly efficient ejection of r-process matter in order to reproduce in total the overall solar r-process abundances, which determine the product of ejected mass and event frequency \citep[e.g][]{hotokezaka15,Rosswog.Feindt.ea:2017}. We will see below that this argument applies surely for the strong or complete r-process stars (r-I and r-II), while the limited-r stars { could} go back to more frequent and less efficient events \citep[for a detailed discussion of these considerations see e.g.][]{Hansen.Primas.ea:2012,Cowan.Sneden.ea:2021}. However, the situation is likely more complex than just a division in limited-r and very rare complete r-process events. If all strong r-process events (with an assumed identical r-process abundance pattern) occur via an inhomogeneous distribution of newly ejected matter from rare events with a low spatial density, a large scatter of [Eu/Fe] is expected in the early Galaxy, because { - opposite to Eu - Fe is mostly due to more} frequent core-collapse supernovae { causing a continuous and steady Fe enrichment}. 

\noindent { The early evolution of the Galaxy has been discussed by a number of authors \citep[see e.g.][]{Audouze.Silk:1995,McWilliam.ea:1995a,McWilliam.ea:1995b} pointing to the breakdown of the instantaneous mixing approximation at [Fe/H]$\approx$-2.5 when the imprint of individual explosive events is seen. \citet{Ryan.Norris.Beers:1996} provide a formula for the amount of swept up ISM by the Sedov blast wave of one supernova with the typical explosion energy of 1 Bethe and an approximate ejected Fe mass of 0.1M$_\odot$, being of the order $7\times 10^4$M$_\odot$. They conclude that in a pristine interstellar medium this leads to a metallicity of [Fe/H]=-2.7, below which one would expect to see abundance patterns
and ratios of two elements stemming from individual events. This pollution approach has also been utilized in simple inhomogeneous galactic evolution modeling \citep[e.g.][]{Argast.Samland.ea:2004,wehmeyer15}. The question remains how much matter of such a remnant will enter 
 the next stellar generation. It can vary from a large fraction in star formation triggered by a nearby supernova or the contribution can also be strongly diluted (after turbulent mixing, triggered  e.g. by galactic spiral arm motion) down to a few percent or even less. Therefore one expects lowest-metallicity stars, affected by these first CCSNe, to possess values down from [Fe/H]$\approx$-3 to about [Fe/H]$\approx$-5 or even less \citep[e.g.][]{Norris.Christlieb.ea:2007,Norris.Christlieb.ea:2012,Frebel.Norris:2015}. In case the latter behavior dominates, one could speculate that events which occur with 1/10 of the frequency of regular CCSNe might first show up at [Fe/H]$\approx$-4 (after typically 10 supernovae enriched the ISM) and events which take only place after about 100 regular CCSNe polluted the ISM are expected to show their impact at [Fe/H]$\approx$-3. This could be consistent with the observations shown in Fig.~\ref{fig:fullEuFe_SrEuFe} for weak and strong r-process events.}  

\noindent { After the discussion in the previous paragraph, we will therefore, in the further analysis, concentrate on the low-metallicity stars with [Fe/H]$<$-2.5, with the initial hypothesis that they experienced probably only one prior r-process "pollution". If only one event contributed to the heavy r-process pattern, one can thus identify correlations as co-production of the observed elements in the same site. Whether this hypothesis can be kept throughout our analysis will be tested in a continuous fashion.} 
What can be realized is that in addition to the observed scatter at low metallicities, there seems to exist also a change of abundance patterns as a function of [Eu/Fe], as seen in Table~\ref{tab:SrEu} which provides the average Sr/Eu ratio as a function of observed [Eu/Fe]. { From} limited-r stars to complete r-process stars there exists a drastic change in the Sr/Eu ratio, indicating a real change in the r-process strength.

\noindent This important indicator for the strength of the r-process is also shown in Fig.~\ref{fig:fullEuFe_SrEuFe} (right panel) for individual stars (with metallicity [Fe/H]$<$-2.5). Table~\ref{tab:SrEu} shows a strong division between limited-r stars 
and complete r-process stars, a ratio of Sr/Eu$\geq$300 is found in stars with [Eu/Fe] up to 0-0.1, a ratio $>$150 up to 0.3 (coinciding with early
definitions of the above mentioned division between limited-r and r-I stars), a ratio of $>$100 up to 0.6, and a ratio $>$60 up to 0.8 (close to the previously mentioned division between r-I and r-II-stars). Given the strong change of Sr/Eu from limited-r to complete r-process stars and the additional fact that in the limited-r stars no elements from the third r-process peak as well as no actinides are { detected, i.e. if existing, their abundances are below the detection limit}, this underlines that the weak r-process pattern requires different processing conditions and argues for a different stellar site { in comparison to} the pattern in complete r-process stars. That for the lowest metallicities in Fig.~\ref{fig:fullEuFe_SrEuFe} (left panel) one finds low [Eu/Fe]$<$0.1, i.e. weak r-process stars, also { could} indicate that these events { might occur} already earlier in galactic evolution. When taking a careful look at Fig.~\ref{fig:fullEuFe_SrEuFe} (right panel) even two subgroups of limited-r stars could { possibly} be identified. For the complete r-process stars only a gradual decline can be observed. It indicates a continuous, but not that drastic, change in r-process strength, and whether the division in r-I and r-II stars is actually related to different sites or only a range of possible conditions in the same site seems not that clear.

\begin{figure*}[h!]
\centerline{
\includegraphics[width=10cm]{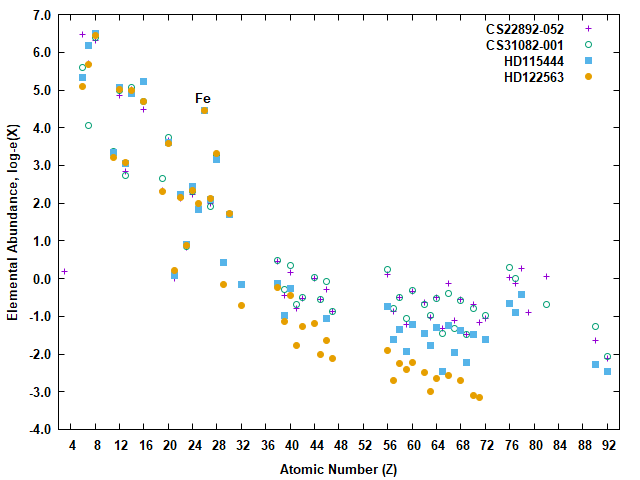}
\includegraphics[width=10cm]{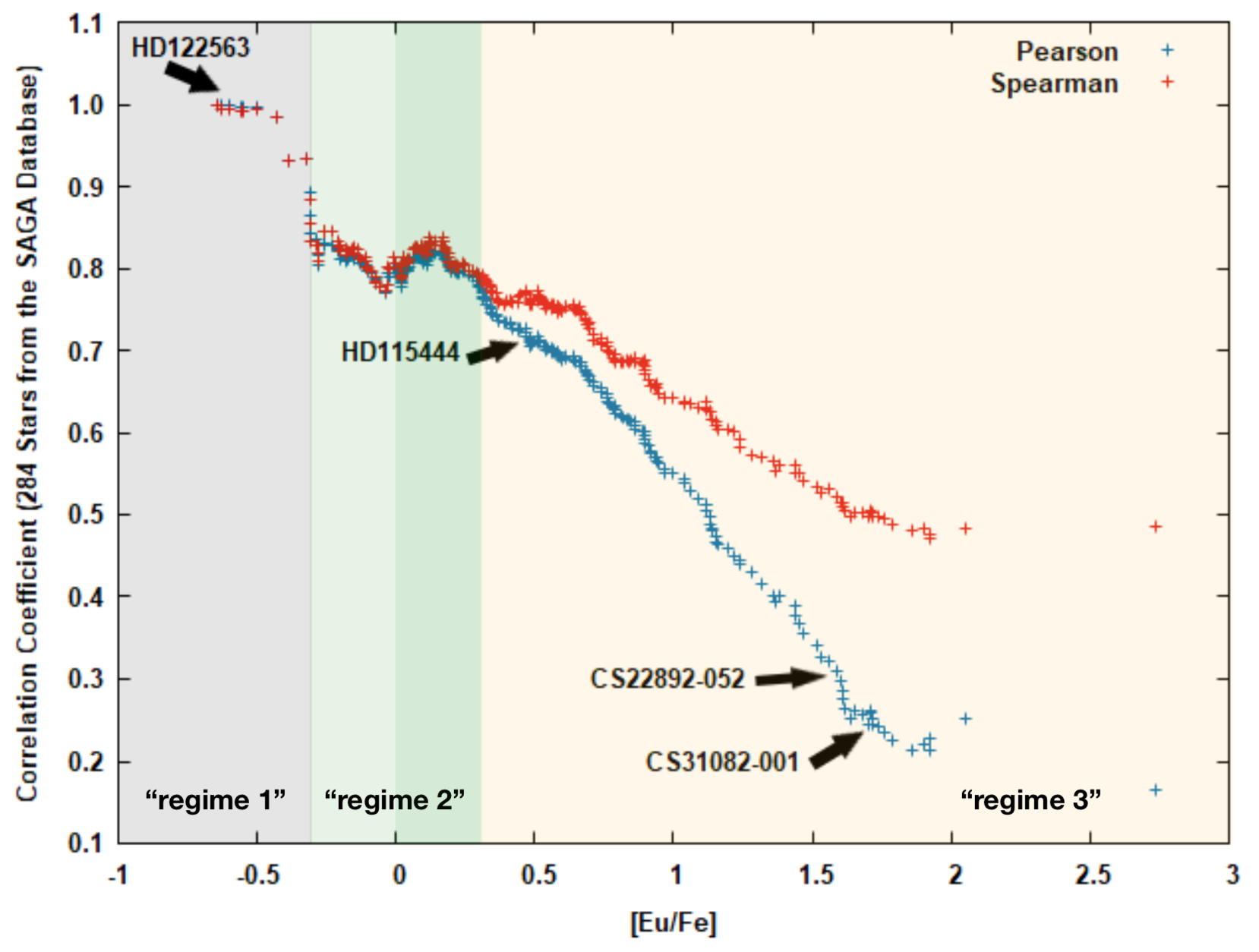}
}
\caption{{ Left panel}: Abundances of the four standard stars: HD122563 ([Eu/Fe]$\simeq -0.64$, limited-r), HD11544 ([Eu/Fe]$\simeq 0.68$, r-I), CS22892-052 ([Eu/Fe]$\simeq 1.53$, r-II) and CS31082-001 ([Eu/Fe]$\simeq 1.62$, r-II and actinide boost), from the JINAbase database and normalized to the absolute abundance of Fe from CS22892-052. log$\epsilon$ of element X is defined via its ratio with respect to the hydrogen abundance (log$\epsilon$$(X)={\rm log}{(N_{X}/{N_H})} +12$). { Right panel}: The Pearson and Spearman correlation coefficients of Fe and Eu in stars with $\mbox{[Fe/H]}\le -2.5$ as a function of the upper limit for the [Eu/Fe] interval utilized. At $\mbox{[Eu/Fe]}\simeq 0.3$ the Pearson and Spearman coefficients start diverging from each other. The position of the four typical stars HD122563 (incomplete, r-poor), HD115444 (complete, r-I), CS22892-052 (complete, r-II) and CS31082-001 (complete, r-II and actinide boost) are indicated.}
\label{4starsplot}
\end{figure*}

\noindent Before going into further details about the variations observed in elemental r-process abundances in very metal-poor stars, we present an overview plot in Fig.\ref{4starsplot} (left panel), showing the abundance pattern of four "typical" stars: an incomplete (also termed r-poor or limited-r) star, and three r-enriched stars of different magnitude (with the above definitions including one r-I and two r-II stars, the second also with an actinide boost).
The figure is normalized to Fe. It is interesting to see that, with the exception of the lighter elements up to N, they show a very similar abundance pattern up to Fe and Ge, 
before diverging strongly for the heavy elements. While the different behavior of the limited-r star in comparison to the complete stars is striking, we can also recognize variations in the r-process strength for the three complete r-process stars. In the remaining paper we will discuss the possible origins of the observed abundance patterns. In the following subsection we will first introduce statistical correlations and concentrate on the correlations of Fe with Eu before extending this analysis across the nuclear chart.

\subsection{A first look at correlations}
In order to understand the difference between the variety of observed abundance patterns we track the linear correlations between two arbitrarily chosen chemical elements X and Y. In appendix \ref{statistics} we describe how to determine Pearson and Spearman correlation coefficients (PCCs and SCCs). They both range from (a) negative values (-1), over (b) 0, to (c) positive values (+1). Case (a) describes a strong anticorrelation, i.e. decreasing values combined with increasing values of either X or Y, (c) stands for a strong correlation, i.e. changes with the same sign for variables X and Y, and a value of (close to) 0  indicates that there is (essentially) no correlation (case b). A The PCC tests a linear relationship of both elements, i.e. a positive value indicates that a straight line with positive slope can be plotted through the data points. A negative value indicates a negative slope. A small absolute value indicates a large scatter of the data points around this straight line, an absolute value of 1 indicates that all data points are located on this line, i.e. a perfect linear relationship. { This means that for +1 the ratio for the two element abundances X and Y is constant, arguing for the pollution by a nucleosynthesis site with these fixed abundance ratios. An overall variation, scaling X and Y, however keeping the same ratio, might just indicate a different amount of pollution by that specific site.} The SCC tests only if there is a monotonic change (increase or decrease) among both elements, which does not have to be linear. It is also based on ranks (the integer numbers in the sequence of increasing values), which judges absolute distances of a data point from a fit in a milder fashion. Therefore, SCCs display typically a better correlation than PCCs. Such correlation tests can provide a deeper insight, in general or e.g. within selected metallicity ranges [Fe/H] or element enrichment intervals [X/Fe], if one wants to analyze e.g. the relation of element X with Fe at different metallicities. It is especially helpful, when analyzing such correlations at very low metallicities, because for these conditions we expect that the nucleosynthesis additions to protostellar clouds (leading to the abundances with which a star is born and which are preserved on its surface) go 
only back to one or very few earlier events
{ (see the discussion in \ref{sec:2.1})}. Such a selection has the advantage that one can find correlations within the abundance patterns which point directly to single contributing sites.
As an example we choose Fe and Eu and calculate the PCCs for different [Fe/H] and [Eu/Fe] ranges. Selected results are given in Tables ~\ref{tab:pcc0}, \ref{tab:pcc1}, and \ref{tab:pcc2}, which also indicate the strengths and weaknesses of utilizing correlations and how to apply them intelligently. \\

\begin{table}
\caption{Available star data from the SAGA database in selected [Eu/Fe]-ranges and their mean of the corresponding (Sr/Eu)-ratios}
\label{tab:SrEu}
\begin{tabular}{rrcc}
\hline
[Eu/Fe] & -range   & \#stars & mean   \\
from   &  to &          & (Sr/Eu) \\
\hline
-0.64 & -0.44 &  4  & 550 \\
-0.43 & -0.23 &  13 & 291 \\
-0.22 & -0.02 &  24 & 331 \\
-0.01 &  0.19 &  45 & 212 \\
 0.20 &  0.40 &  34 & 144 \\
 0.41 &  0.61 &  42 & 114 \\ 
 0.62 &  0.82 &  28 & 124 \\
 0.83 &  1.03 &  18 &  59 \\
 1.04 &  1.24 &  17 &  79 \\
 1.25 &  1.45 &  10 &  40 \\
 1.46 &  1.66 &  11 &  29 \\             
 1.67 &  1.92 &  11 &  34 \\
 \hline
\end{tabular}
\end{table}

\begin{table}
\caption{Selected PCCs for Fe \& Eu  in different [Eu/Fe]  ranges for stars with [Fe/H] $\le$ -2.5.}
\label{tab:pcc0}
\begin{tabular}{rrcr}
\hline
[Eu/Fe] $<$      & PCC & assoc. strength & \#stars\\
\hline
-0.50 & 0.99 &  very strong & 7   \\
 0.00 & 0.88 &  very strong & 52  \\
 0.30 & 0.81 &  strong      & 116 \\
 1.00 & 0.55 &  moderate    & 230 \\
 1.50 & 0.35 &  weak        & 258\\
 2.70 & 0.14 &  very weak   & 282\\
 \hline
\end{tabular}
\end{table}

\noindent In Table~\ref{tab:pcc0} we analyse the (Pearson) correlation between Fe and Eu for all stars with low metallicities ([Fe/H]$<$-2.5), dependent on intervals of [Eu/Fe] bounded by a given upper value but open to low values. It seems that we first see a very high correlation which decreases continuously down to a value of 0.14, i.e. a very weak/negligible (or essentially no) correlation. How can this behavior be explained?
If we have a look at Fig.~\ref{fig:fullEuFe_SrEuFe}
(left panel) one can see that choosing initially quite low upper limits for the [Eu/Fe] interval means that only a small range of Eu/Fe ratios is considered, as there exist essentially no stars with [Eu/Fe]$<$-0.65. This results in a narrow range of Eu/Fe ratios for stars in this selected [Eu/Fe]-interval and therefore an almost linear relationship between these two elemental abundances. If one increases the upper [Eu/Fe]-limit, the interval considered becomes continuously larger, up to the point where the full scatter of more than three orders of magnitude is covered (for the same [Fe/H]-range $<$-2.5). This means that essentially no correlation is found, consistent with the old finding that Eu and Fe are not correlated and the main r-process site produces r-process elements efficiently but no or only negligible amounts of Fe. 

\noindent Only if we have a reasonable argument that different subgroups can be considered separately, we can treat these individually. This is e.g. the case for limited-r stars ([Eu/Fe]$<$0)
with the very high and quite different Sr/Eu ratios as well as 
{ non-detected} third r-process peak elements. This behavior points to a distinct astrophysical site and we can analyze their Eu and Fe correlations separately. Table~\ref{tab:pcc1}, indicates a strong correlation of Eu and Fe, i.e. a co-production of Eu and Fe in limited-r stars (and when considering Fig.~\ref{fig:fullEuFe_SrEuFe}, right panel, one can even argue for a subdivision among stars below and above [Eu/Fe]=-0.3). We also provide correlation values for the [Eu/Fe]-intervals of r-I, r-II, and r-I \& r-II stars. However, beyond the kind of schematic interval division for r-I, r-II (and even r-III) stars, it is still not clear whether there exist convincing arguments that they are related to separate sites, rather than being variations of r-process strength in a typical strong r-process site. Thus, for the moment, we regard the separate table entries for r-I and r-II  as not conclusive and treat the whole set of r-process enriched stars as a site which is essentially not correlated with Fe production. 

\begin{table}
\caption{Selected PCCs for Fe \& Eu in different [Eu/Fe] ranges for stars with $\mbox{[Fe/H]}\le-2.5$.}
\label{tab:pcc1}
\begin{tabular}{rrcr}
\hline
 star type                    &  PCC & assoc. strength &  \#stars\\
 \hline
lim.-r [Eu/Fe]$<$0 .0  &  0.88  &  very  strong      &  52  \\
lim.-r [Eu/Fe]$<$0.3   &  0.81  &  very strong       & 116  \\
r-I                               &  0.65  & strong                & 115 \\
r-II                              &   0.34 & moderate           & 51\\
r-I \& r-II                     &  0.15  & very weak          & 166\\
all stars                      & 0.14   &  very weak         & 282\\ 
\hline
\end{tabular}
\end{table}
\hspace*{0.48cm}
\begin{table}
\caption{Selected PCCs of Fe \& Eu in different [Fe/H] ranges (for all [Eu/Fe]-values).}
\label{tab:pcc2}
\begin{tabular}{rrcr}
\hline
[Fe/H] $<$   & PCC & assoc. strength & \#stars\\
\hline
-2.8 & 0.17 &  very weak &  109 \\
-2.5 & 0.22 & weak          &213 \\
-2.0 & 0.22 & weak          & 362\\
-1.0 & 0.57 & moderate   & 552\\
0.0  & 0.84 & very strong & 1549\\
\hline
\end{tabular}
\end{table}
\begin{figure*}[h!]
\centerline{
\includegraphics[width=9cm]{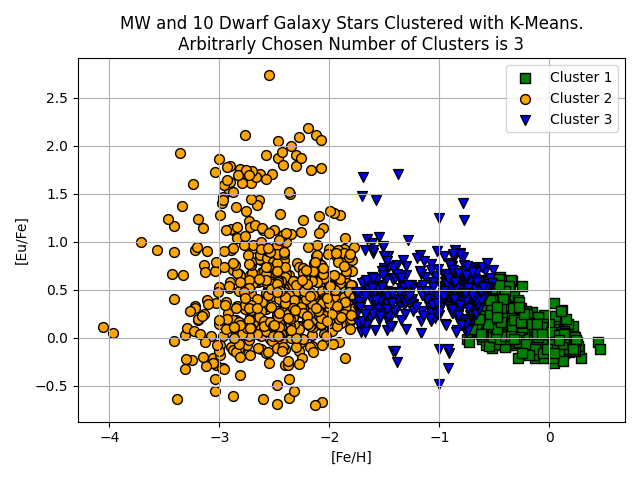}
\includegraphics[width=9cm]{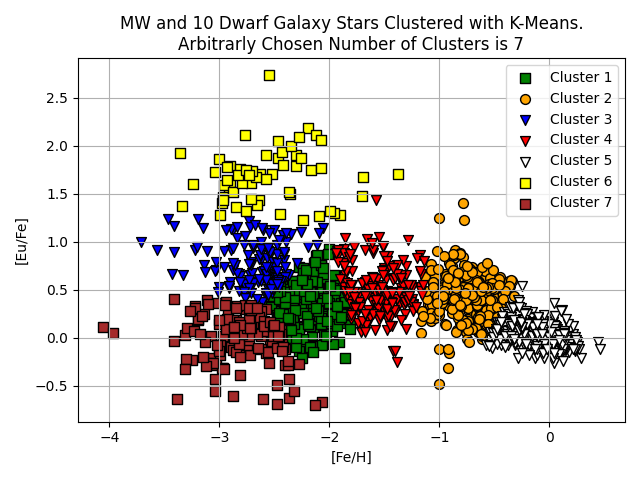}
}
\caption{{ Left panel}: [Eu/Fe] ratios as in Figure~\ref{fig:fullEuFe_SrEuFe} (left panel), but divided into three K-means clusters. It can be seen that essentially the division of Table~\ref{tab:pcc2} into very weak, weak, and moderate to strong correlations is reproduced. But as discussed in the text below, the weak and moderate to strong correlations are no true correlations, they are rather the result of a well mixed medium with quite different stellar origins. { Right panel}: If we choose to divide into 7 clusters, intermediate transition clusters appear around the kink at [Fe/H]=-1 and in the interval -2.5$<$[Fe/H]$<$2, where the domination of core-collapse supernovae turns over to a type Ia supernova domination, as well as where the very weak correlation turns into a weak (but in reality "spurious") correlation. However, for [Fe/H]$<$-2.5 we see the division in limited-r, r-I, and r-II stars, { showing that they indicate different abundance behaviors. The clear distinction of limited-r stars has been discussed before, how r-I and r-II star properties point to possibly different sources will be discussed in later sections.}}
\label{clusterplot}
\end{figure*}

\noindent Table \ref{tab:pcc2} and also Fig.~\ref{fig:fullEuFe_SrEuFe} (left panel) highlight a different aspect of interpreting low-metallicity star observations. The window in [Fe/H], where we can analyse the abundance patterns (and their correlations) of individual (r-process) polluters or nucleosynthesis sites, is probably limited to the interval $<$-2.8 or at most -2.5 (see our discussion of inhomogeneous galactic chemical evolution in subsection \ref{sec:2.1}). For larger [Fe/H] values we see the averaging contributions of many nucleosynthesis sites with a decreasing scatter of Eu/Fe, while Fe/H is enriched mainly by supernovae (early on by massive stars resulting in CCSNe, later on also by type Ia supernovae originating from binary systems). In the metallicity ranges [Fe/H]$>$-2, the metallicity becomes a measure of the time over which galactic evolution has taken place. In parallel to the enhancing [Fe/H] ratios, due to supernovae contributing to galactic evolution, also r-process sites contribute heavy elements, e.g. Eu, as a function of time. In a well mixed interstellar medium, occurring after a while due to many different contributors as well as mixing mechanisms in the Galaxy, an averaged Eu/Fe ratio (with decreasing scatter) emerges, which is close to constant for metallicities ranging from [Fe/H]=-2 up to -1 \citep[see e.g.][]{Cowan.Thielemann:2004,Hansen.Holmbeck.ea:2018}. This constant Eu/Fe ratio is not due to co-production in one specific site with that ratio, it is just a mean measure of what different sources contribute during galactic evolution. Therefore, Eu increases linearly with Fe (and galactic evolution time) under these circumstances, and we see a stronger correlation emerging with increasing metallicity. For [Fe/H]$>$-1 type Ia supernovae start to contribute strongly to the Fe production (and with less co-produced alpha elements than in CCSNe). This has in chemical evolution the effect that [$\alpha$/Fe] decreases by about a factor of 3 with increasing [Fe/H]. The same effect causes decreasing [Eu/Fe] observations in that metallicity interval. But while Eu/Fe is decreasing, because type Ia supernovae lead to a stronger Fe-production, Eu is still also increasing as a function of time or metallicity. The relation between Eu and Fe has a different gradient, but is almost perfectly linear. In principle, this different gradient in the Eu/Fe ratio above [Fe/H]=-1 would lead to a worse linear fit over the whole metallicity range [Fe/H]$<$0 when data points would be equally distributed. But as can be seen from the last entry in Table \ref{tab:pcc2}, 2/3 of the data points are located in the range -1$<$[Fe/H]$<$0, where a strong linear relation exists. For this reason we see also here a continuing increase in linear correlations, { which are, however, spurious correlations, not indicating at all the co-production in a specific site with a constant abundance ratio, but rather an overall constant abundance ratio averaged over many contributing sites.} 

\noindent Fig.~\ref{clusterplot} provides an additional examination of the [Eu/Fe] observations in terms of correlation clusters (see appendix~\ref{kmeans}), which leads to a further understanding of the results of Tables~\ref{tab:pcc0} to \ref{tab:pcc2}. This method includes the possibility to look at correlations as a function of [Fe/H] (as in Table~\ref{tab:pcc2}) as well as a function of [Eu/Fe] (see Tables~\ref{tab:pcc0} and \ref{tab:pcc1}) in a combined way.
It can be noticed that the 3 cluster correlation provides an identical analysis as discussed in Table~\ref{tab:pcc2}, the 7 cluster correlation also exhibits the division into limited-r, r-I, and r-II stars, combined with showing transition regions for the values in Table~\ref{tab:pcc2}.
We will discus these issues in more detail and with more elements in the next sections. It is worth keeping in mind that a correlation does not necessarily - in general - imply causality between two entities, as was already outlined in the 
discussion of Table~\ref{tab:pcc2} { and the discussion at the end of the last paragraph}, when looking at Eu and Fe at higher metallicities or Tables~\ref{tab:pcc0} and \ref{tab:pcc1}, when considering arbitrarily small [Eu/Fe] ranges without convincing reasons why these relate to a category of stars with pronounced features which support to treat them separately.
While the { dominant} astrophysical origin of Fe is well-understood, the origin of r-process elements like Eu is still a matter of debate. This paper attempts to bring more clarity to this issue. Table~\ref{tab:pcc2} made clear that the method of trying to link correlations to co-production of elements only works at lowest metallicities, where we can look at patterns originating from only one (or at most a few) site(s). With this constraint, the method outlined above clearly enables the analysis of correlated
abundance patterns, pointing back to the nucleosynthesis processes of the contributing/polluting site.

\section{Correlations between Fe, Ni, Eu and Th in VMP Halostars}
\label{corrEuFe}

In the following subsections we will first concentrate on the origin(s) of Fe and Eu in the observed low-metallicity stars where both elements are detected. This includes tests for correlations among these elements, before analyzing a different behavior of the Fe-group element Ni, and finally having a look at Th, a representative of the heaviest elements.

\subsection{Is Eu correlated with Fe for the variety of limited-r, r-I as well as r-II stars?}
\label{3-1}

As discussed in section~\ref{application1} we utilize statistical correlation coefficients  to understand the origin of the different observed stellar abundance patterns with known Fe and Eu abundances. The linear elemental abundances of X and Y track linear correlations like the Pearson correlations better than log$\epsilon$(X) and log$\epsilon$(Y), because the logarithm intrinsically distorts a true linear relationship. Similar to Table~\ref{tab:pcc0}, when analyzing correlations among Eu and Fe, we pass through the list of stars from the smallest to the largest [Eu/Fe] values, plotting the correlation values vs. the upper bounds of the corresponding [Eu/Fe]-intervals (which have no lower bounds). Only stars with $\mbox{[Fe/H]}\le -2.5$ are considered, in order to (a) discard the Fe originating from type Ia supernovae, to (b) avoid significant s-process contamination, and (c) preferentially select stars which { according to our hypothesis, outlined in \ref{sec:2.1},} were born with the contamination from only a
single (or at most very few) 
nucleosynthesis site(s). The abundances are taken from the Stellar Abundances for Galactic Archaeology (SAGA) database compilation of stars in the Milky Way \citep[][and http://sagadatabase.jp]{Sagadatabase} or JINAbase of the Joint Institute for Nuclear Astrophysics \citep[][and http://jinabase.pythonanywhere.com]{JINAbase:2018}. In each plot, wherever possible, we indicate the positions of the following standard incomplete and complete stars (HD122563, HD115444, CS22892-052, and CS31082-001). These were already shown with their abundance patterns in Fig.~\ref{4starsplot} and cover the range from limited-r to r-enriched stars, via the r-I and r-II ranges, up to an actinide boost. 

\noindent The SAGA database as well as JINAbase contain more than 200 stars with $\mbox{[Fe/H]}\le -2.5$ where Eu has been measured (avoiding cases where only upper limits are given). 
Fig.~\ref{4starsplot} (right panel)
shows the PCC and SCC curves for { the correlation of the elements Fe and Eu} in those stars, being equivalent to the entries in Table~\ref{tab:pcc0}, which, however, contained only the Pearson correlation. As discussed already before, the  correlation coefficients obtained in such a way include the danger of misinterpretations, because starting with small [Eu/Fe]-intervals (resulting therefore in a close to constant Eu/Fe ratio) leads automatically to high linear correlations, while utilizing the whole [Eu/Fe] range points clearly to vanishing correlations. Only the clear knowledge that a certain subgroup, like e.g. the limited-r stars with their high Sr/Eu ratios (see Fig.~\ref{fig:fullEuFe_SrEuFe}, right panel) must have a different stellar origin than complete r-process stars, permits to employ a PCC
analysis in the corresponding [Eu/Fe]-interval. Having this precaution in mind, we utilize this method for detecting groupings or clusters { also for the higher [Eu/Fe values}. For low, but increasing, [Eu/Fe]-values the correlation coefficients remain initially at a high level in the limited-r stars regime. In fact, we see very high PCCs ($\simeq$0.98) until [Eu/Fe]$\simeq -0.32$. This regime corresponds to the first subgroup already noticed in 
Fig.~\ref{fig:fullEuFe_SrEuFe} (right panel)
with high Sr/Eu abundance ratios, and the relation between Fe and Eu is still almost perfectly linear. Both curves (PCCs and SCCs) are essentially identical and remain approximately also constant at a high level close to 0.8 for the second subgroup up to $\mbox{[Eu/Fe]}= 0.0$ (or even 0.3), with a slight change in the trend around [Eu/Fe]=0. However, beyond that limit they start to diverge. Both coefficients show clearly that the correlation between Fe and Eu decreases with increasing upper [Eu/Fe] limits. The linear relationship between Fe and Eu is very strong for the incomplete (limited-r) stars, but gets gradually weaker for the complete stars, a behavior discussed already to some extent in the previous section. At higher [Eu/Fe]-ratios the two curves for the SCCs and PCCs result in quite different values. Thus, generally three distinct regimes can be deduced from
Fig.~\ref{4starsplot} (right panel):
\begin{enumerate}
\item The first regime with $\mbox{[Eu/Fe]} \leq -0.32$ exhibits a constant high correlation at roughly 1.
\item The second regime with [Eu/Fe] between -0.3 and 0.3 (with slight change in slope around 0) exhibits also a relatively constant and still quite high correlation at roughly 0.80. In regime 1 and regime 2 PCCs and SCCs coincide very well with each other.
\item The third regime with $\mbox{[Eu/Fe]}> 0.3$ is characterized by steadily decreasing and diverging PCC and SCC values. 
\end{enumerate}
For all stars in the limited-r regime (i.e. for regime 1 and 2) 
Eu is increasing in a linear fashion with increasing Fe, while for the complete stars Eu becomes (with strongly decreasing correlations) not related to the Fe abundance, i.e. for that regime 3 Eu and Fe seem to come from different sources which both contributed already at these low metallicites of [Fe/H]$<$-2.5.
We want to address in more detail this striking decrease of the PCC curve in the third regime at $\mbox{[Eu/Fe]} \ge 0.3$. The divergence of PCCs and SCCs can be better understood when having a detailed look at the scatter of the Eu and Fe abundances, which was shown in 
Fig.~\ref{fig:fullEuFe_SrEuFe} (left panel).
As we explained earlier, the PCC tests the linear relationship between two variables, whereas the SCC tests whether a monotonic relationship between two variables exists. 
The SCC is based on the ranked values for each variable (see appendix \ref{appsinglemultible}).
In order to better understand the strong decrease of the (linear correlations measuring) PCCs in the third regime, we show in Table~\ref{eu_fe_ratio} the ratios of maximum and minimum values within the boundaries of the three regimes in [Eu/Fe], i.e. Fe$_{\rm max}$ and Fe$_{\rm min}$, Eu$_{\rm max}$ and Eu$_{\rm min}$, the minimum and maximum of (Eu/Fe), the  the growth rates, and the coefficients of determination (r$^2$, the square of of the PCCs) of the linear regression (see appendix \ref{sec:lin_reg} for a brief summary) for both elements within the three regimes defined above. 

 \begin{table*}[h!]
      \caption{Linear ratios, the growth rates and the coefficients of determination (r$^2$) of the linear regression of Eu and Fe within the three regimes defined above.}
         \label{eu_fe_ratio}
     $$ 
         \begin{array}{ l    l    l     l     l  l l  l}
            \hline
            \noalign{\smallskip}
            \mbox{Regime} & \mbox{Fe$_{\rm max}$/Fe$_{\rm min}$} & \mbox{Eu$_{\rm max}$/Eu$_{\rm min}$} & \mbox{(Eu/Fe)}_{\rm min} & \mbox{(Eu/Fe)}_{\rm max} & \mbox{Growth rate} & \mbox{r}^2  &\#\,\mbox{of Stars} \\
           \hline
            \noalign{\smallskip}
            1 & 18 & 13 & 2.4\times 10^{-8} & 5.01\times 10^{-8}  & 2.1 & 96\% & 9 \\
            2 & 36 & 46 & 5.1\times 10^{-8} & 2\times 10^{-7}  & 3.9 & 70\% & 107\\
            3 & 16 & 240 & 2.1\times 10^{-7} & 1.4\times 10^{-5}  & 65 & 3\% & 170\\
            \noalign{\smallskip}
            \hline
         \end{array}
         $$
   \end{table*}

\noindent The growth rate stands for the ratio of the maximum and minimum value of the ratio of Eu and Fe, and r$^2$ quantifies the quality of the linear fit of Eu as a function of Fe. 
It can also be explained as the proportion of the variation of Eu that can be explained by the variation of Fe in \%. While in the first and second regime (containing limited-r stars) the growth rate is moderate with factors of 2.1 and 3.9 respectively (i.e. changing the linear slope only slightly), it is very pronounced in the third regime and forces the correlation coefficients to dramatically fall from 0.8 to below 0.2. The corresponding r$^2$ is only 3\%, which means that the relationship of both elements in that regime is barely linear. Furthermore, the ratio of Eu$_{\rm max}$ and Eu$_{\rm min}$ is more than one order of magnitude higher than in the first two regimes, whereas that of Fe$_{\rm max}$ and Fe$_{\rm min}$ is similar throughout the three regimes. In summary, while for the limited-r stars of regimes 1 and 2 a high correlation between Eu and Fe exists, { pointing to a co-production of both elements} in the same event with a close to constant ratio, it seems that an additional and very productive source of Eu is contributing in the third regime, which produces huge amounts of Eu and negligible amounts of Fe.

\subsection{A correlation test related to Fe and Ni abundances for all limited-r, r-I, and r-II stars}
\label{FeNi}

\begin{figure*}[h!]
\centering
\includegraphics[width=9cm]{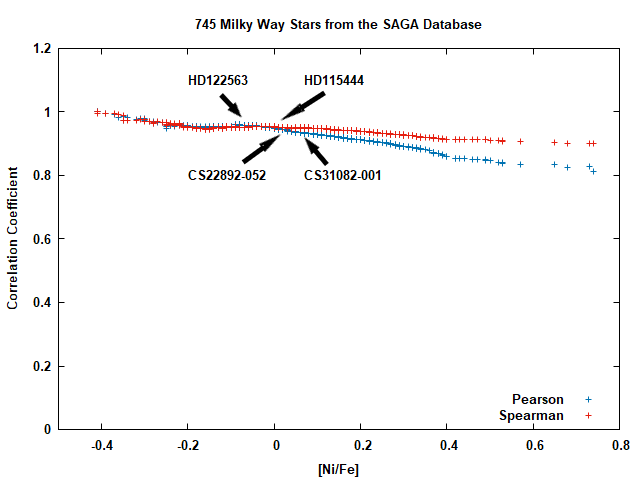}
\includegraphics[width=9cm]{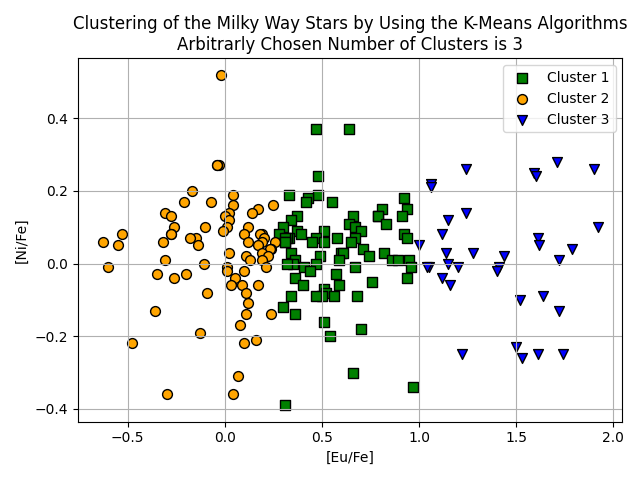}
\caption{{ Left panel:} The PCCs and SCCs of Fe and Ni in stars with $\mbox{[Fe/H]}\le -2.5$. We see a good correlation for the whole Ni/Fe range with a close to linear relation as expected from the same source (core-collapse supernovae?). { Right panel:} Eu/Fe, ranging over three orders of magnitude, is not correlated with the Ni/Fe ratio, thus, not pointing to a regular core-collapse supernova origin. However, in the (K-means) 3 cluster plot we notice the division into limited-r stars, r-I, and r-II stars.}
\label{corr_ni_fe_1}%
\end{figure*}

Similar to our analysis of Fe and Eu abundances we have also calculated the PCC and SCC curves of Fe and Ni, with the result shown in Fig.~\ref{corr_ni_fe_1} (left panel). The relationship of Fe and Ni is very different from that of Fe and Eu. Firstly, both correlation coefficients remain at a high level $(\ge 0.8)$, and secondly the  divergence of PCCs and SCCs for higher [Ni/Fe] values is only small in comparison to the case of Fe and Eu. The linear relationship of Fe and Ni is quite strong for all (i.e. the incomplete as well as the complete) stars. Thus, for all stars at these low metallicities we find in the SAGA Database an average [Ni/Fe] ratio of 0 (with a scatter of up to a factor of 3 in comparison to three orders of magnitude in [Eu/Fe]). This means that Ni increases close to linearly with Fe. This different behavior of Eu/Fe vs. Ni/Fe requires an understanding.

\noindent The ratio Ni/Fe, displayed in Fig.~\ref{corr_ni_fe_1} ({ right} panel), shows a variation/scatter of about a factor of 3 around the solar value. This means that it does not reflect a perfectly linear relation with a  constant ratio (as also indicated by the PCC of 0.8 over the whole range). But for both elements we expect at low metallicities a dominant origin from { explosive Si-burning in} core-collapse events, while for the high Eu/Fe values one requires an additional highly productive r-process source with no or negligible Fe production. Thus, if we observe variations in Ni/Fe this would need to be explained by variations in the ejecta of core-collapse events. Based on recent predictions by e.g. \citet{Curtis.ea:2019} and \citet{Ebinger.Curtis.ea:2020} (see their Fig.9), we expect such variations within a factor of 2.65 around the solar value for CCSNe, if we take into account that elemental Fe is essentially determined by $^{56}$Fe (decay product of $^{56}$Ni) and Ni is dominated by $^{58}$Ni. The amount of $^{56}$Ni varies directly with the supernovae explosion energy.
The whole Fe-group, i.e. also stable Ni isotopes, should vary in sync with the explosion energy and therefore also with $^{56}$Ni, causing a correlation between Ni and Fe. However, $^{58}$Ni shows a slightly erratic behavior and is also varying with the metallicity of the progenitor and might also reflect slightly varying $Y_e$ values, due to weak interaction in the innermost ejecta. (This is probably the reason that we see such variations also at metallicities of [Fe/H]$\approx$-3, when { CCSNe and other core-collapse events should have already} occurred.) Thus, overall we expect the observed variation of Ni/Fe as a result of the range of possible { core-collapse} progenitors with varying initial mass (which determines the explosion mechanism) and metallicity. The relatively high values of the PCCs and SCCs across the observed [Ni/Fe] range, underline that for all stars a close to linear relation exists between the Ni and Fe abundances, which is, however, not perfect, as the production { sites (core-collapse events) come} with (relatively small) variations in the Ni/Fe ratio.

\subsection{Strong and weak Eu contributions}
\label{EuFe}
\begin{figure*}[h!]
\centerline{
\includegraphics[width=10cm]{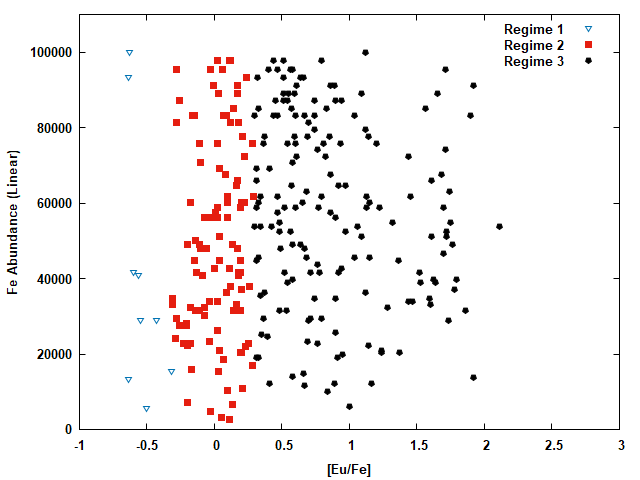}
\includegraphics[width=10cm]{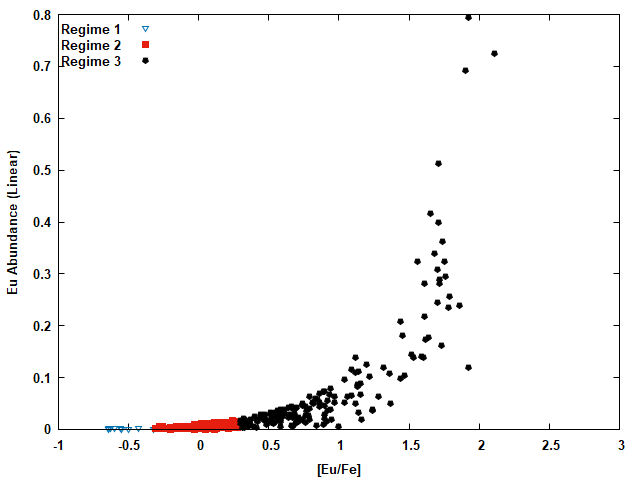}
}
\caption{The left panel shows the linear Fe abundance versus $\mbox{[Eu/Fe]}$ in stars with $\mbox{[Fe/H]}\le -2.5$. The right panel shows the linear Eu abundance versus $\mbox{[Eu/Fe]}$ in the same stars.}
\label{Fe_vers_eu_fe}%
\end{figure*}

\noindent The left panel of Fig.~\ref{Fe_vers_eu_fe} 
shows the behavior of the Fe abundance, when being plotted against $\mbox{[Eu/Fe]}$. 
The Fe abundances { come with a scatter of more than 10, apparently due to the linear scatter in Fe/H for the observed stars in the metallicity range -4$<$[Fe/H]$<$-2.5. They do not show a measurable trend with $\mbox{[Eu/Fe]}$, being almost equally distributed throughout the three regimes. Thus, the} abundance scatter in Fe covers the metallicity range of the sample stars, while the large variations in [Eu/Fe], existing for all values of the Fe abundance, cover the previously discussed three regimes of r-process contributions in the still quite inhomogeneous early galactic environment at low metallicities. 

\noindent The right panel of Fig.~\ref{Fe_vers_eu_fe}
shows the behavior of the Eu abundances when being plotted against $\mbox{[Eu/Fe]}$. The Eu abundances exhibit an increasing trend with $\mbox{[Eu/Fe]}$, especially strong for the third regime.
While for the first and second regime we notice a small, but varying Eu abundance (see also Fig.~\ref{fig:fullEuFe_SrEuFe}, left panel), Eu increases very strongly as a function of [Eu/Fe] in the third regime.
This indicates a strong Eu source { with negligible correlation of Eu and Fe in this regime} (see the discussion related to 
Fig.~\ref{4starsplot}, right panel). 
The scatter at
each [Eu/Fe] value is partially due to the fact that we look at all stars with metallicities smaller than -2.5. The changing Fe
abundances in the interval -3.5$<$[Fe/H]$<$-2.5, would already explain up to a factor of 10 for a constant [Eu/Fe]-ratio, { but in addition Eu from the independent strong r-process source shows strong variations}, due to mixing contributions from the initial production sites and relative locations with respect to the stellar progenitor clouds.

\begin{figure*}[h!]
\centerline{
\includegraphics[width=10cm]{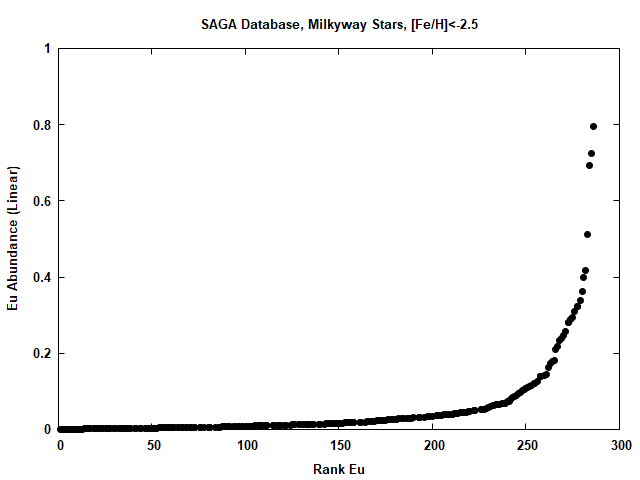}
\includegraphics[width=10cm]{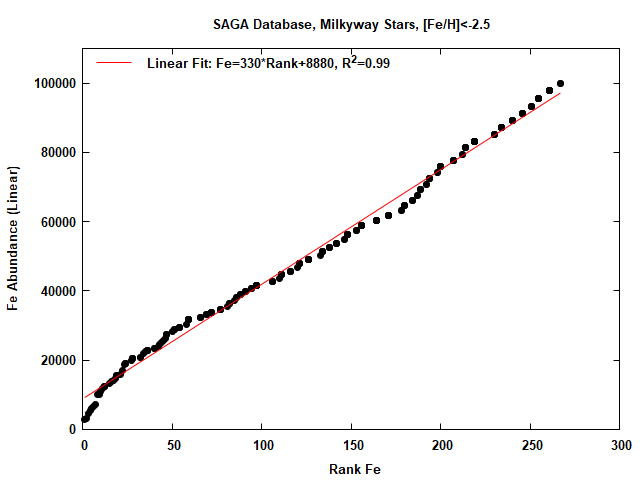}
}
\caption{The left panel shows Eu abundances versus their corresponding ranks. The integer rank passes, with increasing Eu abundances, through all observational points from the smallest to the highest abundance. The fit deviates strongly from a linear behavior at high Eu ranks. As discussed in appendix \ref{appsinglemultible}, this argues for a superposition with an additional Eu source, especially responsible for the high Eu abundances. A missing deviation from the linear slope at small ranks (contrary to the example plot in Fig.~\ref{random} of appendix \ref{appsinglemultible}) underlines that this additional Eu source contains essentially no contributions with negligible Eu production. The right panel shows Fe abundances versus their corresponding ranks. The figure shows an almost linear relation. This would indicate a single production site. The very slight deviation from the linear slope at the lowest ranks might be interpreted as a second source with low/negligible Fe-production, possibly related to a source with high Eu-production.}
\label{eu_vers_rank}
\end{figure*}

\noindent So far we found a remarkable decrease of the correlation coefficients for Eu and Fe in regime 3 (r-enriched stars with [Eu/Fe]$>$0-0.3, as shown in Fig.~\ref{4starsplot}, right panel), consistent with the overall r-I \& r-II entry for regime 3 in Table~\ref{tab:pcc1}, which can be attributed to a very productive Eu source, producing itself no significant amounts of Fe. Otherwise the correlation coefficients would have to remain at a high level as for the case of Ni (shown in Fig.~\ref{corr_ni_fe_1}).
In order to prove that this goes along with an additional strong Eu source in regime 3, 
we plot the observed Eu abundances versus their corresponding ranks, see appendix \ref{appsinglemultible}. As discussed in the appendix, a linear relationship between an abundance X and its rank means, that the production of X is uniform and happens in the same manner by a single type of astrophysical source. If this uniformity is disturbed by a superposition of another source, the linear relationship between the abundances and their ranks are destroyed. This behavior is shown in its generality in Fig.~\ref{random} of appendix \ref{appsinglemultible} for a superposition of two random variables. Fig.~\ref{eu_vers_rank} (left panel) displays such a plot of the Eu abundance vs. its rank, { providing with the strong non-linear trend for high ranks a further indication for} an additional contribution from a strong Eu source in regime 3, i.e. for [Eu/Fe]$>$0-0.3.

\begin{figure*}[h!]
\centering
\includegraphics[width=6cm]{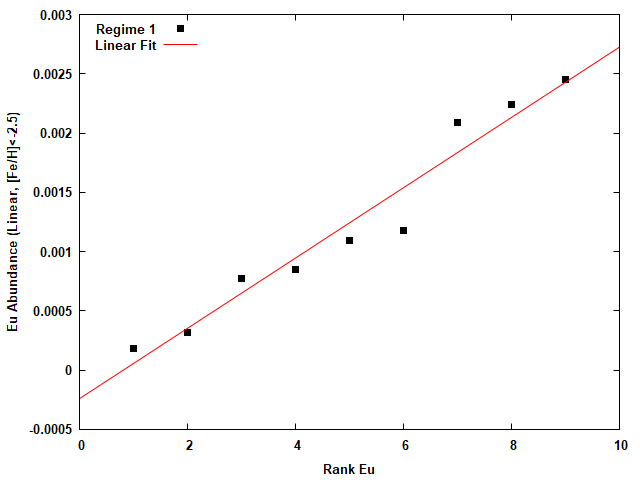}
\includegraphics[width=6cm]{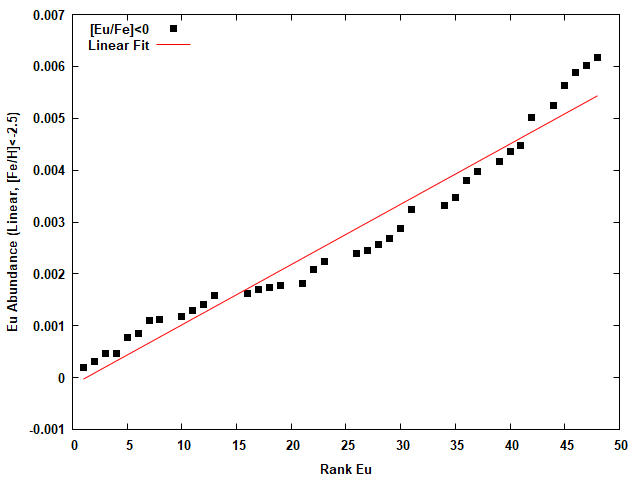}
\includegraphics[width=6cm]{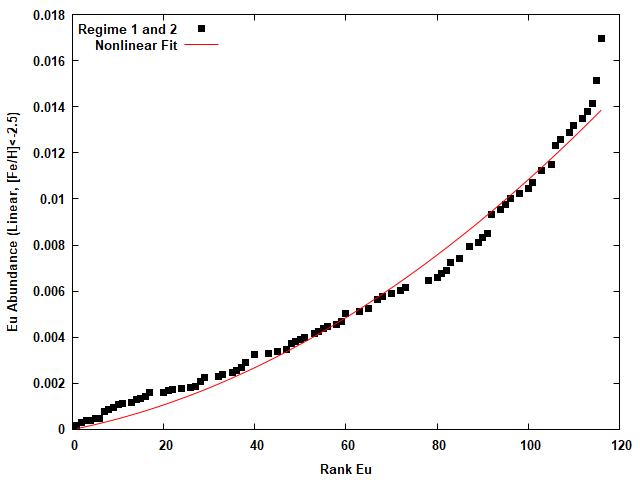}
\caption{(a) Eu abundances versus their corresponding ranks in regime 1, (b) regimes 1 and 2 with an upper limit of $\mbox{[Eu/Fe]}<0$ and (c) with an upper limit of $\mbox{[Eu/Fe]}<0.3$. The fit in (a) is clearly linear, in (b) close to linear and in (c) shows a nonlinear behavior. The linear fit for (b) corresponds to Eu=$9\times 10^{-5}\times$ Rank(Eu).}
\label{eu_vers_rank_2}%
\end{figure*}

\noindent Opposite to the behavior of Eu vs. its rank,
Fig.~\ref{eu_vers_rank} (right panel) gives a linear relation between the observed Fe abundances and their ranks, supporting an essentially unique source of Fe, supposedly related to core-collapse events. The minute offset, visible at small ranks, could perhaps suggest a small superposition of another Fe source, contributing only insignificant amounts of Fe.
When restricting the analysis of Fig.~\ref{eu_vers_rank} only to regimes 1 and 2 (limited r-stars with [Eu/Fe]$<$0-0.3), Fig.~\ref{eu_vers_rank_2} shows initially a close to linear Eu abundance increase with its rank, but develops a slight quadratic modification. This points to similar types of Eu sources for these stars, but indicates also
a superposition of two sources with similar, but slightly different, correlations between regime 1 and 2. Thus, this subsection underlined that while Fe has a single astrophysical source (core-collapse events), Eu has clearly different sources. The latter is based on the fact that 
 Fe and Eu change at highly different rates in regime 3, and in regimes 1 and 2 they change with similar but also slightly different rates.

 \subsection{The weak Eu contribution in limited-r stars and its origin in more than one source}
\label{twoweaksources}
The previous subsections came to the main conclusion that the complete stars (regime 3) require a highly productive Eu source with negligible Fe production, while for the limited-r stars (in regime 1 and 2) the Eu production source is correlated with Fe. However, a more detailed look showed already a slightly changing correlation behavior between regimes 1 and 2 (see also Fig.~\ref{4starsplot}, right panel).
This leads to the question whether limited-r stars go back to a single type of source or whether further events of a different type might contribute.

\begin{figure*}[h!]
\centerline{
\includegraphics[width=10cm]{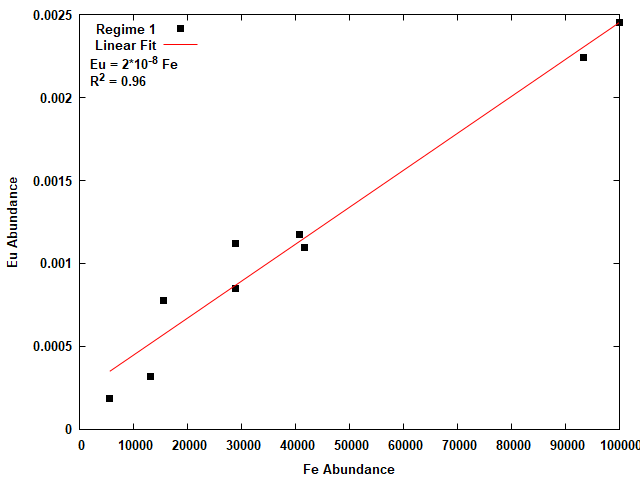}
\includegraphics[width=10cm]{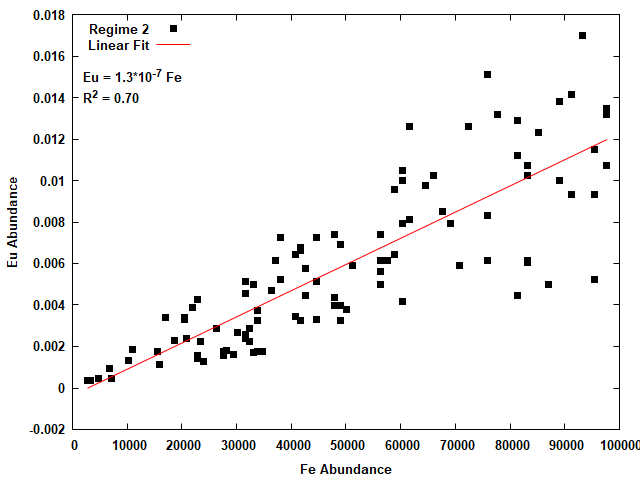}
}
\caption{The linear abundance of Eu versus Fe in the first two regimes ($\mbox{[Eu/Fe]}<0.3$). The left and right panels represent regime 1 ($\mbox{[Eu/Fe]}\le -0.32$) and regime 2 (-0.32<$\mbox{[Eu/Fe]}<0.3$). Two corresponding linear fits are shown as well, { with a large(r) scatter in regime 2}.}
\label{eu_vers_fe_fit}
\end{figure*}

\noindent Fig.~\ref{eu_vers_rank_2} seemed to indicate a close to linear relation between Eu abundances and their ranks for regimes 1 and 2 (with slight quadratic modifications when extending regime 2 up to $\mbox{[Eu/Fe]}<0.3$). 
This is in line with the occurrence of the two plateaus in regimes 1 and 2 as seen in 
Fig.~\ref{4starsplot} (right panel). 
Both SCC and PCC curves are identical for the two plateaus and indicate similar changing rates of Fe and Eu. The SCC and PCC values are roughly 1 in the first plateau and the $\mbox{r}^2$ is 96\%. This means, both elements are in a { virtually perfect} linear relationship. As was shown in Table~\ref{eu_fe_ratio}, the growth rate is about 2 and the number of the considered stars is nine. Therefore, in regime 1 the same astrophysical type of event contributes Fe and Eu with the same ratio. We will refer to this source as category I in the following. In regime 2, the SCC and PCC curves are also identical and their value is about 0.8. 
The growth rate is about 4, which is twice as high as in the first regime. Fe and Eu change at similar rates but the co-production is not perfectly linear as shown in Fig.~\ref{eu_vers_rank_2}, indicating a superposition.

\noindent For this reason we plot the abundances of Eu vs. Fe in
Fig.~\ref{eu_vers_fe_fit} for the first two regimes separately (left and right panel). While stars of regime 1 are perfectly aligned, the stars of regime 2 exhibit a larger scatter, already indicated by the smaller correlation coefficients of approximately 0.8. This can be interpreted as a superposition of a second source, as suggested already above in the discussion of Fig.~\ref{eu_vers_rank_2}. If we also apply a linear fit, the slope of the fit in the second regime is five times higher than in the first regime, leading to the conclusion that the production of Eu is more efficient in the second regime. This suggests that we are dealing with a second astrophysical event, which similar to the first one produces Fe, but is more efficient in co-producing Eu. In the following we will refer to it as a category II event.

\subsection{Th, a strong r-process element and the question of a very early source in the galactic evolution}
\label{veryearly-r}
The production of Fe in the Universe goes back to the early evolution of galaxies via the explosion of massive stars. Signs of { Fe produced by CCSNe can already be found at low metallicites, even for the $\mbox{[Fe/H]}$ range as below -5 to -4, possibly due to an only 1\% to 10\% admixture of a nearby CCSN remnant (see the discussion in section \ref{sec:2.1})}. At higher metallicities, closer to -1, also thermonuclear explosions of SNe Ia start to contribute with a delay due to the longer evolution of low and intermediate mass stars, combined with delayed mass exchange in binary systems after the formation of a white dwarf \citep{matteucci86,Matteucci:2012,Timmes.Woosley.Weaver:1995,Kobayashi.Umeda.Nomoto.ea:2006,nomoto13,Seitenzahl.Timmes.Magkotsios:2014,maoz14,Kobayashi.Karakas.Lugaro:2020}.
At solar metallicity, the fraction of Fe that originated from CCSNe is estimated to be of the order of 40\%. \\
Up to now, all detailed analyses presented in this section utilized stars in the metallicity range [Fe/H]$<$-2.5. Subsection \ref{twoweaksources} showed that in regime 1 and 2 (for limited-r stars) Eu is probably co-produced with Fe in core-collapse events. We have introduced them as category I and II events. However, in regime 3 (for complete r-process stars) a highly productive Eu source, with no or only negligible Fe production, had to be added, which we will give the label of category III events. { The regime 3 stars could be polluted by core-collapse events of possibly different types, producing Fe, but additionally a strong r-process event had to contribute as well, characterized by a negligible correlation with Fe.} The question arises which one of the events discussed in the introduction is responsible, and whether they { occur with a delay after the Fe-contribution from core-collapse events}. A general discussion of r-process contributions in the early Galaxy (with extended literature) can be found in \citet{Cowan.Sneden.ea:2021}. The main question with respect to these sources is related to their (massive) single or binary star origin and a possible delay in their appearance as a function of metallicity. Dependent to some degree on the refinement of the method to perform inhomogeneous galactic chemical evolution simulations \citep[see e.g.][]{wehmeyer15,cescutti15,VandeVoort.ea:2020}, which does not assume immediate extended mixing of ejecta with the interstellar medium on galactic scales, it has also been discussed that a massive star related source is required in order reproduce the very early appearance of strong r-process elements, while neutron star mergers (NSMs) would appear later (at metallicties [Fe/H]$>$-3 to -2.5).
This conclusion could possibly be circumvented by the existence of extensive mixing of matter { or ejecta mass lost in dwarf galaxies, if} all lowest metallicity observations relate to stars which originated in these early substructures and building blocks of galactic evolution \citep[see e.g.][]{ojima18}. 

\begin{figure*}[h!]
\centerline{
\includegraphics[width=10cm]{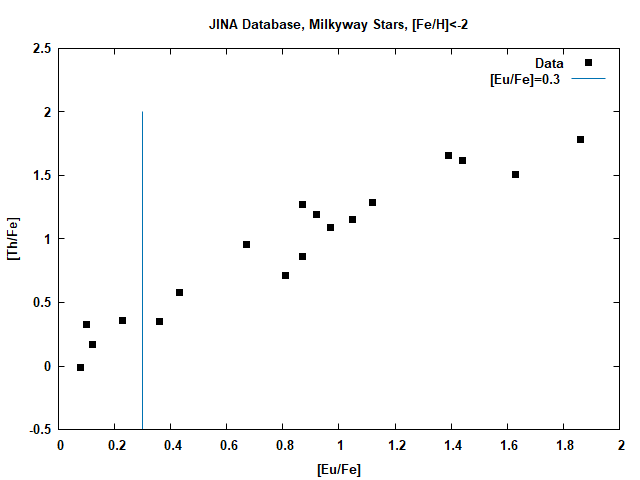}
\includegraphics[width=10cm]{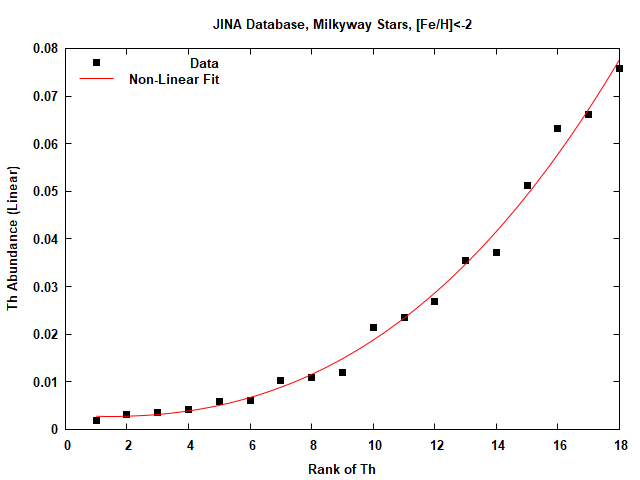}
}
\caption{{ Left panel:} Observed [Th/Fe] as a function of [Eu/Fe] from the JINA database. It can be recognized that already for [Eu/Fe]$>$0 Th can be found, indicating a strong r-process. The blue line shows the previously utilized upper limit for limited-r or weak r-process stars. { Right panel:} Detected Th abundances in low metallicity stars with [Fe/H]$<$-2 plotted vs. their rank. The non-linear behavior indicates a superposition of Th-producing sites among the whole class of category III events responsible for the abundance patterns in complete r-process stars.}
\label{figThEuH}
\end{figure*}

\noindent In order to test whether different events contributed to these regime 3 stars, i.e. whether category III events relate to one or more than one site, we select the actinide element Th which can only be produced in strong r-process events and thus reduces uncertainties introduced by weak r-process components. 
Before discussing this further, we want to point to an interesting additional fact, related to Th observations. Limited-r stars have been initially defined as those with { [Eu/Fe]$<$0.3, this comes typically with a non-detection of elements of the third r-process peak as well as actinides}. When having a look at Fig.~\ref{figThEuH} (left panel), one recognizes that Th can be observed already in stars with [Eu/Fe]$>$0-0.1. Therefore an upper/lower limit for limited-r / complete r-process stars of [Eu/Fe]$\approx$0 seems
more appropriate than the up to now utilized value of 0.3.

\begin{figure*}[h]
\centering
\includegraphics[width=10cm,height=7cm]{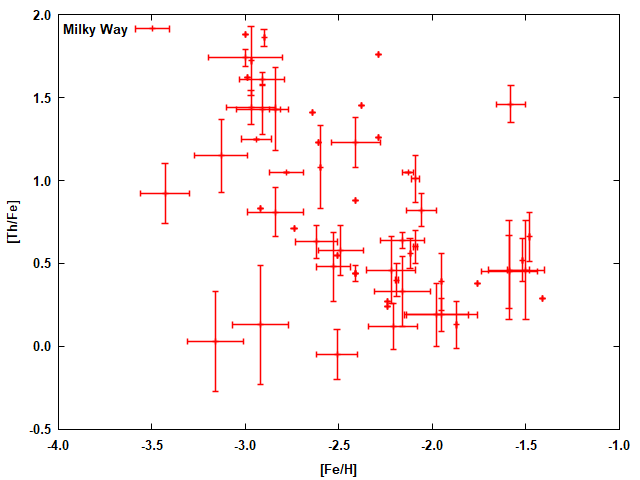}
\caption{Observed [Th/Fe] as a function of metallicity [Fe/H] from the Saga database. For the solar abundance of Th its present-day value {log$\epsilon$(Th)=0.02} \citep{Asplund.Grevesse.ea:2009} from 3D stellar models is utilized. \citet{Lodders19} gives larger values ($\leq$0.03), but this would only lead to a scaling factor.
Considering a half-life of 1.4$\times10^{10}$y, the initial values with which the oldest (lowest metallicity) stars were born could actually be twice as high.}
\label{figThFeH}
\end{figure*}

\noindent After having identified regime 3 more clearly, i.e. already starting at [Eu/Fe]$\approx$0, the question to be answered is whether this strong r-process regime with Th production is related to one or more astrophysical sites. A superposition of sources would be indicated by a non-linear behavior of Th abundances (taken from regime 3 observations) as a function of their rank.
In Fig.~\ref{figThEuH} (right panel)
we show this relation for Th and see clearly a non-linear behavior. Thus, it is obvious that the Th { observed in regime 3 stars requires a superposition of (at least?) two different category III} sources (category IIIa and IIIb, see below). We have neutron stars mergers as a clear candidate for a complete r-process, the question arises which additional site(s) could be responsible, having in mind our earlier discussion related to delays of ejecta from binary stellar systems, while they essentially vanish for collapsing massive stellar sites. In Fig.~\ref{figThFeH} we show observations of [Th/Fe] with respect to metallicity [Fe/H] \citep[with the present-day solar Th abundance taken from][]{Asplund.Grevesse.ea:2009}, indicating quite a number of high values at very low metallicities around [Fe/H]=-3.

\noindent If a strong r-process site related to the core-collapse of massive stars can be considered \citep[like e.g. collapsars or also MHD-jet supernova,][]{Siegel.Barnes.Metzger:2019,Siegel:2019,Winteler.Kaeppeli.ea:2012,Moesta.ea:2015,Nishimura.Sawai.ea:2017,moesta18,Reichert.Obergaulinger.ea:2021}, it would be accompanied by
observed high [Th/Fe]-values at lowest metallicities. This behavior, shown in Fig.~\ref{figThFeH}, might potentially be identified with the contribution of such a site.  
Although one should be careful with interpretations, this could point to one of the (more than?) two strong r-process - and therefore Th (and also Eu) - sources, this one appearing already at lowest metallicities. One option for this source would be collapsars, i.e. stars with masses even higher than those of regular CCSNe, which explode very early in galactic evolution \citep{Siegel.Barnes.Metzger:2019}. For higher metallicities between [Fe/H]=-3 and -2 NSMs will set in \citep{wehmeyer15,VandeVoort.ea:2020}. 
This would consistently lead to a superposition of sources with a non-linear behavior of Th abundances as a function of their rank, as discussed above.
Thus, { regime 3 (or r-process enriched stars)} require a superposition of (at least?) two different sources which we will further down introduce as category IIIa and IIIb. Whether there is room for another source in this superposition \citep[e.g. MHD jet supernovae][]{Winteler.Kaeppeli.ea:2012,Moesta.Richers.ea:2014,Moesta.ea:2015,moesta18,Reichert.Obergaulinger.ea:2021} has to be left open at this point.
In such an interpretation a subclass of category III events, contributing already at lowest metallicities and producing a strong r-process, would come with a small (to negligible) Fe-production (in comparison to solar abundance ratios), while on the other hand those events of category III which contribute with a delay are dominated by sites of a delayed stellar origin, e.g. probably compact binary systems, leading to NSMs and NS-BH mergers with essentially no Fe production at all. We will discuss later { how both types of events relate to the onset of high [Eu/Fe] values in Fig.~\ref{fig:fullEuFe_SrEuFe} (left panel)}.\\

{ \subsection{A generalized approach to low-metallicity abundance interpretations}
As outlined in \ref{sec:2.1}, our initial approach started out with the ansatz that abundance patterns in low-metallicity stars with [Fe/H]$<$-2.5 can be interpreted as the imprint by one contributing nucleosynthesis source to the protostellar cloud of the observed star \citep[based on the analysis of ejecta mixing with the ISM via a Sedov blast wave by][] {Ryan.Norris.Beers:1996}. Thus, if we would have at these low metallicities only one type of contributing nucleosynthesis event, one would find in all corresponding stellar observations the same abundance ratio between to elements (perfect correlation with PCC=+1), although the overall admixtures to the individual stars could vary. If this hypothesis is true (imprint by one single event), but different types of events with different abundance patterns occurred, one can see separate classes of observations. Taking all of them as a global data set would lead to no or a very low correlation, but if one finds a way to separate the classes, one can see the individual correlations, caused by co-production in the events responsible for that class. In this way we could separate the Eu observations in limited-r stars, related to weak-r production sites and r-enriched stars with strong r-process production sites. Among the weak-r production sites we have determined our category I and II events, causing a strong correlation between Eu and Fe, i.e. co-production in these events. If we have a look at the complete (r-enriched) stars, we find a very bad correlation between Eu and Fe, i.e. several types of events must have contributed and overall no or a negligible co-production of Eu and Fe can be concluded. However, we find Fe in all these observations, i.e. that Fe must come from a different site than Eu, and the most probable explanation would be given by regular core-collapse supernovae. This points to the complication that one is already at the stage in galactic evolution where not only one single event is responsible for the abundance observations in a low-metallicity star, but at least two or several events contributed already.
This weakens the interpretation of co-production, valid only for the imprint by a single nucleosynthesis source. Nevertheless detecting strong as well as weak correlations can still give powerful hints for the analysis of a set of stellar abundance observations. Fortunately, an additional powerful tool could be introduced by the rank method, i.e. plotting for a whole set of stars the observed abundances of one element in comparison to the ranks of these data. This method, where a linear relation points to one contributing nucleosynthesis site, while deviations from that point to separate origins, has been utilized in Figs.\ref{eu_vers_rank} and \ref{eu_vers_rank_2}, underlining that Fe goes back to essentially one origin (core-collapse events), while Eu must have several origins (related to weak and strong r-process sites). In Fig.\ref{figThEuH} we could see that even the strong r-process element Th must have two contributing sites.}

\noindent Making use of these insights we could point to the two types of category I and II events ({ due to the correlated Fe production, probably related to specific types of CCSNe) with} different efficiencies for producing r-process elements like Eu, as outlined in subsection \ref{twoweaksources}.
The discussion { in the previous subsection,} led to the question whether a { second category III sub-source is required, which is connected to core-collapse of massive stars, and would also
produce Th and U, while category III in total (IIIa and IIIb) is responsible for the complete (r-enriched) stars of regime 3 with [Eu/Fe]$>$0}. 

We will later return to the question whether the subsets of category III type events discussed above (which in total is responsible for complete r-process enriched stars of regime 3), will also produce Th and Eu in similar { or different} ratios for the majority of complete r-process stars. 

\section{The Fe-Group and Light Trans-Fe Elements:  Asking for a Variety of Core Collapse Events}
\label{Fegroup}
\noindent
In regime 1 and 2 we found a clear correlation between Eu and Fe,
indicating a co-production in a weak r-process by a specific kind of core-collapse supernovae. We also noticed before that all the stars of regime 3 
contain Fe as well. Thus, the question arises whether it was produced in preceding supernovae before NSMs set in with strong r-process contributions, resulting this way in a
small or negligible correlation to Fe. However, a further question needs to be posed, whether some of these r-process enriched stars also go back to sources which co-produced small amounts of Fe combined with "enriched" Eu, as alluded to already in the preceding section. 
For this reason we revisit the correlation of Ni and Fe, shown already in Fig.~\ref{corr_ni_fe_1} for metallicities [Fe/H]$<$-2.5, and extend this analysis also to Zn and other trans-Fe elements.

\subsection{Correlations of Fe with Ni and Zn}
\label{sec:FeNiZn}

Fig.~\ref{corr_ni_fe_1} showed that the Ni/Fe ratio varies for these low metallicities about a factor of 3 around solar ratios. 
The SAGA database finds a relatively stable behavior with a mean value close to solar
as a function of metallicity [Fe/H]. 
A decline of the scatter towards higher metallicities is due to averaging out of individual contributing event characteristics and mixing of the ISM. 

\noindent 

The range of [Ni/Fe] in (simplified) spherical explosion models of CCSNe, e.g. by \citet{Curtis.ea:2019} and \citet{Ebinger.Curtis.ea:2020}, ranges from about -0.35 to +0.4. In case there would be a superposition of CCSNe and other Fe and Ni producing sources at low metallicities, and both sources contribute Ni and Fe in (slightly) different ways, one would expect a superposition in the [Ni/Fe] range.
The predicted CCSN ratios seem to be consistent with the observations within their error bars, however, a larger spread of values due to other sources cannot be excluded at this point. 
In the subsection \ref{FeNi} we discussed the Fe and Ni contributions from explosive Si-burning. Fe is dominated by $^{56}$Fe, a decay product of $^{56}$Ni. Ni has a significant contribution from $^{58}$Ni, being produced under slightly neutron-rich conditions, i.e. $Y_e<0.5$, which can be due to weak interactions during core collapse or also to due metallicity (CNO turns to $^{14}$N in H-burning and via two alpha-captures and one $\beta^+$-decay to slightly neutron-rich $^{22}$Ne). Thus, $^{58}$Ni is a measure of slightly neutron-rich conditions. $^{60}$Ni, on the other hand, is a product of alpha-rich freeze-out of explosive Si-burning, a decay product of $^{60}$Zn, resulting from a further alpha-capture on $^{56}$Ni. Therefore, larger variations of [Ni/Fe] can be due to (a) a slight neutron-richness, and (b) higher entropy explosions, leading to a stronger alpha-rich freeze-out. While probably (a) dominates for regular CCSNe, not leading to too extreme [Ni/Fe]-values, hypernovae/collapsars experience the conditions related to (b) and could be responsible for the highest [Ni/Fe] observations. Such conclusions could be underlined if one sees an accompanying Zn/Fe enhancement, with $^{64}$Zn being the decay product of $^{64}$Ge, resulting also from a very strong alpha-rich freeze-out and discussed often in galactic evolution research as going back to hypernovae/collapsars \citep{Nomoto.Tominaga.ea:2006,nomoto17,Kobayashi.Karakas.Lugaro:2020}.
For this reason we now have a look at [Zn/Fe] as a possible indication for hypernovae/collapsars, which are also considered as a strong r-process source \citep{Siegel.Barnes.Metzger:2019,Siegel:2019} at the lowest metallicities [Fe/H]$<$-3.

\begin{figure*}[h!]
\centering
\includegraphics[width=8cm,height=5.5cm]{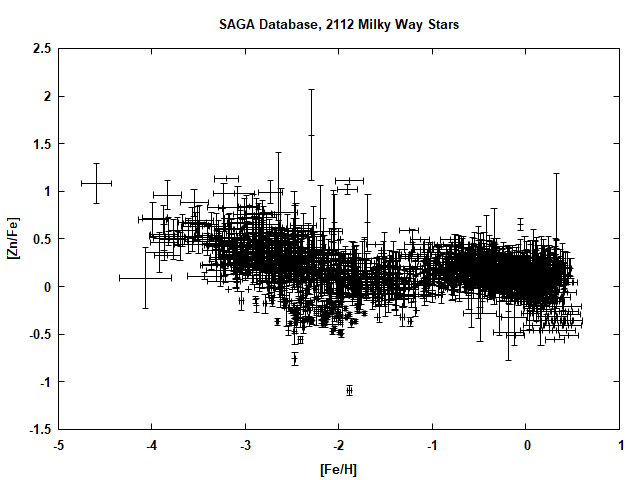}
\includegraphics[width=8cm,height=5.5cm]{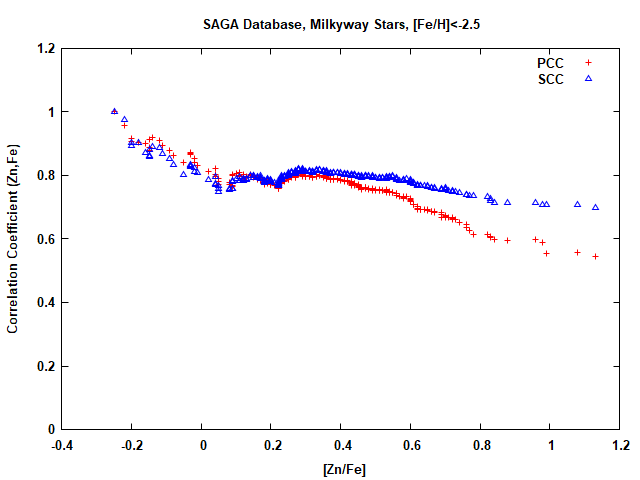}
\caption{{ Left panel:} [Zn/Fe] as a function of metallicity [Fe/H] for Milky Way stars from the SAGA database. Below [Fe/H]=-2 we see an increasing trend, probably related to increasing explosion energies of more massive CCSNe, but possibly also including hypernovae/collapsars. { Right panel:} Correlation of Zn versus Fe for stars with [Fe/H]$<$-2.5. Similar to the earlier correlation test for Ni with Fe, we see in general a good correlation between these two Fe-group elements, resulting from explosive Si-burning in core-collapse events. The higher SCCs than PCCs for [Zn/Fe]$>$0.5 are due to a varying and more extreme scatter.
This could possibly be attributed to an additional source: hypernovae/collapsars.}
\label{corr_zn_fe_low}
\end{figure*}
\noindent As has been discussed above, Zn (via the decay product of $^{64}$Ge) provides a test for a strong alpha-rich freeze-out. The average of [Zn/Fe] observations at metallicities from [Fe/H]=0 down to -2 are of the order 0 (i.e. solar). This is dominated by type Ia supernovae down to -1, and due to a combination from normal freeze-out of explosive Si-burning in Chandrasekhar mass explosions \citep[e.g.][]{Thielemann.Nomoto.ea:1986,Iwamoto.ea:1999,Maeda.Roepke.ea:2010} and an alpha-rich freeze-out in sub-Chandrasekhar He-detonations \citep[e.g.][]{seitenzahl17}. For an overview that a combination of both subtypes of SNe Ia is needed see e.g. \citet{maoz14,Goldstein.Kasen:2018,Livio.Mazzali:2018,Seitenzahl.Chavamian.ea:2019}. Below such metallicities the Zn/Fe ratio is determined by core-collapse events. Down to [Fe/H]=-2 to -2.5
the average [Zn/Fe] is still 0 (solar) and given by a superposition of CCSNe over the whole mass range of an initial mass function \citep{Tsujimoto.Nishimura:2018}. The average rises from [Fe/H]=-2.5 to -3 up to [Zn/Fe]=0.5 with a large scatter and stays at that value for even lower metallicities.\\
In Fig.~\ref{corr_zn_fe_low} (left panel) one can see that for [Fe/H]<-2.5 still [Zn/Fe]$<$0 entries occur. 
At lower metallicities the negative values for [Zn/Fe] disappear, underlining an increasingly alpha-rich freeze-out, which could be interpreted via the fact that at such low metallicities the average mass of exploding CCSNe is shifted to more massive progenitor stars. This leads to increasing compactness and increasing explosion energies. The range of [Zn/Fe]$\approx$0.5 seems to belong to such massive CCSNe \citep{Grimmet.Karakas.Heger.ea:2020}. 
Fig.~\ref{corr_zn_fe_low} (right panel) shows the Pearson and Spearman correlations of Zn and Fe, based on upper limits of the considered [Zn/Fe] intervals for all stars with [Fe/H]$<$-2.5. The entries including [Zn/Fe] intervals over the whole range up to values given on the abscissa lead to a divergence of PCCs and SCCs and to lower correlations above [Zn/Fe] of 0.4 to 0.5. This goes along with a larger scatter in the Zn/Fe ratios and could be interpreted as a sign of larger variations in the Zn and Fe ejecta of highest mass core-collapse sources.
\citet{Tsujimoto.Nishimura:2018} interpret the rise of [Zn/Fe] being due to magneto-rotational supernovae of the type described in \citet{Nishimura.Sawai.ea:2017}, with varying strength of pre-collapse magnetic fields and rotation, while \citet{Nomoto.Tominaga.ea:2006}, \citet{nomoto17}, and \citet{Grimmet.Karakas.Heger.ea:2020} contribute the high levels of Zn/Fe at lowest metallicities to hypernovae/collapsars. 
If we follow our earlier discussion related to Th/Fe
and its possible source in such events \citep{Siegel.Barnes.Metzger:2019,Siegel:2019}, these high Zn/Fe ratios at very low metallicities would 
provide a connection between these abundance features of a high Zn/Fe ratio and a strong (Th and actinide producing) r-process.

\subsection{The light trans-Fe elements, Sr, Y, Zr and correlations with Fe}
\label{sec:lighttransFe}

\noindent As the Fe-group elements 
Ni and Zn are also produced in core-collapse events, we looked for their possible correlations with Fe, especially whether there exists an indication for higher [Zn/Fe] production ratios with decreasing metallicities. The latter pointed to stronger alpha-rich freeze-out conditions in explosive Si-burning which is related to more energetic explosions, expected for massive star explosions with increasing compactness. We found for Zn a superposition of core-collapse events, adding for the lowest metallicities an additional source. This left the question whether there might exist a connection between the Zn enhancement and a second strong r-process source found preferentially at the lowest metallicities, as indicated by the [Th/Fe] enhancement in Fig.~\ref{figThFeH}.

\noindent Next we will also analyze how the light trans-Fe elements Sr, Y, and Zr behave in this respect. It is well known that for higher metallicities, when low and intermediate mass stars started to contribute to galactic evolution, these elements are produced in an s-process. However, observations at low metallicities required a different source, initially dubbed the LEPP \citep[light - heavy - element primary process, see][]{Travaglio.Gallino.ea:2004}. There exist possible explanations
related to core-collapse supernovae with mildly proton-rich and/or neutron-rich conditions \citep{Froehlich.Martinez-Pinedo.ea:2006,Pruet.Hoffman.ea:2006,Kratz.Farouqi.ea:2008,Farouqi.Kratz.ea:2008,Farouqi.Kratz.Pfeifer:2009,Roederer.Cowan.ea:2010,Arcones.Montes:2011,Arcones.Thielemann:2013,Akram.Farouqi.ea:2020}. On the other hand Sr has also recently been identified in the kilonova event following the neutron star merger GW170817 \citep{Watson.Hansen.ea:2019}. Thus, probably weak and strong r-process (and other) sources can contribute and this will be discussed in the following, based on correlations with Fe, in a similar way as performed before in sections \ref{corrEuFe} and in the previous subsection.
\begin{figure*}[h!]
\includegraphics[width=9cm,height=7cm]{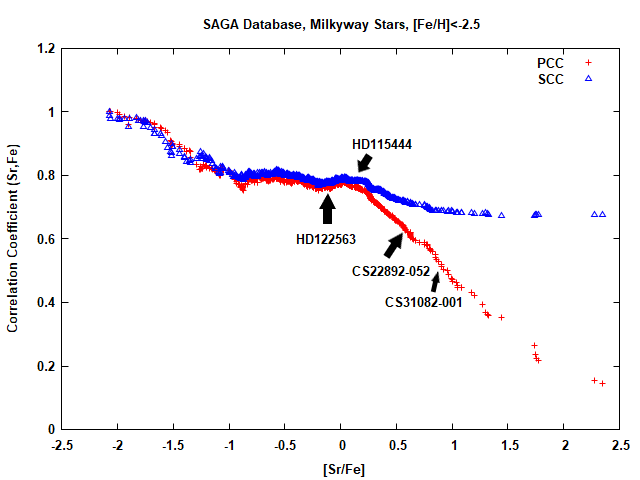}
\includegraphics[width=9cm,height=7cm]{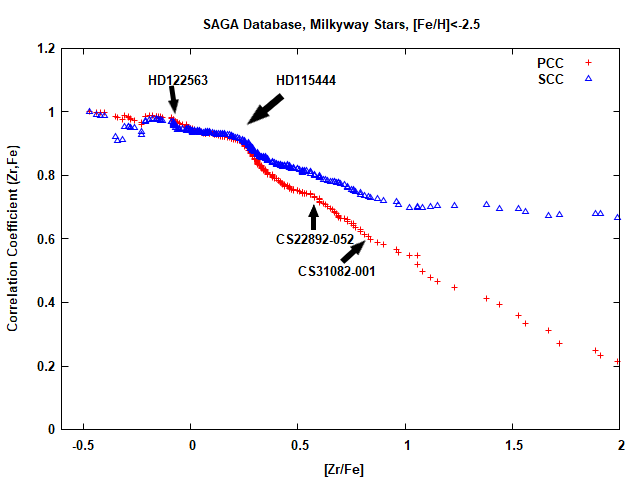}
\caption{The PCCs and SCCs of Fe with Sr, and Zr in stars with $\mbox{[Fe/H]}< -2.5$. Y (not shown here) behaves in a similar way. Contrary to Fig.~\ref{4starsplot} (right panel) the limited-r weak r-process star HD122563 does not appear for the lowest X/Fe ratios, being preceded to the left by other stars which apparently do not show r-process signatures. Otherwise the stars HD115444, CS22892-052, and CS31082-001 appear in the same order as seen in their increasing r-process strength. The ratios below HD122563 can be attributed to (a variety of?) core-collapse supernova events which do not eject r-process elements, but Sr, Y, and Zr. We define them as category 0 events.}
\label{corr_Sr_Y_Zr_Fe}
\end{figure*}

\noindent Similar to Fig.~\ref{4starsplot} (right panel),
we display in Fig.~\ref{corr_Sr_Y_Zr_Fe} the correlations of these elements with Fe as a function of [Sr/Fe] and [Zr/Fe] ([Y/Fe] behaves in a similar way). This extends our PCC and SCC analyses for these correlations, making use of exactly the same method as applied before for Fe, Eu, Ni, and Zn. 
 As in the Fe and Eu case, we see similar (although not identical) features, starting out with two (or better three!) kinds of plateaus with correlation coefficients close to 1 and 0.8, and declining correlations towards higher [Sr/Fe], [Y/Fe], and [Zr/Fe] values, combined with a divergence of the PCC and SCC curves. The PCCs go down to 0.2, i.e. indicate a vanishing correlation, while the SCCs can stay as high as 0.4-0.5 for reasons we discussed earlier, based on metric or rank deviations from the relation between the abundances. 
 
  At first glance it is not obvious how this behavior relates to limited or weak r-process stars and complete or enriched r-process stars, but the location of the four well-known stars HD122563, HD115444, CS22892-052, and CS331082-001 in these plots shows that the increase in [Sr/Fe], [Y/Fe], and [Zr/Fe] goes in parallel with the increase in [Eu/Fe]. While details might be different, the tendency is the same. However, the appearance of additional plateau phases to the left of HD122563, which stands for an extreme limited-r star in these plots (see also Fig.~\ref{4starsplot}, right panel), can be interpreted as stemming from additional sources, not related to an r-process. These must be core-collapse supernova events without r-process production, which we { postulate as category 0 events. Such sources, ranging from contributions of a $\nu$p-process to explosive Si-burning with moderate entropies and/or very modest variations of $Y_e$ around 0.5, have been discussed above.}
  These plateaus at low [X/Fe] values could not have been noted as a function of [Eu/Fe], as such category 0 events do not produce Eu.
  
    \begin{figure*}[h!]
\centering
\includegraphics[width=9.1cm,height=7cm]{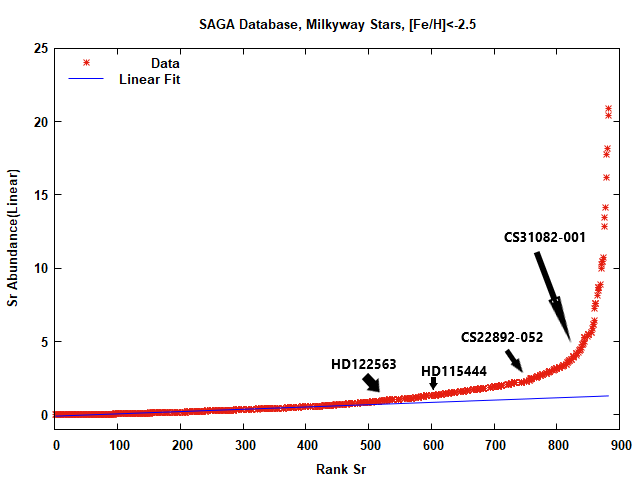}
\includegraphics[width=9.1cm,height=7cm]{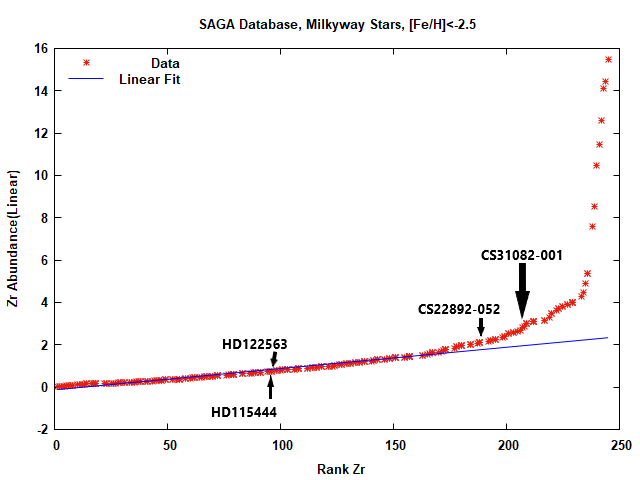}

\caption{Sr and Zr abundances versus their corresponding ranks up to the regime where the PCCs and SCCs indicate only low correlations, i.e. including complete r-process stars of regime 3. Fig.~\ref{corr_Sr_Y_Zr_Fe} showed already
that consideration of the whole range of observed [Sr/Fe] and [Zr/Fe]-values (the same applies for [Y/Fe]), including the high values with their large scatter, requires a (rare) source uncorrelated with Fe. But also the lower values alone (from the plateaus below [X/Fe]$\approx$0 up to close to the HD115444 value), displaying high correlations with respect to Fe, tend to require a superposition. While the behavior of abundances vs. ranks is close to linear up to HD122563, it starts with a non-linear trend for higher [X/Fe] values as a clear indication of a superposition of category 0, in addition to category I, and II events. The early linear trend points to a an apparently unique category 0 source, however, the different deviation points for Sr and Zr might indicate a further sub-category?
}
\label{sr_y_zr_vers_rank}
\end{figure*}

 \noindent The low values of Pearson and Spearman correlation coefficients, when considering the whole range of observed [Sr,Y,Zr/Fe]-values and their scatter, including the higher ones shown in Fig.~\ref{corr_Sr_Y_Zr_Fe}, make also clear that they require additional rare sources of events not correlated with Fe (category III). However, also the stars in the plateaus for the lower [Sr,Y,Zr/Fe]-values below HD115444, which show a high correlation with Fe, require a superposition of different sources, as the plots of their abundances vs. their ranks in Fig.~\ref{sr_y_zr_vers_rank} indicate. It can be noted that Sr shows in Fig.~\ref{corr_Sr_Y_Zr_Fe} an extension to lower [X/Fe] values and an earlier and stronger deviation from the linear fit in Fig.~\ref{sr_y_zr_vers_rank}. As discussed in appendix \ref{appsinglemultible}, the deviation from a linear relationship points to a superposition of sources in this regime. This would be explainable with a superposition of category 0, I, and II events. We have seen such patterns also in Figs.~\ref{eu_vers_rank} and \ref{eu_vers_rank_2} for the relation between Eu abundances vs. their rank, with a strong non-linear behavior in the first case, and a close to (but not perfectly) linear behavior in the second case. Thus, these limited-r or weak r-process regimes 1 and 2 required a superposition (of two sources?), as shown in Fig.~\ref{eu_vers_fe_fit}. For Sr, Y, and Zr we come to the same conclusions, but requiring an additional category 0 source, which does not produce Eu. This leads to the conclusion that there exists a strong source (category III) with no (or a negligible correlation) to Fe, and there exist (more than?) three sources which are correlated with Fe (category 0, I, and II). The strengths with which these sources affect different mass ranges (i.e. trans-Fe vs. Eu) seem different, but the principle effect is similar. It needs to be examined further whether the slightly stronger effect for Sr points to additional sub-categories.
 
 \noindent { Similar to the last paragraph in the previous section, one should point to the more extended interpretations of correlations as in our initial approach where we assumed that at low metallicities $<$-2.5 one sees only the imprint of one nucleosynthesis event and a correlation between two elements means automatically their co-production. We have seen as a first extension that one might see different subclasses, which all still show only the imprint of one type of nucleosynthesis event, where one can analyse the correlations restricted only in these subclasses. The next extension is that in certain observations we see the superposition of several types of events. If this is the case, the hypothesis that high correlations ask for co-production becomes invalid. For that case the rank analysis for each element of interest made it possible to test whether only one or several types of events contributed. We noticed that this is the case for Sr, Y, Zr, Eu and Th. For Th we came to the conclusion that two types of events had to exist among regime 3 (r-process enriched) stars (category IIIa and b), for Eu we noticed three types of events, category I and II for limited-r stars and category III for r-process enriched stars. Now for Sr, Y, Zr we noticed four types of events, category III for r-process enriched stars, category I and II for limited r-stars, and also category 0 (representing regular supernovae) for low metallicity stars which do not even have Eu detected, but show these three elements.
 
 \noindent
 If we summarize these issues with respect to Fe production, we see three types of sources: category 0 (regular CCSNe, producing Fe and light trans-Fe elements, like e.g. Sr, Y, Zr), and category I and II (special core-collapse supernovae with Fe production accompanied by weak r-processing up to the lanthanides, like Eu). Category IIIb events might also produce Fe but according to our existing analysis in negligible proportions in comparison to Eu.
 In \ref{sec:2.1} we have outlined, based on 
 \citet{Ryan.Norris.Beers:1996} that a typical supernova in typical ISM densities leads to an [Fe/H] value in the remnant of about -2.5 to -2.7. This was followed by the conclusion that such low-metallicity stars are probably only polluted by one prior nucleosynthesis event and abundance correlations (i.e. close to constant ratios between two abundances) can be interpreted by co-production. During the course of the present paper, we have by now identified quite a number of low-metallicity observations which point to the pollution by several preceding events. This is not really a surprise, as the metallicity of the ISM in a remnant does not have to coincide with the metallicity of a new-born star, possibly triggered by a nearby supernova. 
 The next stellar generation will form (after turbulent mixing triggered e.g. by the motion of galactic spiral arms) with a probably small contribution of a few percent or even less from this supernova remnant. Therefore one expects lowest-metallicity stars, affected by these first CCSNe, to possess values of about [Fe/H]$\approx$-5 or even less \citep[e.g.][]{Norris.Christlieb.ea:2007,Norris.Christlieb.ea:2012,Frebel.Norris:2015,Nordlander.ea:2017}. For events which occur with 1/10 of the frequency of regular CCSNe one might then expect the first appearance at [Fe/H]$\approx$-4 (after typically 10 supernovae enriched the ISM) and events which take only place after about 100 regular CCSNe polluted the ISM, are expected to show their impact at [Fe/H]$\approx$-3. This seems to be what we observe in Fig.~\ref{fig:fullEuFe_SrEuFe} for weak and strong r-process events, and one should therefore expect that we find already contributions from several events at such low metallicities. Thus, the initial hypothesis that a strong correlation is pointing strongly to co-production becomes in general invalid and our interpretation of co-production of Fe and Eu in category I and II events is not a proven one. However, the rank tool still holds and permits such interpretations. Therefore, for the moment we keep this view and will test it in section \ref{interpr} with existing model predictions.}

\section{Correlations of Actinides with Lanthanides and third r-Process Peak Elements}
\label{corrTh}
Before entering the subject of this section in detail, we want to mention beforehand that { our} work for it included an extended analysis with a closer look at the entire range of elements in the "mass vicinity" of Eu, i.e. from atomic numbers in the range 56 to 66. These elements are here only considered with respect to their observational abundance determinations. It should, nevertheless, be mentioned at this point that modeling of astrophysical conditions, combined with nuclear properties in order to obtain good fits to solar r-abundances, has led in the past to quite a number of publications with different approaches to (and interpretations for) the r-process production of these rare earth elements or lanthanides, especially related to the understanding of the so-called pigmy peak between the second and third r-process peak. 
This is, however, not the focus of this paper, here we want to focus on the interpretation of abundance observations and correlations. 

\noindent Following the earlier detailed analysis of the Eu behavior, in its extension to the rare earth (or lanthanide) elements we realized that they behave similarly with respect to Fe. The correlation coefficients follow similar trends: two plateau phases for regimes 1 and 2 of limited-r or weak r-process stars, and a strongly declining correlation when considering the whole [X/Fe]-range, including the complete or strong r-process stars of regime 3. These trends exist in general, although they can be more or less pronounced for the different elements. The overall low correlation with Fe (PCCs around 0.2) for the complete [X/Fe] range, including regime 3 stars, is again a strong indication for no or negligible co-production of Fe in the responsible category III events. Whether and how the lanthanides correlate with the third r-process peak and the actinides will now be analyzed more extensively in this section. 

\subsection{Th and rare earths elements (lanthanides)}

The large scatter of [Th/Fe] (see Fig.~\ref{figThFeH}), extending over more than two orders of magnitude (if considering that at lowest metallicities the real Th/Fe ratio was higher due to the Th-decay since then), is comparable to the [Eu/Fe] scatter in Fig.~\ref{fig:fullEuFe_SrEuFe}. Thus, considering also that Th is only { detected} in regime 3 stars, i.e. stemming only from category III events, where we find no or a negligible co-production of Eu and Fe, we also find very small global correlations of Th with Fe (PCC$<$0.2). We have linked these events dominantly to NSMs, an exception could be due to very massive stars turning into hypernovae/collapsars or magneto-rotational supernovae with negligible but existing Fe production, as discussed in the previous sections
\ref{veryearly-r} and \ref{sec:FeNiZn}.
Fig.~\ref{figThFeH} 
indicated that such objects might dominate the Th production at very low metallicities and Fig.~\ref{figThEuH} (right panel) showed that a superposition of Th-producing sites (in addition to NSMs) has to exist. Here we will examine the differences in Eu and Th productions and their correlation (which go either back to category I, II, and III for Eu or only category III events for Th).
Independent of this, the Fe in Eu as well as Th containing stars must dominantly originate from core-collapse events { (which can show their imprint already at metallicities as low as [Fe/H]=-5, see last paragraph of the previous section)}. Thus, all or a dominant fraction of Fe must stem from prior CCSN types of the categories 0, I, and II.  

\begin{figure*}[h!]
\centering
\includegraphics[width=9cm,height=5cm]{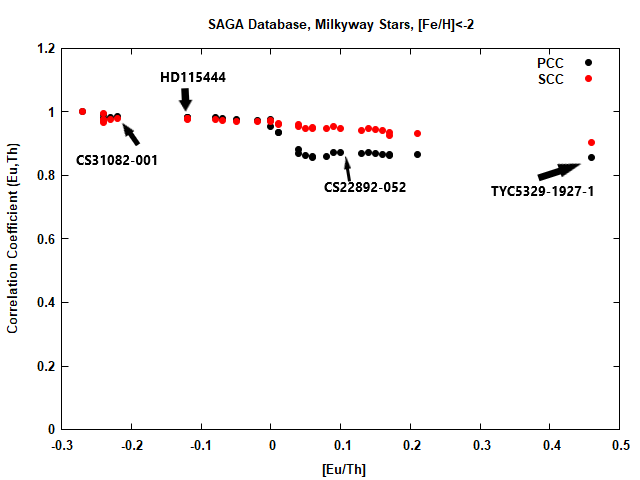}
\includegraphics[width=9cm,height=5cm]{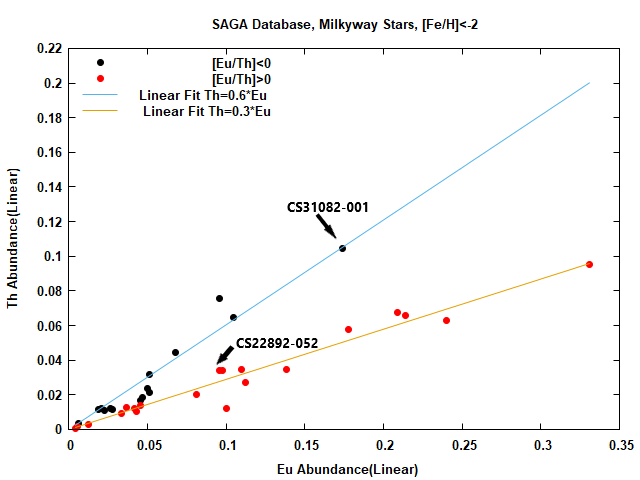}
\caption{Left panel: Pearson and Spearman correlation coefficients for Th and Eu in stars with $\mbox{[Fe/H]}\le$-2 as a function of [Eu/Th]. At $\mbox{[Eu/Th]}\simeq$0 they start diverging in a moderate way, separating two subgroups (actinide-boost and -normal stars). The positions of three typical stars CS31082-001 (complete, r-II, actinide boost), HD115444 (complete, r-I, actinide boost), CS22892-052 (complete, r-II), and TYC5329-1927-1 (complete, r-I) are indicated. It underlines that we can find actinide-boost stars among r-II as well as r-I stars. 
We will show further down that there exists a tendency that most actinide-boost stars are members of the r-II subset ([Eu/Fe]$>$1). But the Eu/Fe ratio can also reflect local inhomogeneous mixing with r-process products of different categories, therefore, not all actinide-boost stars belong to the r-II class. Right panel: Th versus Eu abundances in the two subgroups (actinide-boost and actinide normal) of stars with $\mbox{[Fe/H]}\le$-2 and corresponding linear fits.}
\label{Eu_vers_Th}
\end{figure*}

\noindent Fig.~\ref{Eu_vers_Th} (left panel) shows the PCC and SCC curves of Th and Eu as a function of [Eu/Th]. The linear relationship of Th and Eu is quite strong for the whole set of stars in the database utilized (PCC$\simeq 0.86$ and SCC$\simeq 0.91$).
Only a small divergence between the two statistical methods appears at $\mbox{[Eu/Th]}>0$, unlike in the case of Fe and Eu, which shows a pronounced discrepancy between the two methods when including category III events. This is due to a much smaller scatter in Th/Eu than in Eu/Fe, leading to smaller deviations for fits based on metric or rank displacements \citep[see e.g.,][]{Mashonkina:2014,Holmbeck.Beers.ea:2018}. Furthermore, two subgroups of stars seem to occur, the first one in the regime $\mbox{[Eu/Th]}<0$, where the PCCs and the SCCs are identical and almost equal to 1. This subgroup makes up for the so-called actinde-boost stars (where [Eu/Th] is subsolar) and includes the best known star CS31082-001 with $\mbox{[Eu/Fe]}\simeq 1.62$, $\mbox{[Fe/H]}\simeq -2.92$ and $\mbox{[Eu/Th]}\simeq -0.23$. In the second subgroup, with $\mbox{[Eu/Th]}\geq 0$, the PCCs and SCCs diverge slightly. This is probably an indication that among the r-enhanced stars two slightly different sets of strong r-process conditions prevail, leading to more or less actinide production. The best known star of this category is CS22892-052 with $\mbox{[Eu/Fe]}\simeq 1.53$, $\mbox{[Fe/H]}\simeq -3.1$ and $\mbox{[Eu/Th]}\simeq 0.17$.
{ Plotting the figure as a function of [Th/Eu] instead of [Eu/Th] would lead to a very similar result, just reversing the slightly different behavior in the sub vs. super-solar regime.} Fig.~\ref{Eu_vers_Th} (right panel) shows the plot of Th versus Eu for the two subgroups and the corresponding linear fits. The slope of the regression line in the subgroup of the actinide boost stars is twice as high as for the other subgroup. The corresponding r$^2$ in both fits are very high and almost equal to 1. 

\begin{figure*}[h]
\centering
\includegraphics[width=9cm,height=6cm]{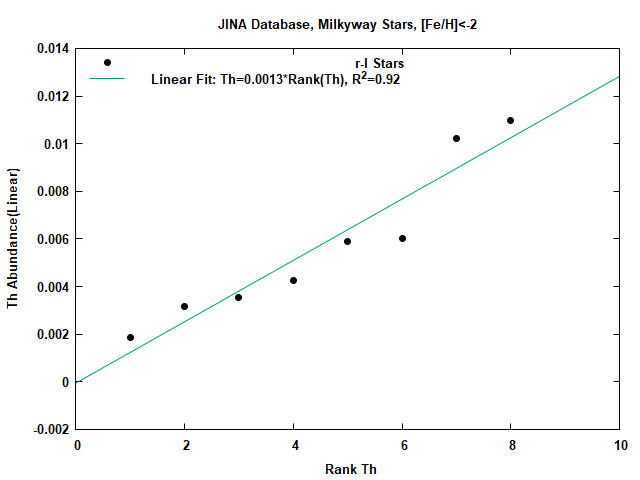}
\includegraphics[width=9cm,height=6cm]{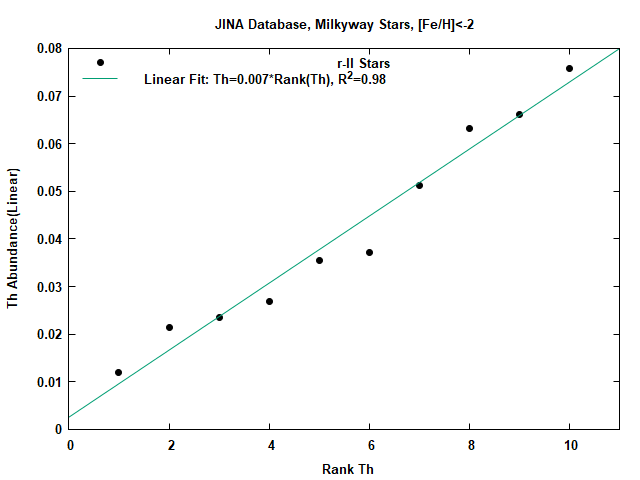}
\caption{Th abundances plotted vs. their ranks separately for r-I and r-II stars. The linear relation shows that each of these groups are polluted by only one type of r-process event, while the overall plot of Th abundances vs. their ranks in Fig.~\ref{figThEuH} (right panel) clearly indicated a superposition of (these two types of) events. Rare outliers seen for the r-I and r-II stars are probably due to the effect that the [Eu/Fe] classification is also affected by early inhomogeneous mixing of category IIIa and IIIb r-process ejecta with the previously existing ISM.}
\label{ranksThrIrII}%
\end{figure*}

\noindent Before entering a more detailed discussion how actinide boosts relate to the different regimes of complete r-process stars, i.e. r-I and r-II stars, we want to first have a look at the Th abundances alone. Fig.~\ref{figThEuH} (right panel) had shown that the observed Th abundances are due to a superposition of (at least) two different types of complete r-process sources, which we introduced as category IIIa and IIIb (only category III produces { detectable amounts of } elements in the third peak and beyond). In Fig.~\ref{ranksThrIrII} it is shown that one finds linear relationships between Th abundances and their the ranks, if one separates them into r-I and r-II stars, i.e. in each of these r-I and r-II subsets Th is apparently only polluted by one of the two IIIa or IIIb sources. One should keep in mind, this is a relation of the Th abundance with [Eu/Fe] ranges (different for r-I and r-II stars), pointing to the fact that high Th abundances caused by IIIb events are also linked to high [Eu/Fe] values of r-II stars. One can explain this behavior if one assumes that category IIIb events produce more main r-process elements than IIIa events by a large factor { (10?, separating r-I with [Eu/Fe]$>$0 and r-II star with [Eu/Fe]$>$1)}. Then, if IIIb stands for an actinide boost (and thus a higher Th/Eu ratio than in actinide-normal IIIa events), one finds in ISM regions polluted by such an event high Th abundances, but combined with high Eu abundances or [Eu/Fe] values as well. This then leads exactly to the linear relation between Th abundances and their ranks for r-II stars. In the same way, if IIIa events produce smaller Th amounts by a large factor in comparison to IIIb events, they will also produce small amounts of Eu (although the Eu/Th ratio is higher than in IIIb events, but by a relatively small factor), and lead to a smaller [Eu/Fe], i.e. the composition of r-I stars. Therefore, it looks that category IIIb events are linked { predominantly} to r-II stars, while categories IIIa events are linked { predominantly} to r-I stars. 

\noindent Outliers from this behavior, as also found in Fig.~\ref{ranksThrIrII}, can occur due to a two phase or inhomogeneous pollution in the early Galaxy. The ISM which eventually formed the low metallicity star we observe, can have experienced subsequent pollutions. If a IIIa event is followed by a category I or II event with a strong nearby contribution, it can lift the [Eu/Fe] highly and causes the star, which carries an actinide-normal behavior, to display an r-II [Eu/Fe] pattern. On the other hand, if a category IIIb event took place quite remotely from the ISM cloud, which led to the observed low metallicity star, it will implant the actinide-boost pattern for Th/Eu, but the mixed in Eu is relatively small in comparison to the pre-existing Fe of prior CCSNe, this causes an r-I [Eu/Fe] pattern. This way, especially in two phase pollution events r-I stars can also carry an actinide-boost pattern and r-II stars an actinide-normal pattern.

\begin{table*}[h!]
\caption{(Present day) Observed Eu and Th abundances in $\log\epsilon(X)= \log(N_X/N_H+12)$ values in low metallicity stars ([Fe/H]< -2.4). The corresponding solar system ratio at the time of its birth (representing the ISM when the solar system got detached) is 0.45 or 0.42
      \citep{lodders09,Lodders19}. Based on their present Th/Eu ratios, the red highlighted stars would be considered as an actinde-boost stars with Th/Eu$>$0.42.}      
      
         \label{th_eu_ratio}
     $$ 
         \begin{array}{ l    c    c     c     c      c     c       c}
            \hline
            \noalign{\smallskip}
\mbox{Star} & \mbox{Eu} & \mbox{Th} & \mbox{(Th/Eu)}  &\mbox {\rm regime 3 (complete r-process) subclass} & \mbox{[Eu/Fe]}\\
            \hline
            \noalign{\smallskip}
{\rm TYC5329-1927-1}	&	-1	&	-1.91	&	0.12	& {\rm r-I} &	0.89	\\
{\rm J0858-0809}	&	-2.41	&	-3.07	&	0.22	&	{\rm r-I}	&	0.23	\\
{\rm HD108317}	&	-1.37	&	-1.99	&	0.24	&	{\rm r-I} &	0.64	\\
{\rm HD110184}	&	-1.91	&	-2.5	&	0.26	&	{\rm r-I}	&	0.08	\\
{\rm HE1523-0901}	&	-0.62	&	-1.2	&	0.26	& {\rm r-II}	&	1.86	\\
{\rm CS22892-052}	&	-0.86	&	-1.42	&	0.28	&	{\rm r-II}	&	1.53	\\
{\rm HD186478}	&	-1.34	&	-1.85	&	0.31	&	{\rm r-I}	&	0.63	\\
{\rm CS29491-069}	&	-0.96	&	-1.46	&	0.32	&	{\rm r-II}	&	1.12	\\
{\rm RAVEJ203843.2-002333}	&	-0.75	&	-1.24	&	0.32 &	{\rm r-II}	&	1.64	\\
{\rm CS29497-004}	&	-0.68	&	-1.17	&	0.32	&	{\rm r-II}	&	1.44	\\
{\rm J1432-4125}	&	-1.01	&	-1.47	&	0.35	&	{\rm r-II}	&	1.44	\\
{\rm HE0240-0807}	&	-1.44	&	-1.9	&	0.35	&	{\rm r-I}	&	0.55	\\
{\rm BD-15_5781}	&	-2.28	&	-2.73	&	0.35	&	{\rm r-I}	&	0.12	\\
{\rm HE2224+0143}	&	-1.02	&	-1.47	&	0.35	&	{\rm r-I}	&	0.87	\\
{\color{red}{\rm HE2327-5642}}	&	-1.29	&	-1.67	&	0.42	&	{\rm r-II}	&	1.07	\\
{\color{red}{\rm HD6268}}	&	-1.56	&	-1.93	&	0.43	&	{\rm r-I}	&	0.54	\\
{\color{red}{\rm HD115444}}	&	-1.64	&	-1.97	&	0.47	&	{\rm r-I}	&	0.68	\\
{\color{red}{\rm HE2252-4225}}	&	-1.3	&	-1.63	&	0.47	&	{\rm r-II}	&	1.12	\\
{\color{red}{\rm CS22953-003}}	&	-1.69	&	-1.92	&	0.59 &	{\rm r-I\ or\ r-II}	&	0.92	\\
{\color{red}{\rm CS31082-001}}	&	-0.76	&	-0.98	&	0.60	&	{\rm r-II}	&	1.62	\\
{\color{red}{\rm CS30315-029}}	&	-2.24	&	-2.45	&	0.62	&	{\rm r-I}	&	0.67	\\
{\color{red}{\rm HE1219-0312}}	&	-0.98	&	-1.19	&	0.62	&	{\rm r-II}	&	1.47	\\
{\color{red}{\rm CS31078-018}}	&	-1.17	&	-1.35	&	0.66	&	{\rm r-II}	&	1.15	\\
{\color{red}{\rm 2MASSJ09544277+5246414}}	&	-1.19	&	-1.31	&	0.76 &	{\rm r-II}	&	1.28	\\
				\noalign{\smallskip}
            \hline
         \end{array}
         $$
   \end{table*}
\noindent At this point we list a number of low-metallicity stars with actinide-boost and actinide-normal patterns in combination with their r-II and r-I nature. Before doing so, one should reflect about what is really meant by a regular (actinide-normal) r-abundance pattern and an actinide boost, considering also that $^{232}$Th has a half-life of 1.405$\times 10^{10}$y.
\cite{lodders09} and \cite{Lodders19} give a solar ratio of Th/Eu=0.45 or 0.42 ($\log_{10}=-0.347$ or -0.377), dated back to the beginning of the solar system 4.57 Gyr ago. This value resulted from Th contributions to the ISM out of which the solar system formed, which requires also to consider the (long) Th half-life since these events took place \citep[e.g.][]{hotokezaka15,Cote.Yague.ea:2019}.
The present terminology puts a star with observed Th/Eu$>$0.42 (the value in the solar system at the date of its birth) in the actinide boost slot. 
Stars were born with higher Th/Eu ratios than listed in Table~\ref{th_eu_ratio}, this difference is actually utilized for age determinations.
When applying the actinide-boost definition of Th/Eu$>$0.42, we see 10 actinide-boost stars with 6 of them being r-II stars. If relaxing the limit between r-I and r-II stars from [Eu/Fe]=1 to 0.9 we find 7 of them being of r-II stars. This goes in line with the previous finding from Fig.~\ref{ranksThrIrII} that actinide-boost stars belong dominantly to the r-II class, but inhomogeneous pollution in the early Galaxy by several overlying events can also lead to high [Eu/Fe] values for stars being born with actinide-normal patterns, as well as low [Eu/Fe] values for an actinide-boost pattern. However, the apparently dominating coincidence of joint r-II and actinide-boost characteristics seems to point to the fact that events responsible for an actinide-boost eject also much more r-process matter in total (determining [Eu/Fe]) than those responsible for actinide-normal characteristics.

\noindent For comparison we have also done the same analysis for other lanthanide elements in addition to Eu, i.e. La, Dy, Er, and Yb. 
They all behave similar to Eu/Th, for the subsolar [X/Th] cases the stars belong to the actinide-boost stars with { [Eu/Th]$<$0}, for the supersolar cases to the "regular" r-processed enriched actinide-normal stars.

\subsection{Th and (close to) third r-process peak elements}

\begin{figure*}[h!]
\centering
\includegraphics[width=9cm,height=6cm]{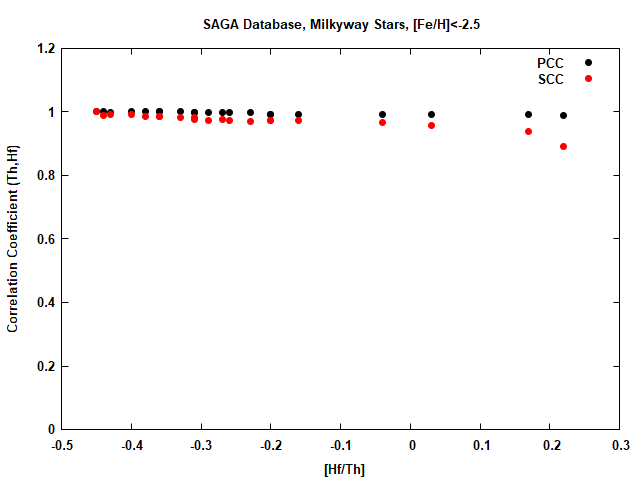}
\includegraphics[width=9cm,height=6cm]{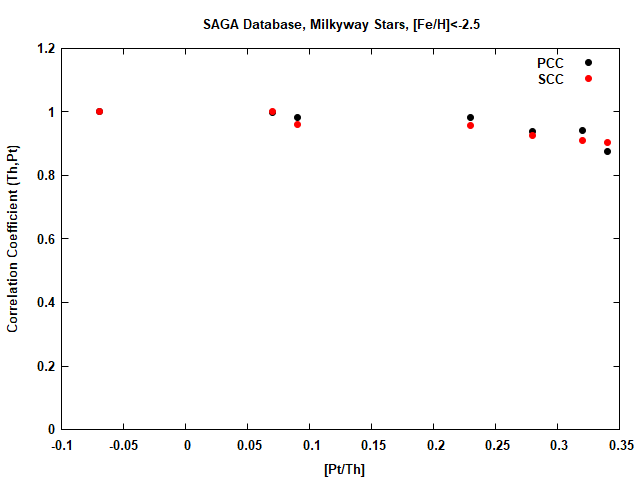}

\caption{The Pearson and Spearman correlation coefficients of Th and element X (Hf and Pt) close to the third r-process peak in stars with $\mbox{[Fe/H]}\le -2.5$ as a function of [X/Th]. The PCCs and SCCs divergence above $\mbox{[X/Th]}\simeq 0$ is much smaller than for the lanthanides. Not shown Os behaves in an almost identical way.}
\label{X_vers_Th_2}%
\end{figure*}

\noindent In the previous subsection we discussed a clear correlation of the lanthanide elements with Th, varying, however, slightly between the actinide boost stars and the "regular" r-enhanced stars, pointing to very similar but slightly different strong r-process conditions for those elements. Fig.~\ref{Eu_vers_Th} had shown this small but measurable difference between actinide boost stars and the regular r-enhanced stars via deviations of the PCC and SCC values, measuring the scatter in different ways. In Fig.~\ref{X_vers_Th_2} we display these correlations with Th for elements being part of or close to the third r-process peak.
We find that opposite to Fig.~\ref{Eu_vers_Th} the slight divergence between PCCs and SCCs becomes { almost} negligible, indicating that the pathway to the actinides passes through the third r-process peak. This can be interpreted in such a way that if r-process environments manage to reach the third peak, they  also lead up to the actinides with similar results.

\begin{figure*}[h!]
\centering
\includegraphics[width=9cm,height=6cm]{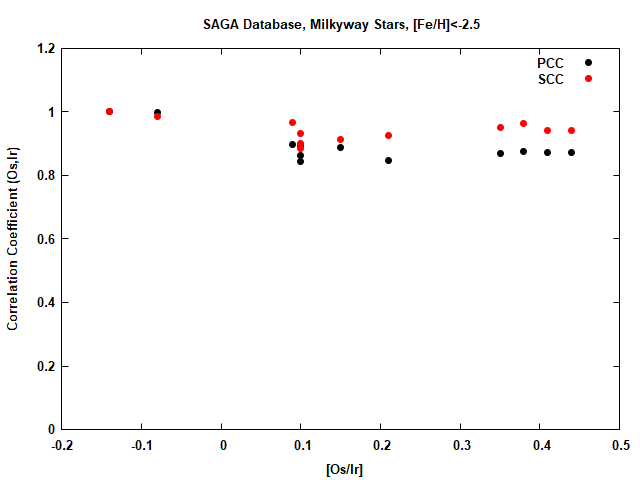}\\
\includegraphics[width=8cm,height=5cm]{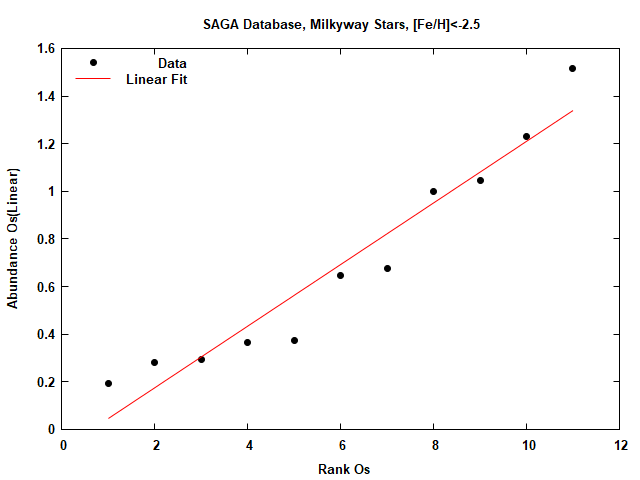}
\includegraphics[width=8cm,height=5cm]{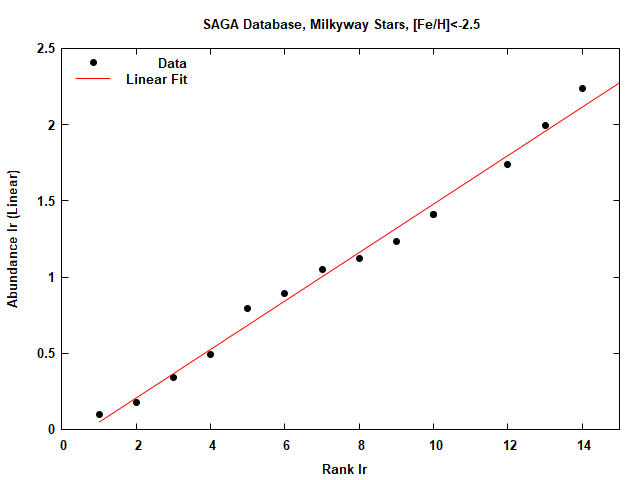}
\caption{(a) Os vs. Ir correlation and (b and c) Os and Ir vs. their ranks. The data set includes stars with $\mbox{[Fe/H]}\le -2.5$ and the corresponding linear fits.}
\label{OsIr}%
\end{figure*}
\noindent However, there remain still small variations between elements at the (low mass) slope of the third r-process peak for actinide boost and "regular" r-process enriched stars, as we can see in Fig.~\ref{OsIr} for Os ("average" A=190) and Ir ("average" A=192). This can again been noticed in the correlation plot vs. [Os/Ir], where the PCCs and SCCs display a small divergence for the higher (Os/Ir) values, belonging to "regular" r-process enriched stars, and the linear correlation seems slightly smaller. When testing the Os and Ir abundances versus their ranks, the relation appears almost linear, i.e. as expected from a single source. However, small deviations from a linear relation are visible, indicating a possible superposition of similar but slightly different conditions. When the abundances of Os vs. Ir are plotted, but split for two different regimes ([Os/Ir]$ < $0.02 or [Os/Ir]$ > $0.02), the slope changes by almost a factor of 2 between these two groups, corresponding to actinide boost and regular r-processed enriched stars.
This behavior (not shown here in a plot) underlines, that the lower mass slope of the third r-process peak is still sensitive to such similar but slightly different conditions. On the other hand, when matter has reached the peak, the flow up to the actinides seems unchanged.

\section{Interpreting the Observational Trends of Correlations in the Framework of Possible r-Process Sites}
\label{interpr}
\subsection{Our current findings from the analysis of observational data}
\label{sec:6.1}
In the preceding sections we have gone through a large set of elements at low metallicities with [Fe/H]$<$-2 or -2.5, initially focusing on Eu, but in addition analyzing Sr, Y, Zr, (Ba, La, Nd, Dy, Er, Yb,) Hf, Os, Ir, Pt, and Th, which (at least) at low metallicities are dominated by (weak or strong) r-process contributions. The results for the elements in brackets from Z=56 to Z=72
have not been presented here, but { we found them to lead} to similar results as discussed for Eu. Sr, Y, and Zr have at low metallicities also contributions from regular CCSNe and not only from r-process sites. If not considering any selection bias in the SAGA database \citep{Sagadatabase}, we can find in total 665 stars with [Fe/H]$<$-2.5, for 379 of them only upper limits with respect to the Eu abundance are known, i.e. they could contain either no or only a negligible r-process contribution.
Of the remaining 286 stars, for which direct detections of the Eu abundance exist, 76 belong to the r-poor or limited-r stars with [Eu/Fe]$<$0-0.1 and 210 to the r-rich or complete r-process stars with [Eu/Fe]$>$0-0.1. { Thus, a sizable fraction of these low-metallicity stars are carrying the imprint of one or { several(?)} r-process site(s). For all of the ones with [Eu/Fe]$<$0-0.1 we found Eu to be strongly correlated with Fe, i.e. a co-production of Fe and Eu was indicated, { if we follow the initial hypothesis that at these low metallicities stars carry only the imprint of one nucleosynthesis event and correlation hints at co-production in these events}. 
This was also seen for Sr, Y, Zr, as well as Ba and the lanthanide elements La, Nd, Dy, Er, and Yb.\\ 
Typically we found two kinds of plateaus in the correlation { patterns} among limited-r stars with Fe, seeing for the first one (a) a clear linear relation with the Fe-abundance, and in the slightly less correlated region (b) an apparently also linear relation with Fe, but with a larger slope and also a larger scatter.
The second case, could point to a further r-process source, also displaying a high correlation with Fe. These limited-r or weak r-process stars with a high correlation to Fe (a core-collapse product) made the connection to core-collapse events, occurring early on in galactic evolution. They seem to produce Fe and r-process matter, but not (in detectable amounts) up to the third r-process peak. We have introduced them as category I and category II events.
Therefore we concluded that category I and II components co-produce Fe and the r-process elements up to the lanthanides, however, category II in a stronger (but more scattered) fashion. For Sr, Y, Zr the (two) plateaus seem to appear a bit more fuzzy or can, if plotted vs. [X=Sr,Y,Zr/Fe] rather than [Eu/Fe], even be interpreted as a three plateau behavior. For this behavior, one can introduce an additional category 0, only producing these light trans-Fe elements, but not contributing up to Eu. Thus, regular CCSNe are a clear candidate.}\\
{ The latter new category of objects could, however, cause a dilemma in our previous interpretations. We started with the hypothesis that low-metallicity stars with [Fe/H]$<$-2.5 carry only the imprint of one nucleosynthesis site and then a correlation, i.e. a close to constant ratio of two elements, points clearly to a co-production in that specific site. If we have a number of contributing categories of nucleosynthesis sites, but each one of them only visible in one specific class of observed stars, one could separate these classes and still follow this interpretation. A problem occurs, if several categories contributed to the same star. We had this problem first when looking at r-process enriched stars which also contain Fe in addition to r-process elements. The strong advantage was that because of a clearly missing correlation, this underlined separate independent contributions and thus made clear that the contributing r-process site(s) must come with no or negligible Fe production. Opposite to that, we have now limited-r stars with a correlation of Eu with Fe, which we identified with category I and II contributions, but in addition other stars without detected Eu are displaying correlations of Sr, Y, Zr with Fe. I.e. in both latter cases (for the last one we introduced category 0, regular CCSNe, as contributing sites) we concluded co-production of Fe with either Sr, Y, Zr or with Eu, but the limited r-stars contain also Sr jointly with Eu in a large Sr/Eu ratio. How could one exclude that the categories 0, I, and II contributed in a combined way to a low-metallicity star without coming to the conclusion that the correlation with Fe is a spurious one not permitting the indication of a co-production.\\
\noindent As we discussed in previous sections, the remnant of a single CCSN should have a [Fe/H] value between -2.5 and -3. However, most probably the star formation triggered by this nearby supernova led probably not to a more than 10\% pollution of the protostellar cloud, if not less. This would point to the lowest metallicity stars in the range [Fe/H]$\approx$-4 or less (we see them already at -6). Thus, at [Fe/H]=-3, we could have already contributions by 10 to 100 CCSNe. The question is how often category I or II events could have contributed as well. If one starts with a suggested site, MHD supernovae - possibly related to magnetar-producing events \citep[e.g.][]{Beniamini.Hotokezaka.Horst.ea:2019}, one would expect that about 10\% of CCSNe would fall into the class of MHD supernovae.
Taking present-day models of MHD supernovae with the ejection of about $10^{-6}$M$_\odot$ of Eu \citep[e.g.][]{Winteler.Kaeppeli.ea:2012,Nishimura.Sawai.ea:2017,Reichert.Obergaulinger.ea:2021}, and 10 regular CCSNe with about 0.1M$_\odot$ of Fe, would lead to a mass ratio of about $10^{-6}$ and an abundance ratio of about $3\times 10^{-7}$ (almost independent of considering also Fe ejecta of such a category II event). This compares already reasonably well with what can be seen in Fig.~\ref{eu_fe_ratio}, right panel, where we see a scatter of a factor of 2 (up and down), allowing for variations in the Fe contributions from CCSNe as well as the Eu contribution from an MHD supernova. This can explain the moderately strong correlation found in PCC value in Fig.\ref{4starsplot} and Table \ref{tab:pcc1}, but puts doubts on a proof for co-production. We should keep this in mind for the further analysis in this section. Nevertheless, we keep for the moment the idea of such category I and II events with co-production of Fe and Eu, leaving direct tests of model predictions for them in section \ref{sec:6.3.1}.}
{ Fortunately the tool of the abundance vs. rank analysis, utilized by now in a number of cases, is not affected by the uncertainties which we revealed above, permitting to identify several r-process contributions for a number of elements in limited-r as well as r-process enriched stars.}

\noindent Among the totality of r-process enriched stars with [Eu/Fe]$>$0-0.1, we noticed up to now a vanishing or at most negligible correlation of r-process elements with Fe (see Fig.~\ref{4starsplot}, right panel and Table~\ref{tab:pcc0}). Their abundances must have originated from events with a strong r-process and negligible Fe production which we introduced as category III events. The existence of Fe, if not correlated with an r-process, points to an almost pure r-process site, polluting matter which had previous Fe contributions from CCSNe. We also noticed a non-linear behavior of abundances vs. their ranks for Eu and Th among r-I and r-II stars combined (i.e. with [Eu/Fe]$>$0-0.1),  when considering all stars with [Fe/H]$<$-2.5. Thus, also category III events include superpositions (see Figs.~\ref{eu_vers_rank}, left panel, and \ref{figThEuH}, right panel). 
\begin{figure*}[h!]
\centering
\includegraphics[width=9cm,height=5cm]{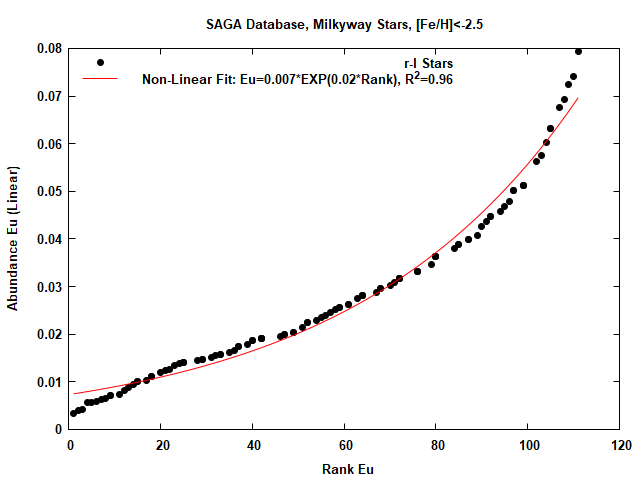}
\includegraphics[width=9cm,height=5cm]{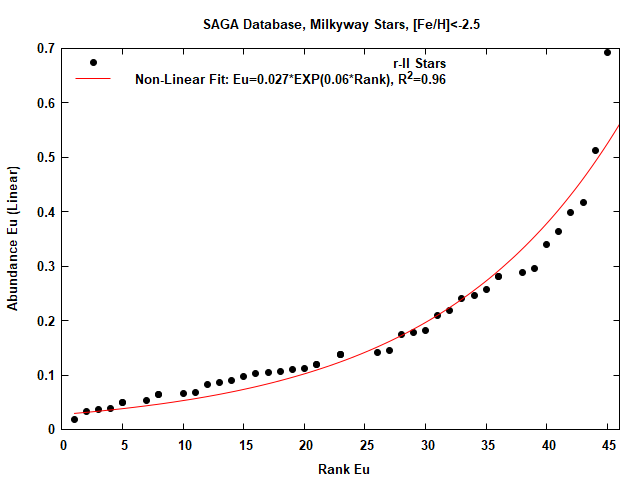}
\caption{Eu abundances vs. their ranks for r-I and r-II stars, separately. Opposite to Th which clearly pointed to two distinct sources for these two regimes, each via a linear relationship, Eu still seems to require other contributions, in addition to the subcategories IIIa and IIIb.}
\label{EurankrIrII}%
\end{figure*}
We have noticed in the case of Th, that this superposition led to a clear division into two types of events, displaying within regime 3 of r-process enriched stars a linear relation for each of the two subregimes of r-I and r-II stars (see Fig.~\ref{ranksThrIrII}). This finding relates to r-I and r-II stars which exhibit Th detections. If we try the same approach for Eu separately for r-I and r-II stars, we do not see clear linear relations, as displayed in Fig.~\ref{EurankrIrII}, i.e. in regime 3. The Eu contributions to r-I as well as r-II stars must, in addition to category IIIa and IIIb sources, experience also a feeding in from category I/II events which dominate in regimes 1 and 2. Thus, subcategories IIIa/b are solely responsible for producing elements in and beyond the third r-process peak (like Th), but the lanthanide element Eu displays also in regime 3 contributions from category I/II events. We know from our earlier analysis that such category I/II events produce Eu. Here we see that these types of events provide also a contribution to regime 3 and not only to limited-r regime 1 and 2 stars. We will attempt to estimate this contribution in the following subsection.

\begin{figure*}[h!]
\centering
\includegraphics[width=9cm,height=5cm]{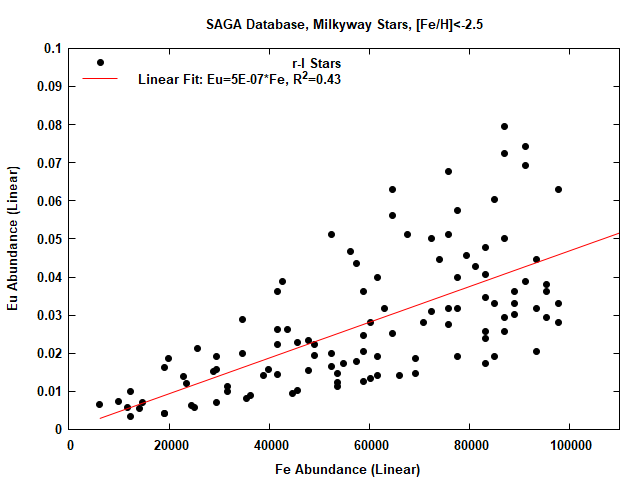}
\includegraphics[width=9cm,height=5cm]{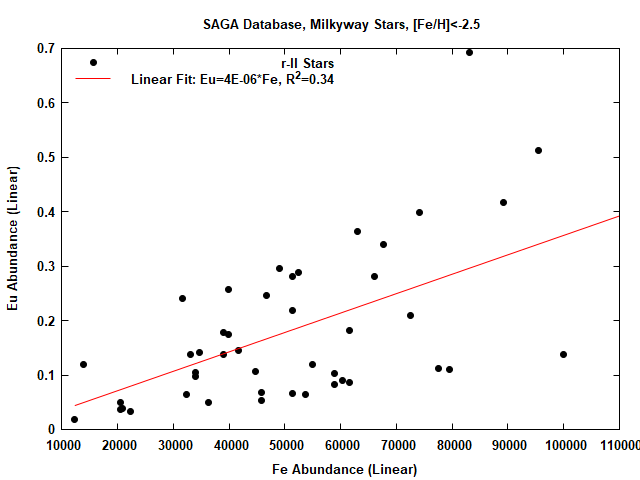}
\caption{Eu abundances vs. Fe for r-I and r-II stars, separately. The PCC values of 0.66 and 0.42, corresponding to the displayed $r^2$'s underline a very moderate or no correlations. The determined mean Eu/Fe ratios will be utilized in section~\ref{sec:Cat_III}.}
\label{EuFerIrII}%
\end{figure*}

\noindent If we attempt a linear fit of Eu vs. Fe abundances, separately for r-I and r-II stars (although knowing from our earlier analysis that over the whole regime of complete r-stars Eu and Fe are uncorrelated), we obtain the following results (Fig.~\ref{EuFerIrII}). The $r^2$ value of 0.43 for r-I stars corresponds to a PCC of 0.66 while 0.18 for r-II stars corresponds to 0.42, underlining the very moderate to poor correlation. But the linear Eu/Fe ratios determined, provide averages for the two samples, equivalent to mean [Eu/Fe] values for the two subregimes. We will apply these means later when trying to identify the origin of the observed abundance patterns.   

\subsection{Attempts to link event categories to contributions in observational regimes and to determine their occurrence rate}
{ We have discussed in the preceding subsection that the interpretation of abundance correlations becomes more complicated if several events have contributed to the elements X and Y, as we know that already at low metallicties supernovae contribute via Fe ejecta (and possibly Sr, Y, Zr) but not necessarily to true r-process elements.} In the following we will test these options in order to see which kind of events (i.e. which of the categories 0, I, II, and III) contributed, and whether it is possible to give an idea of the respective role they play in the overall patterns of r-process elements found in regimes 1, 2, and 3.

\subsubsection{An estimate for the fraction of the elements Sr, Y, Zr, 
and Eu made by non-CCSNe events} 
As already shown in the previous sections, in stars with $\mbox{[Eu/Fe]<0-0.3}$ the rare earth (RE) or lanthanide element Eu can be produced by events of category I and/or II, which we have attributed to specific CCSNe scenarios. For $\mbox{[Eu/Fe]>0-0.3}$, i.e. complete r-process stars of regime 3 (corresponding to r-I and r-II stars), the Eu production has been predominantly assigned to category III events with no or negligible Fe co-production (the latter could still be divided in categories IIIa and IIIb, but we leave them together for the moment in order to simplify things). We will later discuss the astrophysical origin of all these categories. In order to estimate the respective Eu fraction from weak r-process sites (category I/II) and highly efficient r-process sites of category III, we utilize the relationship between the Eu abundances and their ranks.\\ 
As we discovered earlier, also shown in Fig.~\ref{eu_vers_rank} (right panel), for Fe this type of relationship seems perfectly linear throughout all three regimes, hinting at one single type of production site (core collapse of massive stars). In contrast to Fe, however, such a linearity in the case of Eu is close to solid only for regimes 1 and 2 (see Fig.~\ref{eu_vers_rank_2}), corresponding to category I and II events, i.e. events which we interpreted also to result from core collapse via a kind of supernova which co-produces Fe and Eu.
As we have shown in Fig.~\ref{ranksThrIrII},
we have for Th a clear sign of unique contributions from category IIIa and IIIb events and no { detectable} superpositions from supernova-type events which produce a weak r-process. Opposite to this, Eu seems in all regimes to indicate superpositions (see Figs.~\ref{eu_vers_rank},\ref{eu_vers_rank_2},\ref{EurankrIrII}), i.e. also contributions from category I/II events. It is therefore important to also estimate for regime 3 the contributions from weak r-process (category I and II specific CCSN) events, leaving the remaining fraction for category III events. The assumption
is that the trend found before for weak r-process events continues. 
One option is to utilize Fig.~\ref{eu_vers_rank_2} which permits to predict the Eu abundance (calibrated for weak r-process stars with a coefficient of the order $10^{-4}$) as a function of the Eu rank. This method is utilized for the third entry in Table~\ref{eu_amount}. 
For the rank method we use the following relation,
    $\mbox{Eu(category I+II)}=9\times 10^{-5}\times\mbox{Rank(Eu)}$, 
and the amount of Eu which then needs to be attributed to events of category III is given by
    $\mbox{Eu(III)}=\mbox{Eu(tot)}-\mbox{Eu(category I+II)}$. 
\begin{table}[h!]
\caption{Estimate of the Eu amount (in \%) stemming from weak r-process CCSNe (category I+II) in selected stars with $\mbox{[Eu/Fe]}>0.3$.}
\label{eu_amount}
\begin{tabular}{lcc}
\hline\hline
Star & [Eu/Fe] & Eu(CCSNe category I/II)$_{\mbox{rank}}$ \\
\hline
\hline
HD107752	&	0.31&	22.2\\
BS16543-097	&	0.4	&	14 \\
HE0105-6141	&	0.51 &	12.6	\\
HE2206-2245	&	0.6	&	14.1	\\
HD115444	&	0.68 &	14.9	\\
HE0315+0000	&	0.7	&	9.9	\\
CS29499-003	&	0.79 &	7.3\\
CS22882-001	&	0.81 & 8.7	\\
HE1127-1143	&	0.9	&	6.6	\\
HE2138-3336	&	1.09 &	7.1	\\
CS22183-015	&	1.36 &	4.6	\\
HE2208-1239	&	1.52 &	3.6 \\
CS22892-052	&	1.53 &	3.7	\\
CS31082-001	&	1.62 &	3	\\
CS31062-012	&	1.65 &	1.3	\\
HE2258-6358	&	1.68 &	1.6	\\
HE0010-3422	&	1.72 &	1.8	\\
LP625-44	&	1.74 &	1.5	\\
CS29497-034	&	1.79 &	2	\\
HE0243-3044	&	1.9	&	0.8	\\
\hline
\hline
\end{tabular}
\end{table}

\begin{table}[h!]
\caption{Estimate of the Zr amount (in \%) originating from CCSN events in selected stars with $\mbox{[Zr/Fe]}>0.05$ and Zr detections.}
\label{sr_amount}
\begin{tabular}{lcc}
\hline\hline
Star& [Zr/Fe] & Zr(CCSN events)$_{\mbox{rank}}$ \\
\hline
\hline
HD107752	&	0.05&	70\\
BS16543-097	&	0.31	&	45 \\
HE0105-6141	&	0.19 &	55	\\
HE2206-2245	&		&		\\
HD115444	&	0.2 &	66	\\
HE0315+0000	&	0.3	&	49	\\
CS29499-003	&	0.45 &	36\\
CS22882-001	&	0.44 & 43	\\
HE1127-1143	&		&	\\
HE2138-3336	&	0.81 &	30	\\
CS22183-015	&	0.76 &	37	\\
HE2208-1239	&	0.84 &	34 \\
CS22892-052	&	0.62 &	49	\\
CS31082-001	&	0.75 &	41	\\
CS31062-012	&	 &		\\
HE2258-6358	&	0.69 &	31	\\
HE0010-3422	&	1.08 &	17	\\
LP625-44	&	1.52 &	5.4	\\
CS29497-034	&	 &		\\
HE0243-3044	&	1.06	&	11	\\
\hline
\hline
\end{tabular}
\end{table}
\noindent Table~\ref{eu_amount} summarizes for a number of selected regime 3 stars the fraction of Eu predicted this way by weak r-process CCSNe.
When utilizing these estimates and integrating over the total number of VMP stars with $-0.64\le\mbox{[Eu/Fe]}\le1.92$ in the SAGA database, this results in an amount of Eu originating from weak r-process CCSNe of the order 15\%. Thus, the major part of Eu (and very likely the whole RE element or lanthanide region) should be due to category III events, but the category I/II contribution is substantial.\\
\noindent We have utilized a similar approach for
the three adjacent light trans-Fe elements Sr (Z=38), Y (Z=39) and Zr (Z=40).
Here we show Zr as representative case for the neighboring trans-Fe elements and extend our PCC and SCC analyses also for the correlation of Fe with Zr, utilizing exactly the same method as before for Fe and Eu, { applicable here, because Fe as well as Sr, Y, Zr come from the same dominating event, regular CCSNe}. Fig.~\ref{corr_Sr_Y_Zr_Fe} showed the resulting curves for the correlation coefficients versus [Zr/Fe]. As in the Fe and Eu case
(Fig.~\ref{4starsplot}, right panel),
we observe a divergence of the PCC and SCC curves, here for a value of $\mbox{[Zr/Fe]}\simeq 0.3$. The relation between Zr and its ranks is shown in Fig.~\ref{sr_y_zr_vers_rank}. Like in the case of Eu, the best fit between Zr and its ranks is close to linear for the low rank values of regime 1 and 2 stars. 
For a first order estimate we utilize the linear relationship shown in Fig.~\ref{sr_y_zr_vers_rank} to obtain the results shown in Table~\ref{sr_amount} (containing the same stars as Table~\ref{eu_amount} but showing gaps when Zr detections are not known).
 When integrating over the total number of the VMP stars with $-0.5\le\mbox{[Zr/Fe]}\le 1.91$ in the SAGA databse, the amount of Zr originating from CCSN-type events is about 33\%. 
The chemical element Y (Z=39) has only one stable isotope ($^{89}\mbox{Y}$). It shows the same behaviour as [Sr/Fe] and [Zr/Fe], concerning the divergence between the PCC and SCC curves. This happens at $\mbox{[Y/Fe]}\simeq -0.15$ and Y is also the product of different astrophysical sources.
When integrating over the total number of the VMP stars with $-1.49\le\mbox{[Y/Fe]}\le 1.41$ in the SAGA database, the amount of Y originating from CCSNe is about 37\%. 
Concerning the chemical element Sr (Z=38), the divergence between the PCC and SCC curves 
takes place at roughly $\mbox{[Sr/Fe]}\simeq 0.2$ (see Fig.~\ref{corr_Sr_Y_Zr_Fe}). 
Following  the same procedure as for Zr and Y, when integrating over the whole VMP stars with
$-2\le\mbox{[Sr/Fe]}\le 1.32$ in the SAGA database, the amount of Sr originating from CCSN-type events is about 50\% and the remaining part of about 50\% is originating from category III events. In the case of an observational bias (number of observed r-poor vs. r-enriched stars in comparison to a different real distribution) these fractions could increase significantly.

\subsubsection{The frequency of category III events based on the Eu-Fe correlation pattern}
\label{categoryIII}

In the previous sections we discussed the ratio of two elements X/Y imprinted by { explosive events} into the interstellar medium, being then inherited in new star formation environments. We want to address here once more Eu/Fe. If different categories of r-process environments would always come with an Fe co-production and a specific ratio of Eu/Fe, one would see this directly in observations. Instead in reality, one finds certain ranges for r-poor (regimes 1 and 2), and r-enriched stars (regime 3). In order to test realistically how certain classes of event categories contribute to galactic evolution, one should know defined abundance production patterns for all categories, inject them with the appropriate event frequencies { with the delay of stellar (or system) lifetimes before injection}, plus treating the whole system inhomogeneously in a 3D  approach during the early phases of the galaxy. And all of this should preferably also be done as a function of metallicity. Such approaches { have been developed} \citep{wehmeyer15,cescutti15,VandeVoort.ea:2020} but are nevertheless still in their infancy. In addition, the required abundance pattern predictions for all events are still highly uncertain. Therefore, we { have tried} here a reverse engineering approach, based on observations and correlations.  

\noindent Let us in a first simple approach assume that the observed ratios in the three regimes (standing for observed patterns) are consistent with appropriate variations within { all event categories (standing for production sites) and that the observations stand for the co-production the observed elements, i.e. each of the low-metallicity stars carries the imprint of one explosive nucleosynthesis site. After our introductory words in section \ref{sec:6.1} this is probably not true and several events can contribute. But let us
with this approach perform a simple test, in order to check whether this general assumption of e.g. also an overall Fe and Eu co-productions reproduces the correlation pattern of Fe and Eu as shown in Fig.~\ref{4starsplot} (right panel)}.
The method consists of:
\begin{enumerate}
\item Randomly generating equally distributed Fe values (20000) between log $\epsilon\simeq 3.4$ and log $\epsilon\simeq 5$ (log $\epsilon$=${\rm log}{(N_{Fe}/{N_H})} +12 $), which corresponds to metallicities [Fe/H] between -4 and -2.5 of our data.
\item Randomly generating for each of these metallicity values equally distributed (Eu/Fe) values (20000) between $2.4\times 10^{-8}$ and $1.4\times 10^{-5}$ to mimic the ratio of Eu and Fe throughout the three regimes (see Table~\ref{eu_fe_ratio} for more details).
\item Calculating for each of these entries the corresponding Eu values by multiplying the (Eu/Fe) ratios in the list of item 2 with corresponding Fe values in item 1. 
\item Items 1 through 3 result in a list of Eu and Fe abundance pairs of sample stars for which one can test correlations via PCCs and SCCs as a function of [Eu/Fe].
\end{enumerate}

\noindent This procedure was repeated many times, in order to test how robust the calculated PCC and SCC curves are.
It results in the following behavior (not shown here in a figure): When Fe and Eu were always co-produced throughout the three categories with the ratios chosen in item 2, the correlation of Fe and Eu remains at a relatively high level throughout the whole [Eu/Fe]-range at about 0.65 instead of 0.2 for the real case. One finds a linear relation between Eu and Fe abundances with a moderately broad scatter. Furthermore, the PCC and SCC curves do not diverge at all. 
Thus, two interesting features emerge from these tests: (1) Despite largely varying individual Fe and Eu abundances for the sample stars, the correlations between Eu and Fe remain moderately high, as long as Fe and Eu are added jointly to a stellar composition, and (2) PCCs and SCCs do not diverge, contrary to the case when utilizing observational data.
In that case we saw a strong decline for increasing [Eu/Fe]-values and a divergence between PCCs and SCCs. Thus, adding jointly Eu and Fe over the whole observed [Eu/Fe]-interval, which represents the three regimes, is not consistent with reality. This agrees with our previous findings that category III stands for a strong r-process contributor with no or negligible Fe production.\\  
Therefore, we performed an additional test, assuming that Fe and Eu are only co-produced in categories I and II, which should be responsible for reproducing data in regimes 1 and 2 with [Eu/Fe]$<$0-0.3. A fraction $\lambda$ ($0\le \lambda\le 1$) of Eu (here defined as Eu$^{*}$) is produced in addition without Fe in category III. This stands for an independent r-process source with no (or negligible) co-production of Fe. The question is how this reproduces regime 3 stars, as we know from the previous analysis that also category I and II stars contribute to the Eu in regime 3. This means that we do not necessarily find r-process Eu from only a single event which contributed to the ISM before star formation, but we might need to determine Eu$^{*}$ by subtracting the category I/II contribution. Therefore, the second test consists of:
\begin{figure*}[h]
\centering
\includegraphics[width=12cm,height=7cm]{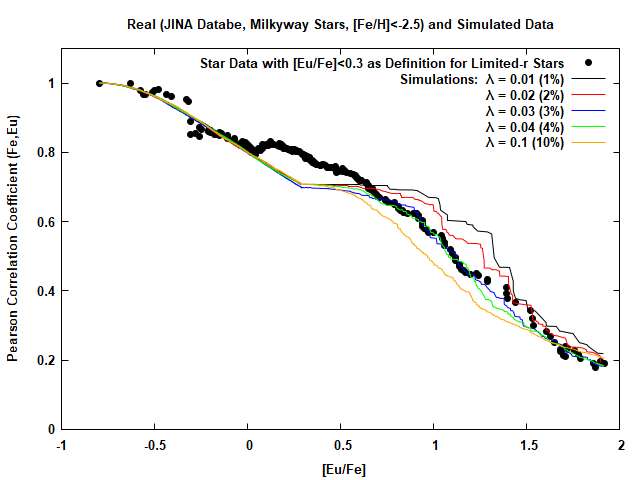}
\includegraphics[width=12cm,height=7cm]{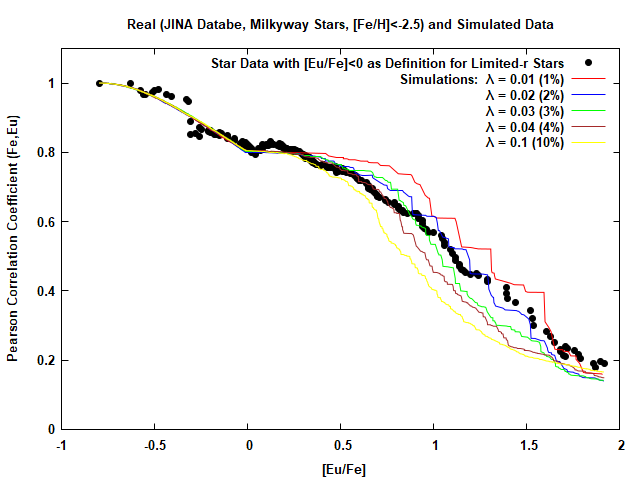}
\caption{PCC values of real and randomly generated Eu and Fe in stars with $\mbox{[Fe/H]}\le -2.5$. See the text for this approach where $\lambda$ relates to the fraction of cases where Eu is added independently from category III without adding any Fe. The top figure utilized [Eu/Fe]$<$0.3 for the upper limit of limited-r or r-poor stars (which defines the division between regime 2 and 3), the bottom figure utilized [Eu/Fe]$<$0 as an upper limit, leading to more consistent results and supporting the earlier made choice for the division between limited-r and r-process enriched stars. A $\lambda$ of about 2\% seems to provide the best fit to the data, representing the contribution of category III events which eject Eu with no or negligible Fe in comparison to category I/II events with co-production of Fe and Eu.}
\label{jinadata-Eu-Fe-Cor-Sim}%
\end{figure*}
\begin{enumerate}
\item As in test 1 choosing the metallicity, i.e. the Fe abundance of a sample star by randomly generating equally distributed Fe values between log$\epsilon$$\simeq 3.4$ and log$\epsilon$$\simeq 5$, which corresponds to metallicities between -4 and -2.5 of our data.
\item Adding Eu from category I/II stars for the same metallicity via randomly generating equally distributed (Eu/Fe) values between $2.4 \times 10^{-8}$ and $2 \times 10^{-7}$ to mimic the ratio of Eu and Fe in the first two regimes (see Table~\ref{eu_fe_ratio} for more details).
\item Adding Eu$^{*}$ from category III stars. This is obtained (in an approximation) in two steps: (a) Randomly generating equally distributed (Eu/Fe) values between $2.1\times 10^{-7}$ and $1.4\times 10^{-5}$ to mimic the ratio of Fe and Eu in regime 3 (see Table~\ref{eu_fe_ratio} for more details). (b) However, Eu$^{*}$ in a regime 3 sample star should be obtained by subtracting the category I/II contribution.
\item Thus, we utilize the Eu$^{\rm I/II}$ (multiplying the (Eu/Fe) values of item 2 with the Fe value for this chosen metallicity of item 1) and subtract it from from the Eu, as obtained in item 3 (a) for the same chosen metallicity. This results in Eu$^{*}$ in regime 3, which is only stemming from category III events without Fe co-production.
\item The remaining question is how frequent category III events are in comparison to category I/II events. Therefore, we multiply Eu$^{*}$ with a free factor $\lambda$ in the range from 0 to 1. The total Eu of the sample star is then $\mbox{Eu}=\mbox{Eu}^{\rm I/II}+\lambda\times\mbox{Eu}^{*}$, providing at the end the pair of Fe and Eu values of this sample star and also its Eu/Fe ratio.
\item This permits to calculate the PCCs and SCCs of the overall Eu and Fe correlation as a function of [Eu/Fe].
\end{enumerate}

\noindent The procedure described above is clearly an approximate one and should in principle be followed at some point by a large-scale chemodynamical inhomogeneous evolution simulation that takes into account the contribution of all discussed sources, their frequency and the individual time evolution. On the other hand it also includes the simplification that one only talks about category I/II events and category III events, rather than dividing also into I and II as well as IIIa and IIIb. The procedure implicitly also assumes that each regime 3 star contains also contributions from category I/II sources. This, however, is strongly supported by the findings of the previous subsection (see Table~\ref{eu_amount}) which actually analyzed the Eu contributions from category I/II events to regime 3 stars. Keeping this in mind, and following the outlined procedure of test 2,  Fig.~\ref{jinadata-Eu-Fe-Cor-Sim} results. We utilized two choices for the division between limited-r (regime 1 and 2) stars and r-process enriched stars (regime 3). The top figure shows observational PCC-values and selected results of PCC-values for the test cases with $\lambda=0.01$ (1\%), 0.02 (2\%), 0.03 (3\%), 0.04 (4\%) and 0.1 (10\%) when utilizing [Eu/Fe]$<$0.3 for the upper limit of limited-r or r-poor stars. We realize, however, in the bottom figure that a break at [Eu/Fe]=0 rather than at 0.3 would be a better choice for dividing among limited-r and r-process enriched stars. This is in line with our previous finding that Th, a strong r-process product, is found already in stars with [Eu/Fe]$>$0 (see Fig.~\ref{figThEuH}). We see that the best agreement with the { correlation pattern of observational data} is achieved for values of $\lambda$ close to 0.02. This means that only about 2\% of all events contributing Eu (without Fe) via r-process sites are sufficient, in order to explain the Eu and Fe correlation pattern. This exercise does not yet point to concrete scenarios for category III events and does not make a difference between IIIa and IIIb events, as long as the Fe co-production is negligible. 

\noindent The two tests performed above simplify somewhat reality, but they come with the result that the strong r-process events of category III, producing Eu without Fe, stand for only a 2\% fraction in comparison to Fe and Eu co-producing core-collapse events, which we introduced as category I/II events. What does this finding say about the ratio of strong r-process events, producing Eu without Fe, to regular CCSNe? In the SAGA database there exist 932 stars with Fe and Sr detections.
Out of these 282 stars show Fe, Sr and Eu. These represent 30\% in comparison to the
650 stars which show Fe and Sr without Eu. The latter could be identified with regular (i.e. category 0) CCSNe which probably co-produce Fe and Sr. Thus, following this interpretation that category III events represent 2\% of the whole sample of category I/II plus III events, and all these represent only 30\% of all category 0, I/II, and III events (dominated by regular CCSNe category 0 events), we find about 6 per mil of category III events in comparison to the overall CCSN events. This is within uncertainties (including possible astronomical selection biases plus the chance that some CCSNe might not produce Sr) close to the 2 per mil of compact binary merger events in comparison to CCSNe found in \citet{Rosswog.Feindt.ea:2017}, consistent with a neutron star merger rate of 3.7$\times 10^{-5}$ { yr$^{-1}$} \citep{Polea:2020} and a CCSN rate in the Galaxy of 1.6$\times 10^{-2}$ { yr$^{-1}$} \citep{Rozwadowska.ea:2020}.

\subsection{Identifying the suggested r-process sources / event categories by comparison with observational regimes}

\noindent In the preceding subsections, we have categorized r-process sources in 
events of category I, II, and III (with the sub categories IIIa and IIIb) 
without trying definitely, yet, to identify them with stellar sites 
(although some of these discussions took place, 
already). When focusing on the Fe-group elements we tried to be already 
more concrete and pointed to stellar core collapse with varying initial stellar masses. We also mentioned that even in the case of regular (neutrino-driven) core-collapse supernovae, due to slight $Y_e$-variations in the ejecta, light trans-Fe elements like Sr, Y, and Zr can be produced. We introduced for these sources category 0 events.
In the following we want to 
discuss possible production sites of the identified r-process categories.

\noindent A minimal requirement for an r-process is a sufficiently large neutron-to-seed-nuclei ratio, which depends on $Y_e$, the entropy, and the expansion velocity of ejected matter. These conditions have to be compared with possible 
scenarios for the different categories of events which we have concluded to exist from observational constraints 
and correlations.\\
In the introduction we have discussed the following possible r-process sources: (a) regular core-collapse supernovae with either no or only a very weak r-process or a $\nu$p-process, possibly producing elements up to Sr, Y, and Zr, (b) electron capture (EC) supernovae in the 8-10 M$_\odot$ mass range (if they exist and are not disqualified by their abundance pattern for realistic $Y_e'$s), (c) magneto-rotational supernovae with a varying (dependent on initial magnetic fields and rotation rates) weak to strong r-process (probably in most cases weak), (d) "quark-deconfinement supernovae" of massive stars that 
explode due a quark-hadron phase transition at supra-nuclear
densities (rather than the commonly assumed  neutrino-powered mechanism),
(e) collapsars (observable as hypernovae) that lead to a black hole plus torus configuration, and finally (f) compact binary mergers (of both double neutron star and
neutron star - black hole systems) driven to coalescence by gravitational wave (GW) emission.
Out of these sites (a) might provide the conditions for a { very(!)} weak r-process or $\nu$p-process, whether only up to Sr, Y, Zr or up to (but not beyond) the A=130 peak is still open \citep{Wanajo.Mueller.ea:2018,Curtis.ea:2019,Fischer.Guo.ea:2020}.
(b) is a class of supernovae whose existence is put into question after recent re-determinations of the electron capture rate of $^{20}$Ne \citep{Kirsebom.Jones.ea:2019,Kirsebom.Hukkanen.ea:2019}, but is not firmly excluded, however, leading to a too strong decline in abundances as a function of $A$ for realistic $Y_e$-conditions \citep{Wanajo.Janka.Mueller:2011}.
(c) could plausibly lead to magnetars, neutron stars with surface magnetic
fields of the order $10^{14}$ Gauss, which form in $\sim$ 1 out of 10 of core collapse
supernovae \citep[e.g.][]{Beniamini.Hotokezaka.Horst.ea:2019}. Dependent on the initial
fields, varying weak (probably dominating) to strong r-process conditions can be obtained, the latter, however, only
for pre-collapse fields  beyond $10^{12}$ Gauss
\citep{Winteler.Kaeppeli.ea:2012,Moesta.Richers.ea:2014,Moesta.ea:2015,moesta18,halevi18,Nishimura.Takiwaki.Thielemann:2015,Nishimura.Sawai.ea:2017,Bugli.ea:2020,Reichert.Obergaulinger.ea:2021}.
Case (d) has been proposed for a while. Dependent on the nuclear equation of state for massive core-collapse events, the collapse of the proto-neutron star to a black hole can be avoided (in a narrow stellar mass range) due to a quark-hadron phase transition with the right properties. The ejecta would experience a weak r-process, but populating even the actinides, however, with negligible abundances \citep{Fischer.Wu.ea:2020}. 
{  Case} (e) has been extensively discussed in the context of long-duration gamma-ray bursts
\citep{woosley93,macfadyen99,macfadyen01}. They involve the collapse of massive
stars that rotate rapidly enough so that an accretion torus can form outside of the
last stable orbit of a forming black hole, and they go along with relativistic polar
 and non-relativistic torus outflows. This scenario has been proposed by
\citet{cameron03} as an r-process site and recently been examined in more 
detail by \citet{Siegel.Barnes.Metzger:2019} and \citet{Siegel:2019}.
The remaining site, (f), is related to compact binary mergers \citep[see][for overviews]{Thielemann.Eichler.ea:2017,Rosswog.Feindt.ea:2017,Cowan.Sneden.ea:2021}.

\noindent In the literature one can find further suggestions for weak r-process sites, i.e. predictions from neutrino winds of magnetized neutron stars \citep{Vlasov.Metzger:2017} and from neutrino winds of regular CCSNe following the explosions \citep[e.g.][]{Wanajo:2013}. However, there exist open questions about these yields and/or their importance. (i) Wanajo utilized a $Y_e$=0.4 for his protoneutron star winds – presently it is not clear whether such low values can be attained \citep{Wanajo:2021} - and the effect comes only for quite massive neutron stars. On the other hand \citet{Bollig.Yadav.ea:2020} show that a long-term accretion flow over seconds hinders the neutrino-driven wind. They suggest that the wind develops only in the lowest-mass cases. 
(ii) The above mentioned investigations for magneto-rotational supernovae predict dynamical ejecta of the order 0.01-0.1 M$_\odot$, while the wind ejecta of \citet{Vlasov.Metzger:2017} for fast rotating magnetized protoneutron stars (which should follow such dynamical ejecta) are of the order $10^{-4}$ to $10^{-3}$ M$_\odot$. Here one might need to include both such ejecta (the dynamical as well as the wind ejecta), which we will discuss when considering case (c).

All sites from (a) to (e) relate to massive stars, and except for (e) they are essentially candidates for a weak r-process. (c) can possibly vary 
from a weak to (in rare cases) a strong r-process, (e) is a convincing case for a strong 
r-process. All of these events will co-produce Fe, but for the strong 
r-process candidate (e) these amounts are on a negligible level in comparison to 
solar ratios, for case (c) this depends on the strength of magnetic
fields and the rotation rate. 
\begin{figure*}[h]
\centering
\includegraphics[width=15cm,height=10cm]{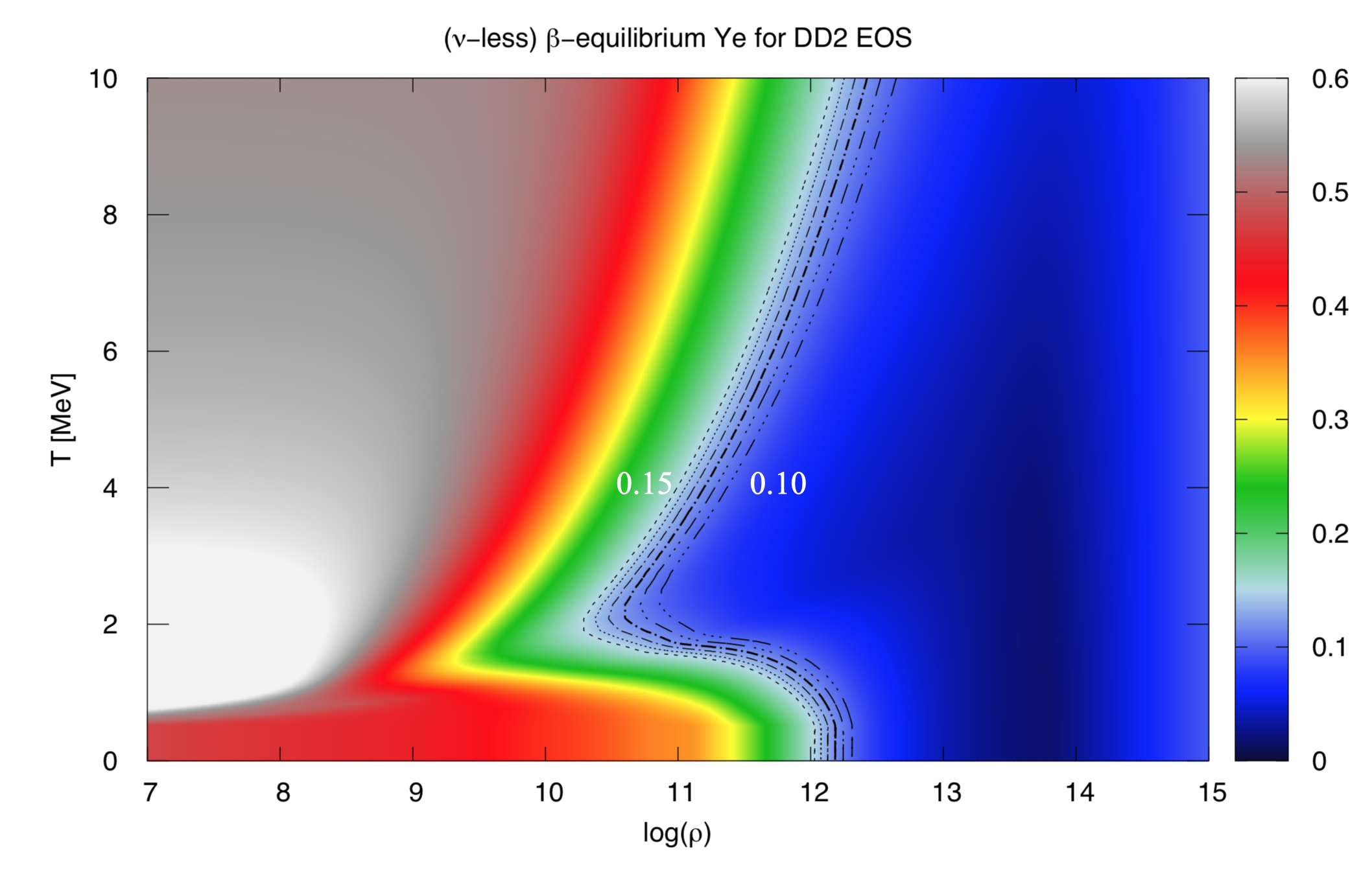}
\caption{$\beta-$equilibrium values for a broad range of relevant densities and termperatures. The colors indicate $Y_e$, contour lines in the range of 0.1 to 0.15 are also displayed.}
\label{fig:Ye_equil} 
\end{figure*}

\noindent The progenitors for the cases (a) to (f) vary substantially in their { initial} electron
fractions: all sites apart from (f) start with a 
pre-explosion $Y_e$-value close to $0.5$ and de-leptonize to low $Y_e$-values just prior to or during the
collapse and explosion (and they need to avoid re-raising $Y_e$ e.g. via
$\nu_e$-capture as in case (c) for low magnetic fields). Only the sites related to (f) start at the opposite end 
with electron fractions close to 0.05 (see Fig.~\ref{fig:Ye_equil}) 
and the ejecta becoming "re-leptonized" by { positron-} and $\nu_e$-captures. 
Extremely low initial $Y_e$-values provide compelling environments 
for a strong r-process.
During the explosive event, which triggers an r-process, it is unlikely to maintain a perfect
$\beta$-equilibrium, but during the expansion a transition from (close to) one
$\beta$-equilibrium to (close to) another one is possible. Matter may, e.g., transition from
cold, catalyzed neutron star matter with a very low $Y_e$ to a higher-$Y_e$ 
$\beta$-equilibrium state that is determined by the  balance of
electron-/positron- and neutrino-/anti-neutrino captures. Thus, when
assuming that the temperatures in the explosion are large enough 
to drive matter (close to) $\beta$-equilibrium, this may provide us with some
hints on the physical conditions at the production site. To fix orders 
of magnitude,
we show in Fig.~\ref{fig:Ye_equil} the $\beta$-equilibrium values (assuming 
vanishing neutrino chemical potentials) for a broad range of physical 
conditions. The concrete values have been calculated using the DD2 
equation of state \citep{hempel10}, but deviations for other equations 
of state are only expected at the highest densities ($\gg 10^{14}$ g cm$^{-3}$)
that are very unlikely to be ejected and therefore are not relevant for our
discussion here. Not too surprisingly, the matter inside a neutron star is extremely
neutron-rich ($Y_e < 0.1$) with $Y_e$-values actually starting to dip down to 
close to zero when approaching $10^{14}$ g cm$^{-3}$.
\\
\subsubsection{Category 0, I, and II events}
\label{sec:6.3.1}
Let us first try to make the connection to category 0, I, and II events which are responsible for either producing only the trans-Fe elements or are responsible for a weak r-process, including Eu production but no {(or negligible amounts of} third r-process peak or actinide elements.
They all showed very strong or strong correlations with Fe, as discussed in earlier sections.
One  major question is whether the non-detection 
of Eu in a large fraction of stars means that (i) all such stars have not experienced any Eu or r-process pollution or (ii) whether this is just due to observational uncertainties. (i) would be consistent with the non-production of even a weak r-process in regular CCSNe, being in line with a strong debate about its occurrence in such events. While the production of Sr, Y, Zr, and maybe even heavier elements below the A=130 peak, might be possible in a combination of a $\nu$p-process \citep[e.g.][]{Froehlich.Martinez-Pinedo.ea:2006,Eichler.Nakamura.ea:2018} and/or a (very) weak r-process \citep{Wanajo.Mueller.ea:2018,Curtis.ea:2019}, for the latter see also \citet{Kratz.Farouqi.ea:2007}, \citet{Farouqi.Kratz.Pfeifer:2009} and \citet{Akram.Farouqi.ea:2020}, Eu seems not to be produced.
This would exclude a weak r-process Eu production in regular CCSNe as sites of category I or II events, within our present understanding. Therefore, we decided to call them
"category 0" events, underlined already to some extent by an apparent three-plateau feature in the correlations for Sr, Y, and Zr (Fig.\ref{corr_Sr_Y_Zr_Fe}). The remaining options for category I and II events among the listed sources would be (b) electron capture supernovae, (c) magneto-rotational supernovae and (d) quark-deconfinement supernovae. 
All limited-r or weak r-process stars are characterized by a high Sr/Eu ratio (see 
Fig.~\ref{fig:fullEuFe_SrEuFe}, 
right panel) and a very strong or strong correlation of Eu with Fe, indicating a co-production of Fe and Eu. This was the reason to identify them with a type of core-collapse supernovae. 
We have summarized presently available predictions for EC supernovae \citep[EC SNe,][]{Wanajo.Janka.Mueller:2011}, MHD supernovae
\citep[MHD1 and MHD2,][]{Nishimura.Sawai.ea:2017,Reichert.Obergaulinger.ea:2021}, and QD supernovae \citep{Fischer.Guo.ea:2020} in Table~\ref{SrEuFeratios}, taken from Fig.~5 in \citet{Wanajo.Janka.Mueller:2011}, Figs.~4 and 5 in \citet{Nishimura.Sawai.ea:2017} (in a restricted range of magnetic field strengths), Table 2 in \citet{Reichert.Obergaulinger.ea:2021} (one out of 4 models), and Table 3 in \citet{Fischer.Guo.ea:2020}, in comparison to the observed ratios in low metallicity stars of regimes 1 and 2. If one would add late wind contributions of \citet{Vlasov.Metzger:2017} to the MHD models, they would at most add 10\% to the Sr and Eu values presented in Table~\ref{SrEuFeratios}. The listed sources are the ones expected to be explained via category I and II events. Note that some of these entries contain a range of values due to a variety of models contained in these publications. For the EC supernova models we utilize the entries for $Y_e$ between 0.2 and 0.25 instead of a more realistic $Y_e=0.4$. This goes beyond uncertainty estimates permitting $Y_e$ values only down to 0.3. However, for such values no Eu would be produced, underlining that EC supernovae (if existing) probably are not good candidates for category I or II events. The MHD1 results represent the range of $i$ entries for intermediate magnetic fields, the MHD2 entries take the only model which produces non-negligible amounts of Eu, but an average over the range of magnetic field strengths and rotation rates would probably be different. Sr ejecta could be lowered by a factor of 10 while Fe ejecta can be higher by up to a factor of 10 than listed. The QD supernova entry gives the range of two model predictions.
The observational entries for regime 1 and 2 are taken from Fig.~\ref{fig:fullEuFe_SrEuFe} (right panel) for Sr/Eu, taking the full range of the observations, for Sr/Fe the upper limits of the second and third plateau in Fig.~\ref{corr_Sr_Y_Zr_Fe}, and for Eu/Fe the average values based on the fits in Fig.~\ref{eu_vers_fe_fit}.
The Eu/Fe ratios which we find in regime 1 and 2 stars are on average of the order $2\times 10^{-8}$ and $10^{-7}$. Thus, in case the observed abundance ratios in those stars would reflect directly the production ratios from the pollution by a single event, we should find consistent and identical ratios in the responsible event ejecta. The last line of Table~\ref{SrEuFeratios} contains also the total amount of $^{56}$Ni (decaying to Fe) ejected in the models/sites considered here for weak r-process events (but, as mentioned before, this can vary strongly in the variety of models considered for these types of events).
\begin{table*}[h!]
\caption{Candidates for category I and II events}
\label{SrEuFeratios}
\centering
\begin{tabular}{ccccccc}
\hline\hline
Element ratio & Obs. Regime 1 & Obs. Regime 2 & EC SNe & MHD1 & MHD2 & QD \\
\hline
\hline
Sr/Eu & 200-750 & 200-800 & 300 -$10^{5}$ & 438-1300 & 340 & 142-468 \\
Sr/Fe & $<2 \times 10^{-6}$ & $<2.5 \times 10^{-5}$ & $ 3.3\times 10^{-2}$ & $7 \times 10^{-3} - 5 \times 10^{-2}$ & $2.6 \times 10^{-2}$ & $(5.1 - 5.8) \times 10^{-3}$\\
Eu/Fe & $10^{-8}$ & $10^{-7}$ & $3 \times 10^{-7}$ -$10^{-4}$ & $1.5-3.6 \times 10^{-5}$ & $7.5 \times 10^{-5}$ & $(1.1 - 4.1) \times 10^{-5}$ \\
Fe [M$_\odot$]& & & $3\times 10^{-3}$ & $3 \times 10^{-2}$ & $ 2\times 10^{-2}$ & $5.5\times 10^{-2}$\\ 
\hline
\hline
\end{tabular}
\end{table*}

\noindent We see for all the candidates in Table~\ref{SrEuFeratios} a reasonable agreement between model predictions and observations for the Sr/Eu ratios, representing a typical weak r-process pattern (maybe with exception of the EC SNe which get close to the observed ratios only with highly reduced $Y_e$ values, stretching the uncertainties possibly beyond permitted limits). Generally, the Sr/Fe and Eu/Fe ratios seem to be 2 to 3 orders of magnitude too high. For MHD SNe this might be due to the fact that we included cases with too high magnetic fields, case $i$ from \citet{Nishimura.Sawai.ea:2017} and 35OC-Rs from \citet{Reichert.Obergaulinger.ea:2021}. When utilizing lower magnetic fields on average, the Sr as well as Eu abundances will be reduced and the Fe abundances enhanced. This can reduce Sr/Fe ratios by one to two orders of magnitude \citep{Reichert:2021}, but a constraint would be that on average the Eu/Fe ratios need to be reduced in similar proportions as Sr/Fe, in order to obtain the same Sr/Eu ratio. This depends on the distribution of such model properties (magnetic field strength and rotation rates) in realistic samples, as the lanthanide fraction beyond the second r-process peak is much more sensitive to such changes than the Sr abundance.

\noindent In all cases the Sr/Fe as well as Eu/Fe ratios produced in these events are exceeding the ones observed in regime 1 and regime 2 stars by a large factor. A solution to this, in line with our previous discussion, is that - already at these low metallicities - we do not see only a single event pollution, but that for these regime 1 and regime 2 stars already many regular supernovae contributed as well, adding essential Fe, but only affecting mildly the Sr/Eu ratio with existing but small Sr production. 
Independent of these considerations, but in line with what was just said, the magneto-rotational supernovae (item c in our list of possible sites) show a highly varying degree of the r-process strength, dependent on the initial magnetic field and rotation. Due to this they will come with a large scatter which we also see for regime 2 in Fig.~\ref{eu_vers_fe_fit}. This might qualify them as good candidates for category II. Weak r-process stars of regime 1 show a highly uniform behavior. Thus, such ejecta properties would also be required from category I events. 
Electron-capture supernovae seemed ideal candidates in this respect, but they have recently been put in question with respect to their real existence \citep{Kirsebom.Hukkanen.ea:2019, Kirsebom.Jones.ea:2019}. While the outcome is still open \citep{Jones.Roepke.ea:2016}, and they could possibly be considered as a well defined event, being progenitors of low mass neutron stars \citep{huedepohl10} with robust abundance features \citep{Wanajo.Janka.Mueller:2011}, the probably much too high Sr/Eu ratio puts them in question.
It remains to be seen whether QD supernovae (d) could be an alternative for category I events or (with possibly improved physics input in future investigations) also regular supernovae (a).\\

\noindent What remains to be said, while MHD1 and MHD2, as well as the QD supernovae, show a reasonable Sr/Eu ratio consistent with regime 1 and/or regime 2 stars, the Sr/Fe and Eu/Fe ratios are by orders of magnitude higher than what is seen in observations. This argues for the additional Fe floor from prior regular CCSNe mentioned above (but not producing Eu, and possibly Sr only to a negligible extent) that has existed before pollution with a category I or II event. This is in line with our previous findings that category 0 supernova ejecta can already be found at metallicities as low as [Fe/H]$<$-5. Comparing the produced Fe from the simulations shown in Table~\ref{SrEuFeratios} to the typical Fe-production in CCSNe would reduce the apparent overproduction. If we would link QD supernovae to category I events (responsible for regime 1) and MHD supernovae to category II events (responsible for regime 2 with the larger scatter),
we find an overproduction of Sr/Fe and Eu/Fe by a factor of 500 and 100 to 1000, respectively. Thus, we would need $n$ additional supernovae with 0.1M$_\odot$ of Fe to reduce these ratios to the ones observed in regimes 1 and/or 2. This leads to $0.055 + n \times 0.1 = 500 \times 0.055$
for QD supernovae or $(0.02\ \rm{to}\ 0.03) + n \times 0.1 = (1000\ \rm{to}\ 100) \times (0.02\ \rm{to} \ 0.03)$ for MHD supernovae. Taking this at face value would require the addition of 275 CCSNe with their Fe to each QD supernova in order to reproduce the observations of regime 1 and 30 to 200 CCSNe in addition to each MHD supernova to reproduce the observations of regime 2. But permitting Sr/Fe and Eu/Fe ratios for MHD supernovae, averaged over the full range of magnetic field strengths and rotation rates, could reduce these ratios by up to a factor of 10, resulting in $n$ values as small as 10, consistent with the number of supernovae from massive stars which lead to magnetar formation.  {\it To summarize: about 287 CCSNe per QD supernovae, and possibly a number as small as 10 regular CCSNe per MHD supernova are required to explain the observations for regime 1 and 2, respectively}. This has to be taken with care, requiring that these simulation models are approaching reality!
The number for the ratio of regular CCSNe (with typically 0.1~M$_\odot$ Fe production)to the category I as well as II events would make sense, as these are at the higher end of the IMF and for QD supernovae in a very narrow stellar mass range or require for MHD supernovae specific conditions for pre-collapse magnetic fields and rotation rates (possibly consistent with the fraction of CCSNe resulting in magnetars).

\subsubsection{Category III events}
\label{sec:Cat_III}
\begin{figure*}[h!]
\centering
\includegraphics{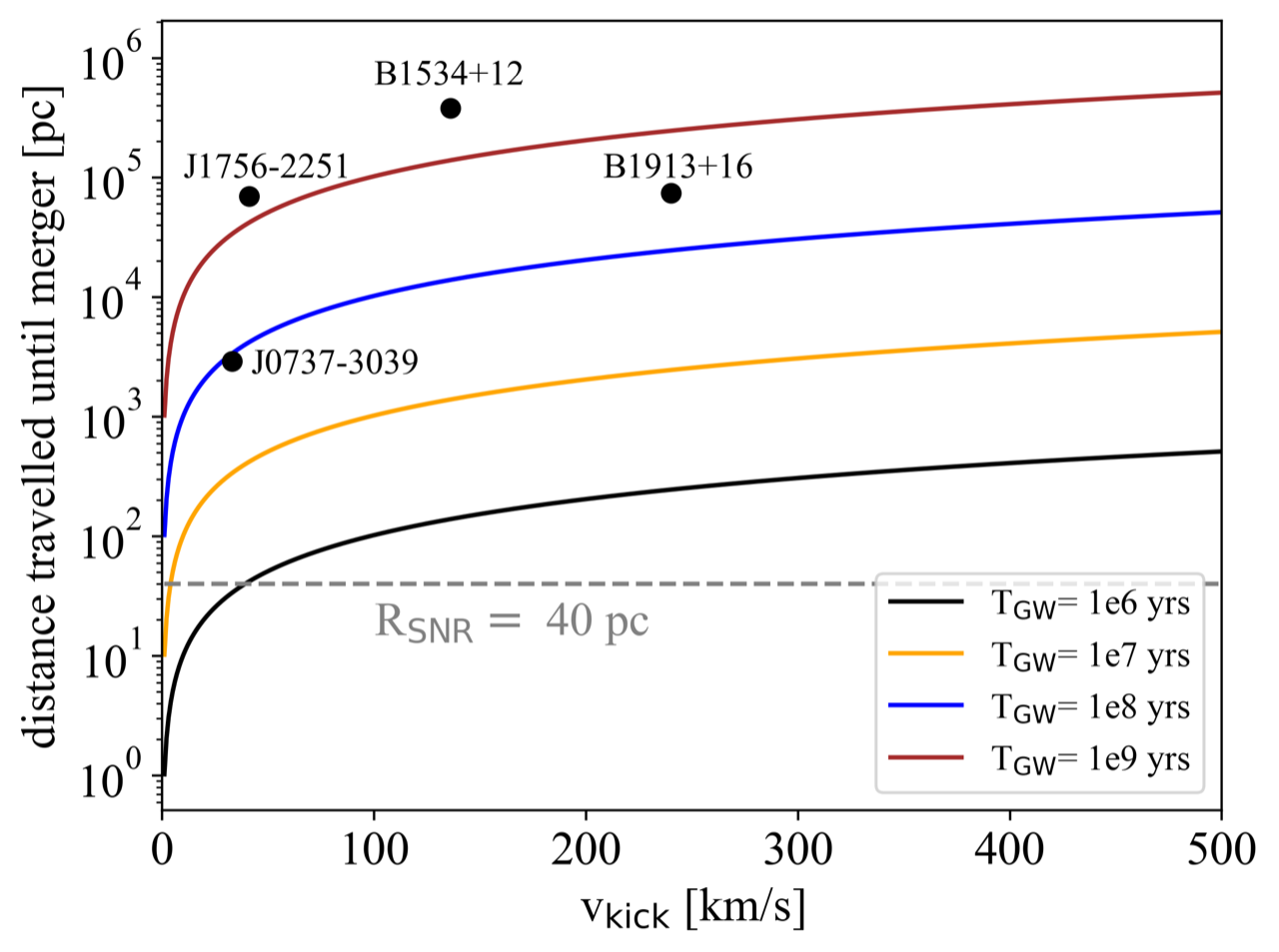}
\caption{A few neutron star binaries with sufficiently known details to determine the gravitational wave inspiral time $T_{GW}$ and their kick velocities. This information suffices to know the distance this system will move before the neutron star merger and compare it to the typical size of a supernova remnant (SNR). Even the closest one ( "double pulsar" J0737-3039) will merge about 100 SNR radii from the site of the last supernova. 
The observational data are taken from Table 3 in \citet{Tauris.Kramer.ea:2017}.}
\label{fig:merger} 
\end{figure*}

Finally we have also to identify the astrophysical counterparts of category IIIa/b events.
Let us begin with sites of type (f) compact binary mergers, more specifically the mergers 
of two neutron stars, since they are the only events that 
have been directly observed to synthesize  (both strong and lighter) r-process nuclei, 
as discussed in the introduction. 
Fe could not be detected,
underlining a non-correlation of r-process elements with Fe. However, the two neutron
stars originate from earlier CCSN explosions which produced Fe and all r-process enriched stars in regime 3 contain Fe as well. This is due to the pollution by the merger, but also by preceding supernovae. Whether the Fe stems from the two supernovae of the progenitor system or prior frequent CCSNe needs to be discussed. It relates to the question whether the neutron star mergers take place within the supernova remnant(s), spanned out after the preceding supernovae explosions, or the merger might take place outside the remnants due to neutron star kick velocities.   
Here we show those neutron star binaries from the overview table 3 in \cite{Tauris.Kramer.ea:2017} where gravitational wave inspiral times $T_{GW}$ and system velocities are known. With this information we know which distance this binary moves before the merger and we can compare it to the typical size of a supernova remnant (SNR). Of those systems none will merge near their supernova remnants. In Fig.~\ref{fig:merger} even the closest one, the "double pulsar"
J0737-3039, will merge about 100 SNR radii from the site of the last supernova. Thus, at least for these selected well-known systems the neutron star merger ejecta will not easily mix on short time scales with{ their} prior supernova ejecta. 

\noindent The question remains whether the cases presented in Fig.~\ref{fig:merger} are a representative sample for the whole merger population or are affected by an observational bias. It is easier to detect long-lived systems, as for higher eccentricities the merging event will come much faster \citep[see, e.g. Fig.1 in][] {Rosswog:2015}).
Nevertheless, it needs a $v_{kick}$ as small as 30 km/s (while these kick velocities are typically as high as several hundreds of km/s) in order to inspiral and merge after a short period of $10^6$ years within a range of about 30 pc (smaller than a typical supernova remnant).
The presence of r-process-enriched ultra-faint dwarf galaxies (UFDs) with a shallow gravitational
 { potential} 
suggests small natal kick velocities for binary
neutron stars at birth \citep[$<$ 15 km/s,][]{Beniamini.Hotokezaka.Piran:2016}. In fact, the
analysis of observed binary neutron stars in the Milky Way suggests that
the majority of the systems receive small kicks of  $<$ 30 km/s \citep{Beniamini.Piran:2016}, although a high velocity population of
$>$ 150 km/s exists as well \citep[about 20\%,][]{Behroozi.Ramirez-Ruiz.ea:2014}.
Thus, to summarize: we expect a low-kick population, but even such a
population would need an extremely short inspiral time of only $10^6$ years in
order to be able to permit mixing of the merger ejecta with Fe from the preceding supernova remnants. In addition a high-kick population exists, which is very 
unlikely to merge within the supernova remnant.
In the second case the question arises how enriched the ISM was already with Fe ejecta of other independent supernovae at the point when the merger takes place. This is independent of the preceding double supernova system, responsible for creating the binary neutron star system.
In all cases NSMs fulfill the requirements of category III events, i.e. no or negligible correlation of r-process elements with Fe. Nevertheless, as we know from observations, the ISM around the merger apparently contains already pre-existing Fe.

\noindent We have also discussed case (e), i.e. 
collapsars, as contributors to category III. In addition, those magneto-rotational supernovae of case 
(c) with highest magnetic fields, permitting a strong r-process, might also be possible contributors to category III events, although \citet{moesta18} argue that the high fields required are unlikely and in most cases (in 3D simulations) a kink instability prevents a strong r-process, causing only a weak one. Both types of events, i.e. collapsars and extreme cases of MHD supernovae, would come with some co-production of Fe, displaying maybe a small but possibly negligible correlation. Such explosions related to the most massive stars occur already in the earliest phases of galactic evolution. i.e. for sure at [Fe/H]$<$-3. 
A further analysis with 
respect to the appearance of strong r-process events at very low metallicities might 
also help to test whether not only core-collapse events of massive stars can
be responsible at such low metallicities, see Fig.~\ref{figThFeH}. \citet{Wehmeyer.ea:2019} investigated neutron star - black hole mergers in chemical evolution studies, systems which would also lead to a black hole torus, similar to collapsars. Also neutron star mergers, with a large combined mass and fast black hole formation, can lead to black hole torus systems. How in those cases the prior binary evolution retards the merger event in galactic evolution with respect to metallicity remains an important issue.
\noindent The above considerations lead to the question:
{\em How to identify category IIIa and IIIb events, and how are they possibly related to an actinide boost?}\\

An interesting question is related to the division 
between actinide boost stars and normal r-enriched stars 
which display a strong r-process. In Fig.~\ref{figThEuH} (right panel) we have seen that the actinide production (Th) is due to a superposition of events. In 
Fig.~\ref{ranksThrIrII} we saw a clear difference for two distinct subgroups in regime 3, related to two subcategories of category III (IIIa and IIIb), explaining r-enriched stars responsible for solar-type strong r-process abundances as well as actinide boosts. We could also show that this division is { somewhat} related to the observational classes of r-I and r-II stars. Thus, two questions have to be answered: (i) what are the environment conditions permitting such an actinide boost, and (ii) which astrophysical objects experience these physical conditions.  
(ii) could possibly lead to further questions whether environments different from typical neutron star mergers could be responsible, i.e. neutron star - black hole mergers or possibly collapsars? But let us first concentrate on (i), the required environment conditions.

\begin{figure*}[h]
\centering
\includegraphics[width=2\columnwidth]{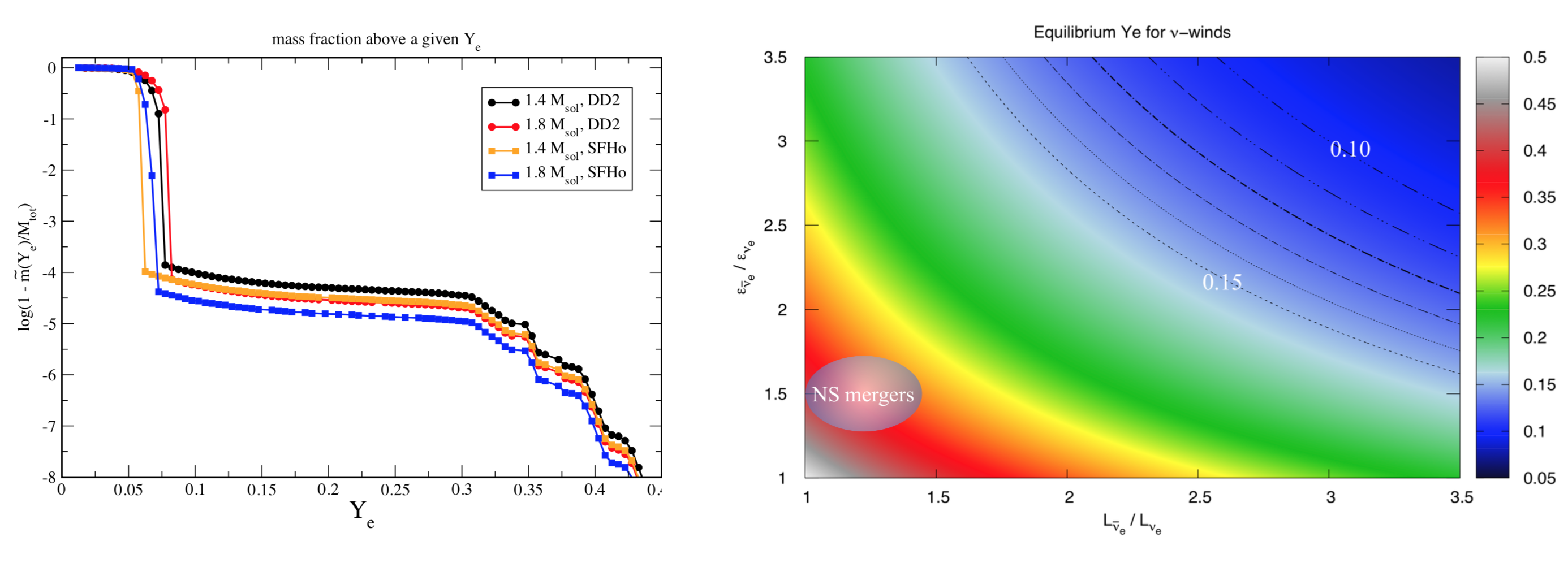}
\caption{Left panel: Mass fractions in neutron stars above the given electron fraction $Y_e$ of the x-axis. Rather insensitive to mass and equation of state, the bulk of a neutron star has an electron fraction below $\approx 0.08$, only a tiny mass fraction of $\approx 10^{-4}$ is above this value. Right panel: equilibrium $Y_e$ in a neutrino-driven wind as a function of the ratio of (anti-)neutrino energies and luminosities (see text for details). Marked is the $Y_e$-range favorable for an actinide boost (0.1 ... 0.15) and the regions expected for neutron star mergers.}
\label{fig:mass_above_Ye_nu_wind} 
\end{figure*}

\noindent Recent studies \citep{Holmbeck19a} based on one hydrodynamic 
trajectory from tidal dynamical 
ejecta\footnote{Trajectory from the simulations described 
in \cite{rosswog13a}.} conclude that actinides are substantially
overproduced relative to lanthanides for $Y_e$-values in the range 0.1-0.15, due to the influence of fission cycling. 
This is consistent with \citet{Wu:2017} and a recent study of \citet{Eichler.Sayar.ea:2019}, which  
finds, with a variety of nuclear mass models, that slightly 
larger electron fractions in the range
of $\sim 0.15$ are most favorable to explain "actinide boost" matter. Interestingly, the initial neutron
stars, or also massive stars before core-collapse, are practically free of such matter. In  
Fig.~\ref{fig:mass_above_Ye_nu_wind} (left panel) we have binned  
the masses according to the $Y_e$-values of initial neutron
stars. The models were calculated by solving the 
Tolman-Oppenheimer-Volkoff equations
together with realistic equations of state (DD2 and SFho). We 
show the results for two masses, 1.4 M$_\odot$ representing
the probably most common neutron star mass, and 1.8 M$_\odot$ 
at the higher mass end, but consistent with the recent LIGO 
detection GW190425 \citep{abbott20a}, where at least one 
neutron star was very heavy. Plotted is the mass fraction 
inside the neutron star that has an electron fraction above 
the $Y_e$-value given on the x-axis. To fix ideas, let us
focus on the black line (1.4 M$_\odot$ and DD2-EOS): the
curve starts trivially at a value 1, since all the matter has
a non-zero $Y_e$. The curve shows that essentially the whole 
star has an electron fraction below $\sim 0.08$ and only a
tiny mass fraction of $\sim 10^{-4}$ has a $Y_e$ above 
this value. In other words: there exists hardly any matter in the original neutron stars in the 
range that is required to produce actinide boost stars! This
result is rather robust against changes of the mass and the 
EOS. Similarly, massive stars before becoming collapsars/hypernovae do not contain matter with the required properties before core-collapse.\\
\noindent Given that "actinide boost" stars contribute a substantial fraction to the
r-process enriched stars, but the observed abundance patterns require 
a dominant contribution from a very narrow $Y_e-$range,
indicates that nature robustly produces a restricted range of
conditions where such $Y_e-$values occur. The question
is how? To date this question is not settled and we want to
discuss here a new, admittedly somewhat speculative
possibility.\\
A binary merger, where a central neutron star survives, 
drives, due to the intense neutrino irradiation,  
the electron fraction of the secular ejecta values
well beyond the upper limit of 0.15
for an actinide boost \citep{perego14,martin15,sekiguchi16}. 
To corroborate this, we plot in Fig.~\ref{fig:mass_above_Ye_nu_wind},
right panel, the equilibrium electron fractions for neutrino-driven winds \citep{Qian.Woosley:1996}
\begin{equation}
Y_e^{\rm eq} \approx \left[ 1 + \frac{L_{\bar{\nu}_e} 
\left( \epsilon_{\bar{\nu}_e} - 2 \Delta + 1.2\Delta^2/\epsilon_{\bar{\nu}_e} \right)}
{L_{\nu_e} \left( \epsilon_{\nu_e} + 2 \Delta + 1.2\Delta^2/\epsilon_{\nu_e}\right)}\right]^{-1}
\approx \left[1 + \left(  
\frac{ \epsilon_{\bar{\nu}_e} }{ \epsilon_{\nu_e} } \right) 
\left( \frac{L_{\bar{\nu}_e}}{L_{{\nu}_e}}  \right) \right]^{-1}.
\end{equation}
Here, $\Delta= 1.293$ MeV is the neutron-proton mass 
difference, $\epsilon= \langle E^2 \rangle/\langle E\rangle$ relates to neutrino energies, and $L$ to their luminosities.
In the approximation on the right we have assumed that neutrino
energies are large enough so that, to acceptable accuracy, 
the terms containing $\Delta$ can be neglected. This 
approximation is shown in the right panel of Fig.~\ref{fig:mass_above_Ye_nu_wind}.
We mark in the plot the relevant contours $Y_e=0.10$ and
0.15, and we have indicated the equilibrium $Y_e$ region that is expected, 
based on the neutrino properties from neutron star merger simulations \citep{Dessart.Ott.ea:2009,rosswog13a,perego14,sekiguchi16,Foucart.Oconnor.ea:2016},
assuming such neutrino-dominated environments will also attain a weak equilibrium, which may only be the case for long-lived central neutron stars.
If so, neutrino-driven winds from neutron star 
mergers with surviving central remnants are not viable
candidates for "actinide boost matter". In reality simulations contain a spectrum of ejecta, starting from ejected tidal tails of essentially pristine 
neutron star matter up to the high $Y_e$-values related to the discussed neutrino-driven
winds which reached a weak equilibrium. However, it is not expected that in this
spectrum the preferred narrow $Y_e$-interval around 0.1-0.15 plays a dominant role.
\\
If instead a black hole is present or forms, the neutrino
irradiation is dramatically quenched and the gas flow around 
the BH robustly regulates itself into a state of mild
electron degeneracy. Due to negative feedback between
electron degeneracy and neutrino cooling 
(higher degeneracy leads to fewer electrons and positrons, 
therefore reducing the neutrino emission, which leads to a higher
temperature and thus to a lowering of the degeneracy),
the disk midplane settles inside of the inner 
$\sim 10 \; GM_{\rm BH}/c^2$  to electron fractions of $Y_e\approx 0.1$ \citep{Beloborodov:2003}.
Interestingly, this occurs once the accretion rates exceed 
an "ignition value" \citep[that depends on the BH spin, see][]{chen07}
and the corresponding accretion rates are those that are needed to power 
(long or short) GRBs \citep{Lee.Ramirez-Ruiz:2007}. While discovered in 
semi-analytic models, this self-regularization to low $Y_e$-values
in the disk midplane is also found in full-fledged numerical 
(magneto-) hydrodynamic simulations, see e.g. \citet{Siegel.Metzger:2018} and \citet{Fernandez.Tchekhovskoy.ea:2019}.\\
\noindent The simulation of such neutrino-cooled accretion flows is a major challenge
since models should include the (potentially self-gravitating) MHD flow
around a rapidly spinning black hole, (semi-transparent) neutrino 
transport and effects of composition, degeneracy and nuclear
recombination. To make things even harder, one needs to resolve 
the small length scales of the magneto-rotational instability
\citep{chandrasekhar60,balbus98} and to evolve the black hole
torus system for a very large number of dynamical time scales (up to
several seconds, while the dynamical time scales are $\sim$ ms). Therefore,
it is not entirely surprising that the exploration of this topic 
is still in initial stages, that large parts of the relevant parameter
space are not explored yet and, where parameters are comparable, the results
do not yet agree (at least not concerning the ejecta composition).
The currently existing GRMHD explorations \citep{Siegel.Metzger:2017,Siegel.Metzger:2018,Miller.Sprouse.ea:2020,Fernandez.Tchekhovskoy.ea:2019}
agree that a large fraction ($\sim 40$ \%) of the initial torus 
mass becomes unbound, but to date there is no agreement about 
the resulting $Y_e$ and composition of the ejecta\footnote{For 
example, \citet{Fernandez.Tchekhovskoy.ea:2019} find $Y_e$ values around 0.12,
those of \cite{Siegel.Metzger:2018} peak around $\sim 0.14$ while \citet{Miller.Sprouse.ea:2020}
find a broad distribution between 0.2 and 0.4.}. Despite the current 
lack of consensus about the ejecta properties, we find the black hole 
torus idea for the source of "actinide boost" material compelling 
since
\begin{itemize}
\item[a)] it is the only convincing scenario that we are 
aware of that regulates itself robustly into the needed $Y_e$ range, 
\item[b)] the progenitors are known to exist, 
\item[c)] there are good physical reasons why torus black hole 
systems form in this parameter range. For example, 
compact binary mergers have, maybe within a factor $\sim 3$, 
total masses of a few M$_\odot$. They start out with a 
huge orbital angular momentum reservoir which is continuously
diminished by GWs until disruption occurs close to the last 
stable orbit. The angular momentum left at this stage is still very  
large and chances are good to form a substantial accretion torus. 
In double neutron star merger cases where a black forms, the 
large inherited angular momentum ensures that the post-merger 
black hole has a substantial dimensionless spin of $\chi\approx 0.8$ \citep{kiuchi09,rezzolla10}. If the merging system consists of
a neutron star and a black hole, the latter cannot be too massive, otherwise
the neutron star is swallowed as a whole and no substantial torus forms
\citep[for Schwarzschild black holes their mass must be $\leq 8$ M$_\odot$, see e.g. Fig. 18 in][]{rosswog15c}.
Substantially spinning BHs ($\chi \geq 0.8$) are permitted to have larger masses. In any
case, the resulting BH-torus system has similar parameters. Single stars 
have much less of a reason to form BH-torus systems in the suitable parameter
range, however, if they really are the causes of long GRBs, they must accrete
at rates in the range from $\sim 5 \times 10^{-3}$ to $\sim$ 0.1 M$_\odot$/s
and could therefore be in a similar parameter range 
\citep{Siegel.Barnes.Metzger:2019}. And last, but not least, 
\item[d)] they have additional signatures --gamma ray bursts-- 
that are regularly observed.
\end{itemize}
The progenitor systems of actinide boost material could 
then be either neutron star binaries that form massive enough 
accretion disks and black holes, or neutron star black hole 
systems that form substantial tori (either from low mass black holes
or large BH spins) and, potentially, also collapsar accretion disks.
There are good reasons to believe that the relativistic jets
needed for GRBs are triggered when a black hole forms
\citep{mckinney13,mckinney14,ruiz16,murguia17,ruiz19}\footnote{But see e.g. \cite{moesta20} for a possible alternative.}.
If black hole torus systems indeed manage to eject matter with properties 
similar to what they produce robustly in their inner torus regions, 
and a black hole is needed to launch an (either long or short) GRB
(rather than, say, a magnetized neutron star), then it would be the 
GRB engines that produce the "actinide boost" matter. 
Neutron star mergers, where instead a central stable or meta-stable massive neutron star survives long enough, 
eject very low $Y_e$ matter in tidal ejecta together with a broad range 
of $Y_e$ due to the exposure to the intense neutrino field. Such remnants
may be responsible for the more regular r-process enriched stars.
In this picture, GW170817 could have produced the early blue component
by polar dynamical ejecta together with neutrino-driven winds from
initial stages where a massive neutron star was still present. The ensuing 
collapse would have triggered the GRB launch, the torus would regulate
itself to $Y_e \sim 0.1$, eject a fair fraction of this material
in the form of actinides which would be consistent with the decay time scales 
inferred from late time observations of AT2017gfo
\citep{Wu.Barnes.ea:2019,Kasliwal.Kasen.ea:2019}. Opposite to the earlier discussed BH-torus systems (NS-BH mergers, collapsars), here the early phases related to dynamical ejecta and the neutrino-driven wind (during the period when a meta-stable massive neutron star still existed, before turning into a black hole) contribute sizable fractions of the
ejecta and lead to a broad range of $Y_e$-values, less dominated by the narrow interval responsible for an actinide boost.\\
\noindent Thus, after having gone through the options of category III events, which lead to a strong r-process, it seems that we found two types of subclasses: category IIIa, probably including neutron star mergers with combined masses which permit the formation of a stable or meta-stable massive neutron star for extended times after the merger and during the matter ejection and neutrino wind phase. In addition, a subdivided subclass of category IIIb events seems to include black hole torus systems (with massive neutron star binaries leading to fast black hole formation, neutron star - black hole mergers, and collapsars/hypernovae from very massive single stars). The latter would tend to be characterized by an actinide boost, while IIIa events would produce a strong r-process as well, but no actinide boost.\\

\noindent After this attempt of trying to identify the possible astrophysical sources for the different categories of events, deduced from abundance observations of low metallicity stars, we will address more quantitative aspects with respect to linking also the here discussed category IIIa and IIIb events to the different classes of r-process enriched stars of regime 3.
This leads to the question: {\em Do category IIIa and IIIb events dominate the abundances observed in r-I and r-II stars?}\\
\noindent It remains to be seen how the subregimes r-I and r-II of regime 3 stars can be explained with category IIIa and IIIb events. In Fig.~\ref{ranksThrIrII} we have shown that Th in complete r-process stars of regime 3 comes from two distinct event categories which essentially coincide with regime r-I and r-II stars.  Fig.~\ref{EurankrIrII} underlined, in addition, that yet another superposition is necessary to explain the Eu abundances in r-I and r-II stars, and we concluded that the limited-r or weak r-process events of category I/II have to contribute as well. In Tables~\ref{eu_amount} and \ref{sr_amount} we attempted to quantify this contribution. While both category IIIa and IIIb events are expected to have either no co-production of Fe at all or a negligible co-production, nevertheless, we see a "scattered" relation between Eu and Fe in r-I and r-II stars (see Fig.~\ref{EuFerIrII}). The "average ratios" for Eu/Fe of $3\times 10^{-7}$ and $3\times 10^{-6}$ determined from the sample of r-I and r-II stars, correspond to a mean [Eu/Fe] value for these subregimes.
Let us see how such ratios could be obtained e.g. in collapsars/hypernovae or how such values could be obtained with the Eu input from neutron star mergers.
We take these two cases: (i) collapsar input from \citet{Siegel.Barnes.Metzger:2019}, (ii) the average Eu ejecta from neutron star mergers \citet{Cote.Belczynski.ea:2017}. They are not necessarily accompanied by ejected Fe. We discussed before whether neutron star kicks avoid that the merger ejecta are mixed with the preceding supernova remnants (containing Fe) or the merger takes place within the remnant material, polluted by the Fe ejecta from the two preceding core-collapse supernovae (producing the two neutron stars) of 0.1 M$_\odot$ \citep{Ebinger.Curtis.ea:2020,Curtis.ea:2019}. 

\noindent The collapsar models of \citet{Siegel.Barnes.Metzger:2019} predict about $10^{-1}$M$_\odot$ of r-process matter and typically about 0.5M$_\odot$ of $^{56}$Ni (decaying to Fe). This leads to about $10^{-4}$M$_\odot$ of Eu and an Eu/Fe mass fraction ratio of $2 \times 10^{-4}$ and an abundance ratio of $7.4 \times 10^{-5}$. For neutron star mergers the typical Eu ejecta mass is $10^{-5}$M$_\odot$ \citep{Cote.Belczynski.ea:2017}. If we would assume that this Eu is ejected into the interstellar matter polluted with the preceding two supernovae, it would amount to about 0.2M$_\odot$ of Fe. In the opposite case the ISM, into which the Eu is ejected, would not be polluted by the Fe of the prior supernovae from the preceding binary system, but it could contain Fe from other independent prior supernova events. Let us, for the moment, just assume the mixing with 0.1M$_\odot$ of Fe from one supernova, independent of its origin, just to get an idea of the Eu/Fe ratio resulting from such an assumption. This leads to Eu/Fe of $10^{-4}$ or an abundance ratio of $3.7 \times 10^{-5}$. If we follow the previous discussion that r-II stars are related to an actinide boost and an actinide boost is related to black hole torus systems, then the r-II observations should be compared to the collapsar case, which leaves the neutron star mergers to be related to r-I stars (but keeping in mind that according to Table~\ref{eu_amount} category II events, i.e. core-collapse with Fe co-production contribute as well, and are more frequent than compact binary mergers).

\begin{table*}[h!]
\caption{Candidates for category IIIa and IIIb events for complete r-process stars of regime 3}
\label{EuFeratios}
\centering
\begin{tabular}{ccccc}
\hline\hline
Element ratio & Obs. average r-I & Obs. average r-II & neutron star mergers (when including Fe from one SN) & collapsars \\
\hline
\hline
Eu/Fe & $3 \times 10^{-7}$ & $3 \times 10^{-6}$ & $3.7 \times 10^{-5}$ & $7.4 \times 10^{-5}$\\
\hline
\hline
\end{tabular}
\end{table*}

\noindent If we take the resulting ratios at face value, the observed ratio of Eu/Fe is 25 times smaller than in the produced ejecta for the collapsar case and 124 times smaller than in the ejecta (if including Fe from one supernova) for the neutron star merger case. Thus, one would need to add additional Fe from other preceding core-collapse supernovae. This would require n $\times$ 0.1 M$_\odot$ of Fe from additional supernovae to obtain
(0.5+ n 0.1) = 25 $\times$ 0.5 for the collapsar case and (0.1 + n 0.1) = 124 $\times$ 0.1 for the neutron star merger case. This results in $n_{coll}$=120 and $n_{NSM}$=123 times 0.1M$_\odot$ of additional Fe from preceding core-collapse supernovae in order to arrive at the observed ratios. This would require the Fe of about 120 additional supernovae. 

\noindent Turning it around means that on average the ISM out of which r-I as well as r-II stars formed experienced 1 category IIIa or IIIb event combined with about 120 core-collapse supernova events, producing each 0.1M$_\odot$ of Fe. This stands for about 8 per mil and is nicely consistent with our previous considerations related to Fig.~\ref{jinadata-Eu-Fe-Cor-Sim}. On the one hand the large scatter in the observed ratios of the order 10 to 20, indicates the inhomogeneity of matter in the early galaxy. Nevertheless, the result also shows that at the low metallicities considered here, we apparently do see the imprint of about one category III event mixed into an ISM containing already the pollution of about 123 core-collapse supernovae. When looking at Fig.2 from \citet{Rosswog.Feindt.ea:2017}, the ratio $n_{NSM}$=123 is consistent with the requirements for dominant r-process producing events that eject about $10^{-2}$M$_\odot$ of r-process matter per event. On other hand, the ratio of $n_{coll}$=120 is somewhat on the low side for events which eject about 0.5M$_\odot$ of r-process matter per event. A number of about 500 would be rather required, which might be marginally consistent with observations when considering the uncertainties of Fe and r-process predictions in collapsars. But we should notice that Fig.2 from \citet{Rosswog.Feindt.ea:2017} looks at the overall statistics in order to explain solar system r-process abundances. We look here at low metallicity stars, i.e. the early inhomogeneous evolution of galaxies, when only the most massive stars have contributed, yet, twisting the ratio to the fast evolution of the most massive stars, i.e. collapsar progenitors. The attempt of identifying alternatively collapsars with r-I stars and neutron star mergers with r-II stars would lead to numbers which do not necessarily exclude these interpretations, but the taken choice looks more reasonable.

\section{Conclusions}
\label{concl}
We have utilized statistical correlation methods ({ see appendices \ref{statistics}, \ref{appsinglemultible}, and \ref{kmeans}}) for r-process elemental abundance patterns in low metallicity stars with [Fe/H]$<$-2.5.
{ The initial analysis was based on the assumption that at such low metallicities one sees only the imprint of a single  nucleosynthesis event. This expectation is based on the estimate
that a typical core-collapse supernova ejects 0.1M$_\odot$ of Fe 
which subsequently mixes with a few times $10^4$M$_\odot$ of pristine ISM, so that the remnant has a [Fe/H]-value close to -2.7
 \citep[e.g.][]{Ryan.Norris.Beers:1996}.
 
This sets the stage for the first generation of stars that contains Fe and the abundance
ratios one would observe are specific to the involved explosion. If at early stages
already several explosions have enriched the ISM locally, but the ISM is not yet fully
mixed, one would see "pockets" of stellar abundance patterns. Analyzing these pockets separately, also permits to interpret correlations of two elements as co-production.
The situation is more complicated when already several explosive events have
contributed to a single newborn star, say, several core-collapse supernovae
have enriched the gas, but also an additional r-process event. In such situations
the relation between abundances and their ranks, described in  appendix \ref{appsinglemultible}, can provide information on the elemental properties of contributing sites. The analysis of observed abundance patterns by means of such
statistical tools has lead us to suggest four to five different r-process sites.}
In a slightly speculative, but promising, manner we identify these with 
\begin{enumerate}
\item Category 0 events are regular core-collapse supernovae. They can contribute, with a possible combination of a very weak r-process and a $\nu$p-process, to trans-Fe elements (possibly not much beyond Sr, Y, and Zr, in any case ending before the second r-process peak). [This follows e.g. from an analysis of Fig.~\ref{4starsplot} (right panel) in comparison to Fig.~\ref{corr_Sr_Y_Zr_Fe}. In the first case we see for Eu high or relatively high correlations with Fe for [Eu/Fe] values in the range between the stars HD122563 and HD115444, which indicates a co-production of Fe and Eu in specific/special core-collapse supernova events where HD122563 marks the lowest observed Eu/Fe ratio. Opposite to this behavior one can notice in Fig.~\ref{corr_Sr_Y_Zr_Fe} Sr/Fe or Zr/Fe ratios with a high correlation to Fe even below HD122563, indicating even weaker r-process or alternative events which produce light trans-Fe elements. We identify them with regular core-collapse supernovae, i.e. category 0 events.]
\item Category I events always co-produce Eu together with Fe in an apparently unique way, but under weak r-process conditions. Our initial idea was that EC-supernovae would be a good candidate, although their occurrence has been put in doubt (but not definitely ruled out) with recent electron-capture data on $^{20}$Ne. However, they have problems in producing sufficient amounts of Eu for realistic $Y_e$-values. An alternative site could be QD-supernovae, which (if the chosen equation of state properties are realistic) derive from a narrow initial mass region of massive stars and will thus lead to a very narrow range of explosion conditions. [This would be consistent with the observed tight correlation between Eu and Fe, see Figs.~\ref{eu_vers_rank_2}(a), \ref{eu_vers_fe_fit} (left panel) and the discussions related to Table~\ref{SrEuFeratios}.]
\item Category II events stand for another weak or limited r-process site. A strong candidate are magneto-rotational MHD supernovae which, dependent on pre-collapse rotation and magnetic fields, can produce quite varying r-process abundances, but with average Sr/Eu ratios consistent with weak r-process patterns. [This would be
in line with the observed large scatter in Fig.~\ref{eu_vers_fe_fit} (right panel)
and the rank relation in Fig.~Figs.~\ref{eu_vers_rank_2}(b). Both, category I and category II, events are dominantly responsible for limited-r or r-poor stars in regime 1 with [Eu/Fe]$<$-0.3 and regime 2 with [Eu/Fe]$<$0 (see Fig.~\ref{4starsplot}, right panel and the high Sr/Fe ratios in Fig.~\ref{fig:fullEuFe_SrEuFe}, right panel)].
\item Category III events consist of strong r-process sites which produce all heavy elements up to the third r-process peak and the actinides. Neutron star mergers belong surely to this category. Further options are neutron star - black hole mergers. Also, there exists evidence that at low metallicities a site related to massive single stars contributes as well. We categorize the compact binary options as IIIa and the massive collapsing star option as IIIb. A further question exists how actinide boost stars fit into this scheme. Our suggestion is that those sites which lead quickly to a black hole accretion disk system, and are not too affected by neutrino winds (increasing $Y_e$), are the best candidates for actinide boost events which can be found dominantly in observations of r-II stars.
[The motivation behind this suggestion is that these events have a good physical reason to preferentially produce matter in the required electron fraction range of
$Y_e\approx 0.10...0.15$, see the discussion in Sec.~\ref{sec:Cat_III}.
Such events include collapsars/hypernovae, neutron star - black hole mergers, and also neutron star mergers which lead to an early black hole formation, rather than a stable or meta-stable massive neutron star. The regular neutron star mergers (category IIIa) seem to be dominantly related to r-I-stars, while r-II stars are dominantly related to category IIIb stars, i.e. systems with black hole accretion disk tori (see Figs.~\ref{Eu_vers_Th}, \ref{ranksThrIrII}, and Table~\ref{th_eu_ratio}.]
\end{enumerate}
\begin{figure*}[h]
\centering
\includegraphics[width=18cm]{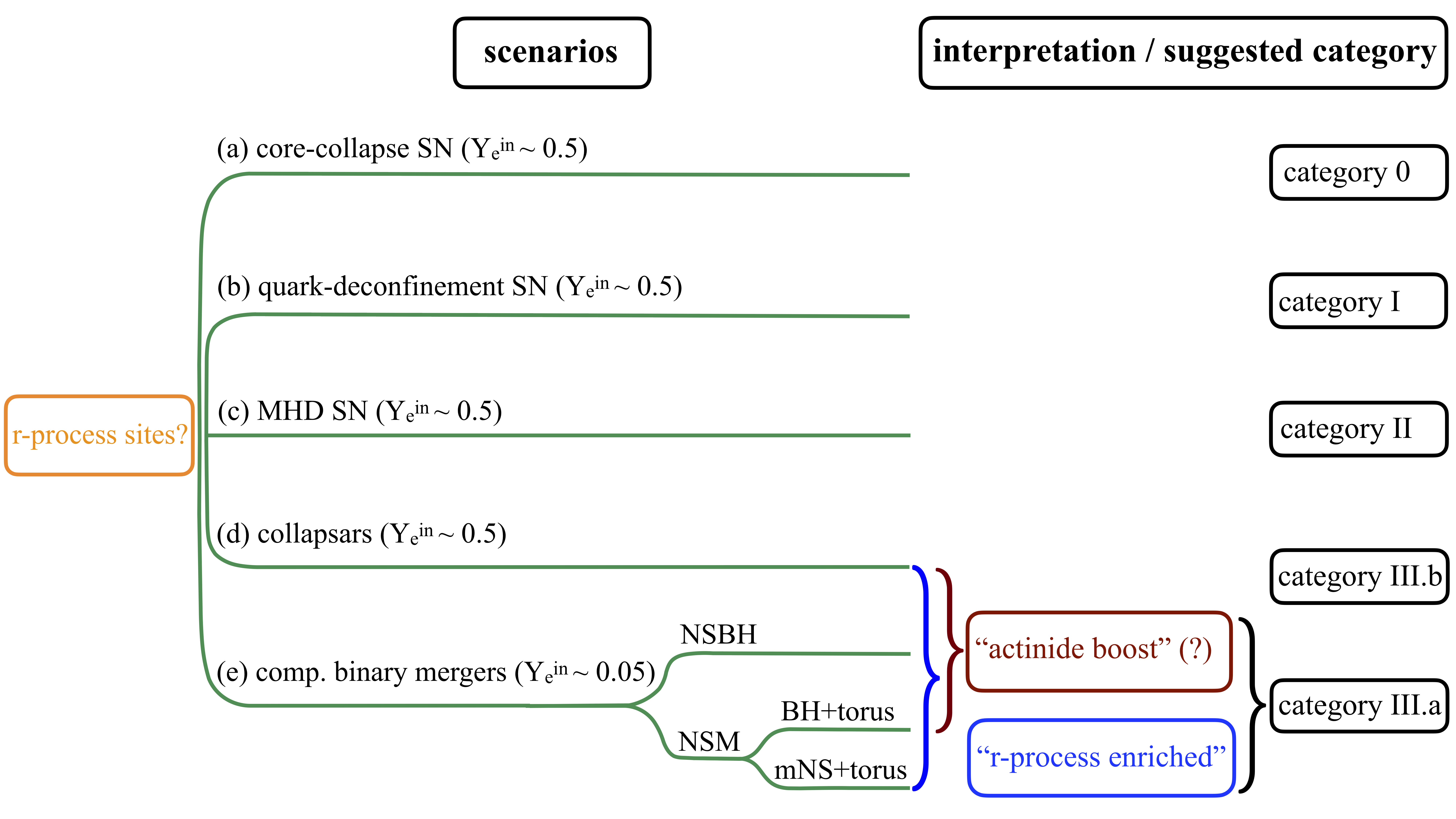}
\caption{Suggested event categories which are responsible for the observed limited-r (regime 1 and 2) and r-process enriched (regime 3 with subregimes r-I and r-II) stars. There is a close connection to the event categories I, II, and IIIa and IIIb, although not a one-to-one connection. This is due to the fact that the observational regimes are determined by [Eu/Fe] intervals ($<$-0.3, $<$0, $<$1, and $>$1, which are not only determined by the ejecta of event categories, but also due to inhomogeneous ejecta mixing with the ISM in the early Galaxy. While the actinides (e.g. Th) are only found in r-I and r-II stars and only produced in category IIIa and IIIb events, lanthanides / rare earth elements have also various contributions from category I and II events. Among the compact binary mergers all systems which contain initially a black hole or lead to fast black hole formation with BH accretion tori fall into category IIIb, while those which contain for a long duration a stable or meta-stable massive neutron star fall into category IIIa. The initial $Y_e^{\rm in}$ (before collapse or merger) is indicated as well, but will be altered due to weak interactions during the explosive events.}
\label{fig:interpretation} 
\end{figure*}
\noindent We combine all items, given above, in a display of all suggested event categories responsible for an understanding of the observed abundance patterns in low metallicity stars (Fig.~\ref{fig:interpretation}). This also includes as category 0 regular core-collapse supernovae which only contribute Fe and trans-Fe element events but no heavy r-process elements.
Besides this attempt to identify the four to five categories 0, I, II, and IIIa/b of stellar explosive events which contribute to weak and strong r-process nucleosynthesis, we found further constraints for these events:
\begin{itemize}
\item We found two good reasons to move the division between limited-r and r-enriched stars (in our terminology between regime 2 and 3) down from [Eu/Fe]=0.3 to 0. [This is based (a) on the existence of Th in stars with [Eu/Fe]$>$0 (see Fig.~\ref{figThFeH}, left panel) and (b) the best reproduction of the Eu vs. Fe correlation curve in Fig.~\ref{jinadata-Eu-Fe-Cor-Sim}.]
\item We found also that the category III strong r-process events { make up in number} for about 6 per mil of all core-collapse supernovae. [This is consistent with findings for the frequency of binary merger events in comparison to CCSNe), based on Fig.~\ref{jinadata-Eu-Fe-Cor-Sim}. A similar result was obtained from Table~\ref{EuFeratios}, hinting at the fact that r-I stars are dominated by neutron star merger events, but the ISM out of which the stars formed had been previously already polluted by $\approx$123 core-collapse category 0 events, i.e. mergers would amount to an 8 per mil contribution.]
\item In a similar way we analyzed the Eu and Fe contributions to r-II stars and their relation to collapsars, standing for category IIIb. [Here we found a similar result of about 120 core-collapse category 0 contributions to each collapsar event (see also Table~\ref{EuFeratios}), which is a bit on the low side to explain solar system abundances if such hypernovae eject typically $10^{-1}$M$_\odot$ of r-process matter or $10^{-4}$ M$_\odot$ of Eu \citep[see][]{Rosswog.Feindt.ea:2017}. However, the observed r-II stars at very low metallicites do not yet sample a full statistical spectrum, in fact the ratio is probably tilted towards the most massive contributors in a still inhomogeneous ISM, i.e. collapsars.]
\item Finally we found constraints for the limited-r stars of our regimes 1 and 2 and the connection to category I and II contributions. [If one identifies category I events with QD supernovae, which should occur in a very narrow mass range, we find that their ejecta had to mix with those of about 275 prior supernovae in order to explain the very linear, i.e. high, correlations of Eu vs. Fe in regime 1. If one identifies category II events with highly fluctuating MHD supernova contributions a mix with as low as 10 regular category 0 supernovae could be realistic, which would be consistent with the fraction of magnetars resulting from core-collapse events (see Table~\ref{SrEuFeratios}).]
\end{itemize}

\begin{figure*}[h]
\centering
\includegraphics[width=20cm]{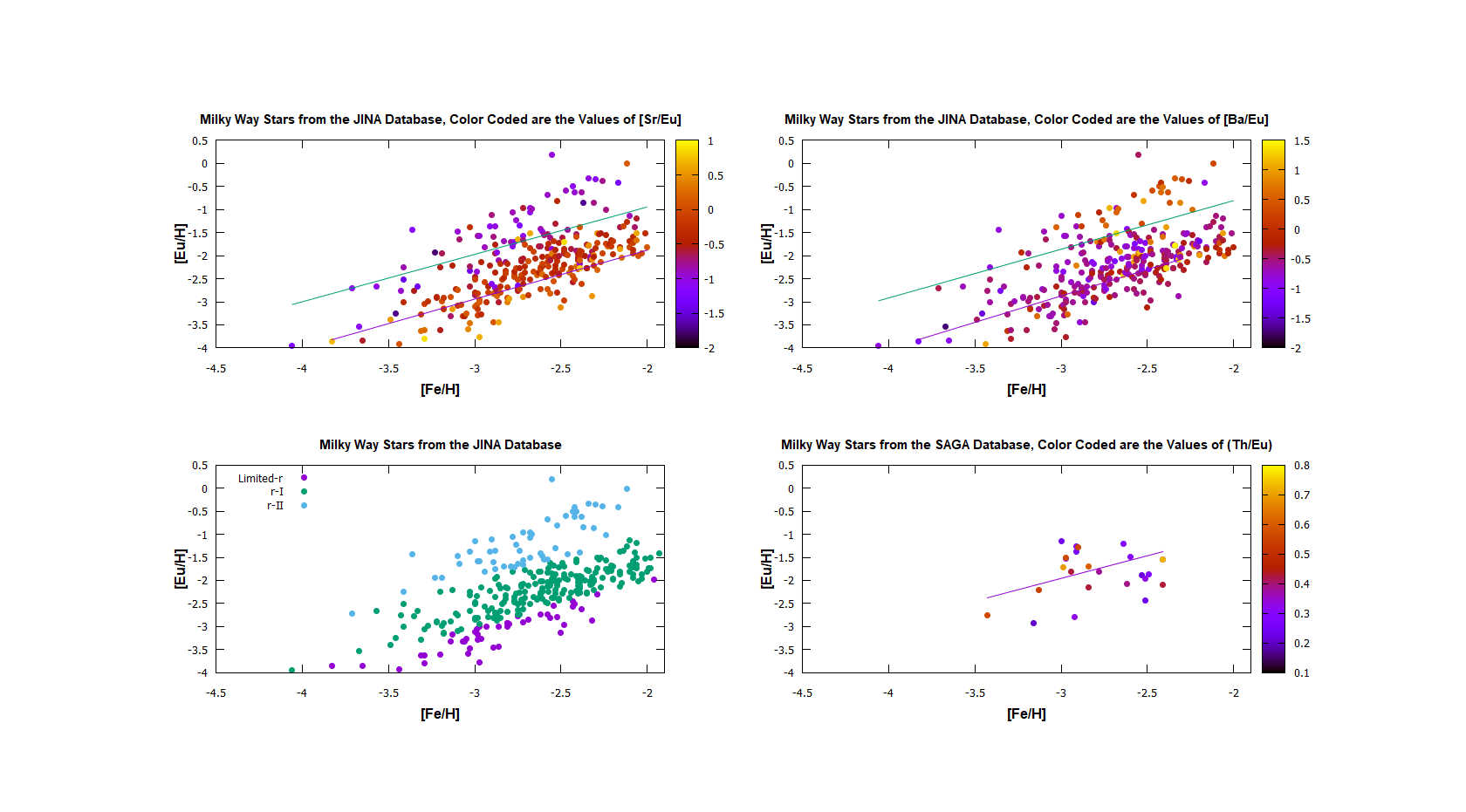}
\caption{{ Properties of low metallicity limited-r, r-I, and r-II stars with [Fe/H]$<$-2 with respect to [Eu/Fe] vs. [Fe/H]: Bottom left the division into the three subregimes, top left color-coded with [Sr/Eu], bottom right color-coded with [Th/Eu] (for those stars with detected Th), and top right color-coded with respect to [Ba/Eu].}}
\label{reffig} 
\end{figure*}

\noindent 
We want to stress that the above interpretations are "suggestions" and should be understood as "guided speculations", based on a variety of strong hints from statistical data mining of the observed abundance patterns of low metallicity stars. They should provide the motivation for further thorough investigations and maybe approval or contradiction. But they are an attempt to bring some structure in the present understanding of the role of different r-process sites with the help of { statistical methods and observed correlation patterns}. One of the major results which became clear in our investigations is the following: our initial assumption that low metallicity stars with metallicities [Fe/H]$<$-2.5 have been polluted by only one or at most very few events might be valid for the r-process contributions, but many prior regular CCSNe of category 0 must have contributed earlier, starting { possibly} already at metallicities as low as [Fe/H]$<$-5 with their Fe and some trans-Fe elements like Sr, Y, Zr and possibly beyond. We could make { existing nucleosynthesis model predictions} only consistent with observations, if the interstellar medium, which led to the formation of stars with observed r-process elements, experienced already many Fe contributing CCSNe. These include category 0 but also category I and II events. The latter (category I/II events) are responsible for a sizable fraction of Eu (lanthanide) contributions to r-process enriched (regime 3) stars, although those are dominated by the ejecta of category III events. On the other hand, we found results consistent with earlier findings that category III events have an occurrence rate on the per mil level in comparison to regular CCSNe.

\noindent { This analysis can be concluded with another interesting graph, supporting our interpretations from Fig.\ref{fig:interpretation}. It was our ambition from the beginning how to explain the huge inhomogeneities in abundance patterns at low metallicities ([Fe/H]$<$-2.5), where there exists some chance to interpret abundance patterns of r-process contributions as resulting from individual events. As discussed above, our present understanding is that this still applies to some extent to r-process elements (but probably not for Fe). In Fig. \ref{reffig} we include [Eu/Fe] vs. [Fe/H] plots for low metallicity stars from the SAGA database in the range [Fe/H]$<$-2, separating by lines the limited-r, r-I, and r-II stars. The bottom left plot shows only the separation into the three categories of stars, with the side effect that apparently the metallicity at the time of first occurrence seems to delay in steps from limited-r to r-I and r-II, but still leaves open whether these regimes are really related to different nucleosynthesis contributing sources. The limited-r stars, although showing the lowest (and thus hardest to detect) Eu content, extend to the lowest metallicities, while the higher Eu content baring r-I and r-II stars (which would be easier to detect at lowest metallicities) come in in a delayed fashion. When looking at the Sr/Eu color-coded figure (top left), it indicates that especially the limited-r stars show a highly supersolar Sr/Eu ratio, while the r-II stars show a strongly sub-solar ratio and the r-I stars a close to solar (or moderately subsolar) ratio, permitting the later s-process contributions to Sr at higher metallicities to approach the solar ratio. This quite extremely different behavior (also shown in Fig. \ref{fig:fullEuFe_SrEuFe}, right panel, of this paper) points to apparently different stellar nucleosynthesis sources (and not only a formal division in Eu/Fe categories). The bottom right insert displays color-coded Th/Eu ratios in those stars where Th was detected. The supersolar (or actinide-boost) behavior is strongly prevailing in the r-II stars and the upper end of the r-I stars. This supports our interpretation, based on Table \ref{th_eu_ratio}, Figs.\ref{Eu_vers_Th} and \ref{ranksThrIrII}, and the comparison analysis of different model sources given in section \ref{sec:6.3.1} and \ref{sec:Cat_III}. We also added a figure with a color-coding of Ba/Eu (top right). As Eu is essentially originating from the r-process while Ba has a strong s-process contribution, solar Ba/Eu contains an s+r/r comparison. In our figure we see especially above -2.5 an apparently strong s-process contribution, but also to some extend at lower metallicities. This might be due to primary s-process sources of fast rotating and very massive “spin stars” which due to rotational mixing produce primary $^{14}$N and $^{22}$Ne \citep{Frischknecht.Hirschi.ea:2012,Cescutti.Chiappini.ea:2013,Frischknecht.Hirschi.ea:2016}, and are ideal candidates for collapsars, further supporting our interpretation of r-II stars. (The transition of r-process to the dominant s-process contributions from low and intermediate mass stars seems to occur only around [Fe/H]=-1 to -1.5 from the figures in \citet{Battistini.Bensby:2016}).}

\appendix

\section{The Pearson and Spearman Correlation Coefficients}
\label{statistics}

We summarize here briefly concepts that we are applying frequently throughout this paper. None
of these  concepts are our original work, for more information we refer to statistics text books \citep{tamhane00,spiegelhalter19}.

\subsection{The Pearson correlation coefficient}
Assume that we have two random variables $X$ and $Y$ that can have discrete 
values $\left\{x_i\right\}$ and $\left\{ y_j \right\}$. If the value $x_i$ is found with probability 
$p_i$, the {\em expectation value} is
\begin{equation}
\mu_X \equiv \sum_i p_i x_i \equiv E(X).
\label{eq:mean}
\end{equation}
The fluctuations around the mean value are characterized by the {\em variance}
\begin{equation}
\sigma_X^2 \equiv \sum_i p_i (x_i - \mu_X)^2 \equiv VAR(X).
\label{eq:var}
\end{equation}
Obviously, $VAR(X)$ carries the dimension of $X^2$ and it may be more convenient to
deal with a quantity of the same dimension as $X$ which leads to the {\em standard deviation}
\begin{equation}
\sigma_X = \sqrt{\sum_i p_i (x_i - \mu_X)^2}.
\end{equation}
The notion of variance can straight forwardly be generalized to the {\em covariance} of two variables $X$ and $Y$
\begin{equation}
\sigma_{XY} \equiv \sum_i p_{XY,i} \; (x_i - \mu_X) (y_i - \mu_Y) \equiv \rm{COV(X,Y)},
\label{eq:def_COV}
\end{equation}
where $p_{XY,i}$ is the probability that $x_i$ and $y_i$ occur together. A positive value indicates 
the tendency of one variable to increase when the other variable does so, a negative value indicates 
that one variable decreases when the other increases, therefore the name "co-variance".
Obviously, $VAR(X)= COV(X,X)$. From its definition, Eq.~(\ref{eq:def_COV}), one can straight forwardly show that
\begin{equation}
COV(X,Y)= E(XY) - \mu_X \mu_Y
\end{equation}
and this implies that the covariance vanishes for independent variables, i.e. for the case that 
 $E(XY)= E(X) E(Y)$.
Since the covariance vanishes when the variables are independent of each other, it obviously describes
 how they are {\em correlated}. The dimension of $COV(X,Y)$ is the same as the one 
of $XY$ and one may prefer to work with a dimensionless quantity. One can straight-forwardly obtain a dimensionless
version of the covariance by normalizing it with $\sigma_X$ and $\sigma_Y$
\begin{equation}
\begin{split}
\mathcal{P}_{XY} & \equiv \frac{COV(X,Y)}{\sigma_X \; \sigma_Y}= \frac{\sigma_{XY}}{\sigma_X \; \sigma_Y} \\
 & = \frac{\sum_i p_i (x_i - \mu_X)(y_i-\mu_Y)}{\sqrt{\sum_j p_j (x_j - \mu_X)^2}  \sqrt{\sum_k p_k (y_k - \mu_Y)^2}},
\end{split}
\label{eq:def_dimless_cov}
\end{equation}
which, for the case of equal probabilities, becomes the {\em Pearson correlation coefficient} (PCC)
 \begin{equation}
r_{XY} \equiv 
\frac{\sum_i (x_i - \mu_X)(y_i-\mu_Y)}{\sqrt{\sum_j (x_j - \mu_X)^2}  \sqrt{\sum_k  (y_k - \mu_Y)^2}}.
\label{eq:Pearson}
\end{equation}
Apart from de-dimensionalizing the covariance, the division by $\sigma_X \; \sigma_Y$ in Eq.~(\ref{eq:def_dimless_cov})
has the additional consequence of restricting the range of the Pearson correlation coefficient to values $-1 \le r \le 1$. \\
Some commonly used guidelines for the interpretation of numerical values of $r_{XY}$ are summarized in Table~\ref{tab:pcc}.
  \begin{table}[h!]
      \caption{Guidelines for interpreting the Pearson Correlation Coefficient  $r$}
         \label{tab:pcc}
     $$ 
         \begin{array}{p{0.5\linewidth}l}
            \hline
            \noalign{\smallskip}
            Strength of Association      &  \mbox{PCC} \\
            \noalign{\smallskip}
            \hline
            \noalign{\smallskip}
            Very weak & 0.0 \le |r| <  0.2     \\
            Weak           & 0.2 \le |r| <  0.4 \\
            Moderate & 0.4 \le |r| <  0.6            \\
           Strong & 0.6 \le |r| <  0.8           \\
           Very strong & 0.8 \le |r| \le 1.0            \\
            \noalign{\smallskip}
            \hline
         \end{array}
     $$ 
   \end{table}
It is worth stressing under which conditions the PCC should be applied:
a) the data sets should approximate  normal distributions, b) the errors should be similar  for different 
values of the independent variable  (so-called "homoscedascity"), c) the data should be related linearly
and d) it should be continuous within the considered interval and e) no outliers should be present in the 
data set. In general, a data point that is more than $\pm$ 3.29 standard deviations away, is considered 
an outlier. Scatter plots are a reasonable first step to inspect how well the assumptions are fulfilled.

%
%
\subsection{Linear regression and the coefficient of determination}
\label{sec:lin_reg}
There is a close relation to "linear regression", i.e. to fitting the data by a straight line, $\tilde{y}(x)= ax+b$,   that is optimal in a least squares sense.
If one adopts as the error measure 
\be
\chi^2 \equiv \sum_i e_i^2 = \sum_i \left[ y_i - \tilde{y}(x_i)\right]^2,
\ee
where the $e_i$ are the residuals between the data, $y_i$, and the linear model values  $\tilde{y}(x_i)$, and
 minimizes it with respect to $a$ and $b$, $\partial \chi^2/\partial a= 0 = \partial \chi^2/\partial b$, one finds
\be
a= \frac{COV(X,Y)}{VAR(X)} \quad {\rm and} \quad b= \mu_y -  a \mu_x,
\ee
i.e. the least squares estimators are simple functions of the means, variances and covariances. Straight forward
algebra shows that the slope can also be expressed in terms of the Pearson correlation coefficient $r_{XY}$ as
\be
a= r_{XY} \sqrt{\frac{VAR(Y)}{VAR(X)}},
\ee
which leads to the interpretation of the square of the correlation coefficient. If one uses the properties of the variance and covariance,
one finds that the variance of the deviation of the linear model from the data is
\be
\begin{split}
VAR(e) & =  VAR(y - (a x + b))= (1 - r_{XY}^2) VAR(Y)\\
&           = VAR(Y) - r_{XY}^2 VAR(Y),
\end{split}
\ee
or, turned around, we have
\be
r_{XY}^2= 1 - \frac{VAR(e)}{VAR(Y)}.
\ee
In other words, the square of the Pearson correlation coefficient measures which proportion of the total variance
of $Y$ can be explained by the linear regression. The quantity $r_{XY}^2$ is often called the "coefficient of determination". For a correlation 
coefficient of $r_{XY}= \pm1$, the variance of the residuals vanishes. i.e. the linear model is a perfect fit to the data. 

\subsection{The Spearman correlation coefficient}
At the end of section A1 we had summarized the conditions under which  Pearson's correlation coefficient should be
applied. If one of the conditions should be violated, in particular if the relation is non-linear, one can still apply the 
Spearman rank correlation coefficient (SCC). The idea is to rank the variables (i.e. sort them according to size and assign
them the integer of their position in the sorted list; use (non-integer) averages in the case of ties) and  apply 
Pearson's formula, Eq. (\ref{eq:Pearson}), to the {\em ranks} $\mathcal{R}(x_i)$ and $\mathcal{R}(y_i)$ (rather than the data values)
 \be
\rho_{XY}\equiv
\frac{\sum_i (\mathcal{R}(x_i) - \mu_{\mathcal{R}(X)})(\mathcal{R}(y_i)-\mu_{\mathcal{R}(Y)})}{\sqrt{\sum_j (\mathcal{R}(x_j) - \mu_{\mathcal{R}(X)})^2}  \sqrt{\sum_k  (\mathcal{R}(y_k) - \mu_{\mathcal{R}(Y)})^2}}.
\label{eq:Spearman}
\ee
For a perfectly monotonic relation  one obtains a value of +1 (highest value of $X$ is associated with highest value of $Y$ etc. down to lowest values)
or -1 (highest $X$-value associated with lowest $Y$-value etc.). If high  $X$ values are not preferentially related to high (or low) $Y$-values and instead
the ranks are not correlated, the contributions approximately cancel and one obtains a value $\rho_{XY}\approx 0$. If $\mathcal{R}(x_i)$ and $\mathcal{R}(y_i)$ 
are integers, the Spearman rank correlation can also  conveniently be expressed as
\be
\rho_{XY} = 1 - \frac{6 \sum_{i=1}^N \left[\mathcal{R}(x_i) - \mathcal{R}(x_i)\right]^2}{N(N^2-1)}.
\ee
Apart from requiring fewer assumptions, the major difference to the PCC  is that the  
SCC measures the strength of a {\em monotonic} association
while  the PCC measures the strength of a {\em linear} association.

\section{Relationship between a Variable and its Rank}
\label{appsinglemultible}
\begin{figure*}[h]
\centering
\includegraphics[width=12cm,height=8cm]{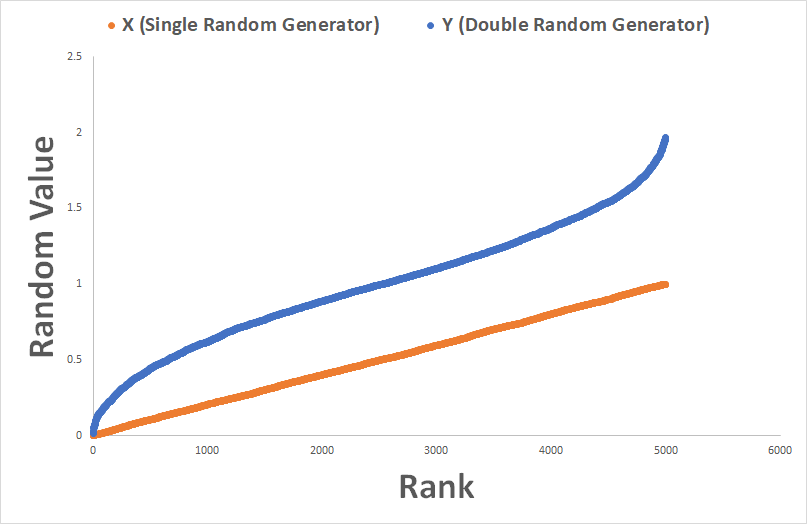}
\caption{Randomly generated variables $X$ and $Y$. $X$ is generated by a single random generator in the interval [0,1], $Y=X_1+X_2$ by two random generators. Rank 1 indicates the smallest abundance in the whole data. We see a non-linear behavior for $Y$, especially for the lowest and highest ranks.}
\label{random}%
\end{figure*}

We can order continuous variables according to either rising or decreasing values. If we list them in such an ordered sequence and define the integer number in the list as its rank, this variable can be plotted as a function of its rank.
If one throws a dice for a sufficient number of times (i.e. sufficient statistics), one will find all numbers from 1 to 6 resulting for an equal amount of times, and plotted as a function of their ranks, a step function will result (in this case of an integer random variable). Switching over to a random number generator for continuous variables, a linear relationship results between the variable and its rank, which is identical with the averaged straight line through the step function for an integer variable.\\
In an astrophysical environment of stars being formed from the local interstellar medium (ISM), the pollution with a single element $X$ from an (explosive) astrophysical site depends on the question how much the ejecta of this event were mixed into the gas cloud from which the star was formed. We can think of the extreme case of no pollution in the early galaxy
if no such nearby event took place, yet, and otherwise of a strong pollution, if the proto-stellar cloud was fully mixed with the ejecta of such an event. The amount of the contribution of element X from one type of event can thus be represented by a random number generator from zero to a maximum, which would be observed as a linear relation between abundance and the related rank. Therefore, 
a linear relationship between an abundance $X$ and its rank is expected if one type of an astrophysical source contributes. \\
If another astrophysical source contributes to the same element X, this can be represented by a random number generator as well for the amount of element $X$ mixed into the
protostellar cloud by this second type of event. If we add these contributions for each random event of type A and type B, we can - for the extremes - have cases from negligible contributions of both types of events to maximum contributions from both types of events, and in general also many cases with varying and different contributions from each event type. When ordering these summed abundances according to their ranks, a different pattern appears, and the linear relationship between the values and their ranks is destroyed. We show this behavior in Fig.~\ref{random} with two random variables $X$ and $Y$, where $X$ is generated by only one and $Y=X_1+X_2$ by two random number generators. 

As discussed above, the relationship between $X$ and its rank is perfectly linear, whereas $Y$, a superposition of two random variables does not show this linearity. In the case displayed in Fig.~\ref{random} both variables lie in the interval [0,1], if the variables in the superposition are of different size (or even a further source is added), a deviation from a straight line will remain, but it can experience further distortions.\\

\section{Cluster Analysis}
\label{kmeans}
Cluster analysis, or "clustering" for short, is a method of unsupervised learning 
where the aim is to identify patterns in data without any predetermination.
It is frequently used for statistical data analysis in a variety of fields \citep[see e.g.][]{Everitt.Landau:2011}. It allows  to identify groups of similar objects ("clusters") that are more related to each other than to objects in other groups. Clustering algorithms can also provide valuable insights for the interpretation of our data by identifying groups in our data points.\\
Several dozens of clustering algorithms have been published, they can generally be divided into the following categories:
\begin{itemize}
    \item {\em{connectivity-based}} clustering: it is based on the idea of objects (with their parameters) being more related to objects with similar/close-by parameter values than to objects with more different parameter sets.
    \item {\em{centroid-based}} clustering:  a central vector represents the average properties of the whole cluster, which is with its parameters not necessarily identical to a single member of the data set.
    \item {\em{distribution-based}} clustering: clusters can be defined as objects belonging most likely to the same distribution.
    \item {\em{density-based}} clustering: clusters are defined as areas of higher density in the data space than surrounding regions of the data set.
    \item {\em{grid-based}} clustering: it is used for multi-dimensional data set.
\end{itemize}
In this paper we used the probably best-known clustering algorithm, $k$-means, which is  centroid-based. It has the advantage that it is algorithmically simple
and computationally fast.
$k$-means is often referred to as Lloyd’s algorithm and it contains three basic steps. One begins by choosing a predefined number of initial guesses for
the centroid positions and then iterates over the remaining two steps:
\begin{itemize}
    \item assign each sample to its nearest centroid
    \item update the centroid positions based on the mean value of the samples assigned to the previous centroid
\end{itemize}
until convergence.
The procedure is considered converged once the distances between two subsequent
updates drop below a predefined tolerance, i.e. once the centroid positions have stopped moving.
$k$-means has the advantage of being computationally efficient, as all necessary steps consist of computing the distances between points and group centers. But it has the disadvantage that one needs to preselect the number of cluster divisions in the data beforehand. This is not always trivial as ideally a clustering algorithm should also provide the appropriate number of clusters in order to gain insight from the dataset.

\begin{acknowledgements}
The investigations of this paper would not have been possible without the publicly available observational SAGA and JINA data bases \citep{Sagadatabase,JINAbase:2018}, and we want to express out thanks to the authors of these tremendous research resources. { We also thank the referee for his/her very careful reading of the manuscript, for spotting some logical glitches in the original version, and for suggesting a presentation which led to Fig.\ref{reffig}}. Furthermore, the COST actions ChETEC (Chemical Elements as Tracers of the Evolution of the Cosmos, CA16117), GWvese (Gravitational waves, black holes and fundamental physics, CA16104), and Pharos (The multi-messenger physics and astrophysics of neutron stars, CA16214) provided an inspiring atmosphere for thoughts along the lines discussed here.
SR has been supported by the 
Swedish Research Council (VR) under grant numbers 2016-03657\_3 and
2020-05044, by the Swedish National Space 
Board under grant number Dnr. 107/16, the research
environment grant “Gravitational Radiation and
Electromagnetic Astrophysical Transients (GREAT)” 
funded by the Swedish Research council (VR) under 
Dnr 2016-06012 and by the Knut and Alice Wallenberg Foundation (KAW 2019.0112). 

\end{acknowledgements}

\bibliography{paper}

\end{document}